\documentclass[12pt]{article}
\usepackage{amscd,amsmath,amssymb,amstext,amsthm,exscale,latexsym}
\usepackage{graphicx,floatflt}
\newcommand {\al}   {\alpha}       \newcommand {\bt}  {\beta}
\newcommand {\g }   {\gamma}       \newcommand {\G }  {\Gamma}
\newcommand {\dl}   {\delta}       \newcommand {\e }  {\epsilon}
        
\newcommand {\ve}   {\varepsilon}  
\newcommand {\lm}   {\lambda}      \newcommand {\m }  {\mu}
          
\newcommand {\s }   {\sigma}      
         
\newcommand {\vf }  {\varphi}      
         \newcommand {\om}  {\omega}
\newcommand {\Lm}   {\Lambda}      \newcommand {\Om}  {\Omega}
\newcommand {\Th}   {\Theta}       
\newcommand {\pl}   {\partial}     \newcommand {\nb}  {\nabla}
\renewcommand {\sin}{{\sf\,sin\,}}       \renewcommand {\cos}{{\sf\,cos\,}}
\renewcommand {\ln}{{\sf\,ln\,}}         
\renewcommand {\det}{{\sf\,det\,}}       
\renewcommand {\dim}{{\sf\,dim\,}}       
\newcommand   {\const}{{\sf\,const}}     \newcommand   {\diag}{{\sf\,diag\,}}
\newcommand {\ex}{{\sf\,e}}
\newcommand {\arctg}{{\sf\,arctg\,}}
\newcommand {\ML}  {{\mathbb L}}
\newcommand {\MM}  {{\mathbb M}}   \newcommand {\MG}  {{\mathbb G}}
\newcommand {\MO}  {{\mathbb O}}   \newcommand {\MR}  {{\mathbb R}}
\newcommand {\MS}  {{\mathbb S}}   \newcommand {\MP}  {{\mathbb P}}
\newcommand {\CC }  {{\cal C}}     
\newcommand {\Go}  {\mathfrak{o}}   
   
\newcommand {\Gs}  {\mathfrak{s}}   
\newcommand {\Br}  {\boldsymbol{r}}   \newcommand {\Bb}  {\boldsymbol{b}}
   \newcommand {\Sd}  {{\textsc{d}}}
\newcommand {\Sl}  {{\textsc{l}}}   \newcommand {\St}  {{\textsc{t}}}
\newtheorem{theorem}{Theorem}
\begin{document}
\title     {Geometric theory of defects}
\author    {M. O. Katanaev
            \thanks{E-mail: katanaev@mi.ras.ru}\\ \\
            \sl Steklov Mathematical Institute,\\
            \sl Gubkin St. 8 \\
            \sl Moscow, 119991, Russia}
\date{31 March 2005}
\maketitle
\begin{abstract}
A description of dislocations and disclinations defects in terms of
Riemann--Cartan geometry is given, with the curvature and torsion tensors
being interpreted as the surface densities of the Frank and Burgers vectors,
respectively. A new free energy expression describing the static distribution
of defects is presented, and equations of nonlinear elasticity theory are used
to specify the coordinate system. Application of the Lorentz gauge leads to
equations for the principal chiral $\MS\MO(3)$-field. In the defect-free
case, the geometric model reduces to elasticity theory for the displacement
vector field and to a principal chiral $\MS\MO(3)$-field model for the spin
structure. As illustrated by the example of a wedge dislocation, elasticity
theory reproduces only the linear approximation of the geometric theory of
defects. It is shown that the equations of asymmetric elasticity theory for
the Cosserat media can also be naturally incorporated into the geometric
theory as the gauge conditions. As an application of the theory, phonon
scattering on a wedge dislocation is considered. The energy spectrum of
impurity in the field of a wedge dislocation is also discussed.
\end{abstract}
\section{Introduction}
Many solids have a crystalline structure. However, ideal crystals are
absent in nature, and most of their physical properties, such as
plasticity, melting, growth, etc., are defined by defects of the crystalline
structure. Therefore, a study of defects is a topical scientific question of
importance for applications in the first place. A broad experimental and
theoretical investigations of defects in crystals started in the 1930s
and continues nowadays. At present, a fundamental theory of defects is
absent in spite of the existence of dozens of monographs
and thousands of articles.

One of the most promising approaches to the theory of defects is based on
Riemann--Cartan geometry, which involves nontrivial metric and torsion.
In this approach, a crystal is considered as a continuous elastic medium with
a spin structure. If the displacement vector field is a smooth function, then
there are only elastic stresses corresponding to diffeomorphisms of the
Euclidean space. If the displacement vector field has discontinuities, then
we are saying that there are defects in the elastic structure. Defects in
the elastic structure are called dislocations and lead to the appearance
of nontrivial geometry. Precisely, they correspond to a nonzero torsion tensor,
equal to the surface density of the Burgers vector.

The idea to relate torsion to dislocations appeared in the 1950s [1--4].
\nocite{Kondo52,Nye53,BiBuSm55,Kroner58}
This approach is still being successfully developed (note reviews [5--11]),
\nocite{SedBer67,Kleman80A,Kroner81,DzyVol88,KadEde83,KunKun86,Kleine89}
and is often called the gauge theory of dislocations. A similar approach is
also being developed in gravity \cite{HeMcMiNe95}. It is interesting to note that
E Cartan introduced torsion in geometry \cite{Cartan22} having the
analogy with mechanics of elastic media in mind.

The gauge approach to the theory of defects is being developed successfully,
and interesting results are being obtained in this way [14--17].
\nocite{Malysh00,Lazar00,Lazar02,Lazar03}
We note in this connection two respects in which the approach proposed
below is essentially different. In the gauge models of dislocations based
on the translational group or on the semidirect product of the rotational group
with translations, one usually chooses the distortion and displacement fields as
independent variables. It is always possible to fix the invariance under
local translations such that the displacement field becomes zero because it
transforms by simple translation under the action of the translational group.
In this sense, the displacement field is the gauge parameter of local
translations, and physical observables are independent of it in
gauge-invariant models.

The other disadvantage of the gauge approach is the equations of equilibrium.
Einstein type equations are usually considered for distortion or
vielbein, with the right hand side depending on the stress tensor. This
appears unacceptable from the physical point of view because of the following
reason. Consider, for example, one straight edge dislocation. In this case,
the elastic stress field differs from zero everywhere. Then the torsion
tensor (or curvature) is also nontrivial in the whole space due to the
equations of equilibrium. This is wrong from our point of view. Indeed, we
can consider an arbitrary domain of medium outside the cutting surface and
look at the creation process for an edge dislocation. The chosen domain was
a part of the Euclidean space with identically zero torsion and curvature
before the defect creation. It is clear that torsion and curvature remain
zero because the process of dislocation formation is a diffeomorphism for
the considered domain. In addition, the cutting surface may be chosen
arbitrary for the defect creation, leaving the dislocation axis unchanged.
Then it follows that torsion and curvature must be zero everywhere except
at the axis of dislocation. In other words, the elasticity stress tensor
can not be the source of dislocations. To avoid the apparent contradiction,
we propose a radical way out: we do not use the displacement field as an
independent variable at all. This does not mean that the displacement field
does not exist in real crystals. In the proposed approach, the displacement
field exists and can be computed in those regions of medium that do not
contain cores of dislocations. In this case, it satisfies the equations
of nonlinear elasticity theory.

The proposed geometric approach allows considering other defects that do not
relate directly to defects of elastic media.

The intensive investigations of other defects were conducted in parallel
with the study of dislocations. The point is that many solids have not only
elastic properties but also a spin structure. For example, there are
ferromagnets, liquid crystals, spin glasses. In this case, there are defects
in the spin structure which are called disclinations \cite{Frank58}. They
arise when the director field has discontinuities. The presence of
disclinations is also connected to nontrivial geometry. Namely, the curvature
tensor equals the surface density of the Frank vector. The gauge approach
based on the rotational group $\MS\MO(3)$ was also used for describing
disclinations \cite{DzyVol78}. $\MS\MO(3)$-gauge models of spin glasses
with defects were considered in \cite{Hertz78,RivDuf82}.

The geometric theory of static distribution of defects which describes both
types of defects -- dislocation and disclinations -- from a single standpoint
was proposed in \cite{KatVol92}. In contrast to other approaches, it involves
the vielbein and $\MS\MO(3)$ connection as the only independent variables.
The torsion and curvature tensors have direct physical meaning as the
surface densities of dislocations and disclinations, respectively.
Covariant equations of equilibrium for the vielbein and $\MS\MO(3)$ connection
similar to those in a gravity model with torsion are postulated. To define
the solution uniquely, we must fix the coordinate system (fix the gauge)
because any solution of the equations of equilibrium is defined up to
general coordinate transformations and local $\MS\MO(3)$ rotations.
The elastic gauge for the vielbein \cite{Katana03} and Lorentz gauge for the
$\MS\MO(3)$ connection \cite{Katana04} were proposed recently. We stress
that the notions of a displacement vector and rotational angle are completely
absent in our approach. These notions can be introduced only in those domains
where defects are absent. In this case, equations for vielbein and
$\MS\MO(3)$ connection are identically satisfied, the elastic gauge reduces
to the equations of nonlinear elasticity theory for the displacement vector,
and the Lorentz gauge leads to the equations for the principal chiral
$\MS\MO(3)$ field. In other words, to fix the coordinate system, we choose two
fundamental models: the elasticity theory and the principal chiral field model.

To show the advantages of the geometric approach and to compare it with
the elasticity theory, we consider in detail a wedge dislocation in the
frameworks of the elasticity theory and the proposed geometric model. We show
that the explicit  expression for the metric in the geometric approach is
simpler and coincides with the induced metric obtained within the elasticity
theory only for small relative deformations.

As an application of the geometric theory of defects, we consider two examples
in the last sections of the present review. First, we solve the problem of
phonon scattering on a wedge dislocation. The problem of phonon scattering
is reduced to the integration of equations for extremals for the
metric describing a wedge dislocation (because phonons move along extremals
in the eikonal approximation). As a second application, we consider the quantum
mechanical problem of impurity or vacancy motion inside a cylinder
whose axis coincides with a wedge dislocation. The wave functions and
energetic spectrum of the impurity are found explicitly.

The presence of defects results in nontrivial Riemann--Cartan geometry.
This means that for describing the phenomena that relate ingeniously to
elastic media, we must make changes in the corresponding equations. For
example, if a phonon propagation in an ideal crystal is described
by the wave equation, then the presence of defects is easily taken into
account. For this, the flat Euclidean metric has to be replaced by
a nontrivial metric describing the distribution of defects. The same
substitution must be made in the Schr\"odinger equation to describe other
quantum effects. It is shown nowadays that the presence of defects essentially
influences physical phenomena. The Schr\"odinger equation in the presence
of dislocations was considered in [25--46] for different problems.
\nocite{FurMor94,FudaMoBeBe94,Moraes95B,AzeMor98,BaJoMoMo98,FurMor99,BaScTu98,%
BaScTu99,FurMor00,AzeMor00A,AzePer00,FuBeMo00B,deFuMo01,VieAze01,Azeved01B,%
FuBeMo01,Azeved01B,deFuBeMo01,FudeAz02,Azeved02A,Azeved02B,Azeved02C,%
Azeved03}
Problems related to the wave or Laplace equations were considered in [47--54].
\nocite{DeBeBe99,AzFuMo98,Moraes95A,BeBeGr98,Azeved01A,FuBeMo00,Tanaka00,DeLMor02}
The influence of the nontrivial metric related to the presence of defects
was investigated in electrodynamics \cite{AzeMor00B} and hydrodynamics
\cite{MirMor03}. Scattering of phonons on straight parallel dislocations
was studied in [57--59].
\nocite{Moraes96,dePaMo98,KatVol99}

Another approach to the theory of defects based on affine geometry with
nonzero nonmetricity tensor was considered recently in \cite{MirRiv02}.
\section{Elastic deformations                          \label{seldef}}
We consider infinite three dimensional elastic media. We suppose that the
undeformed medium in the defect-free case is invariant under translations
and rotations in some coordinate system. Then, the medium in this coordinate
system $y^i$, $i=1,2,3$, is described by the Euclidean metric
$\dl_{ij}={\diag}(+++)$, and the system of coordinates is called Cartesian.
Thus, in the undeformed state, we have the Euclidean space $\MR^3$ with a given
Cartesian coordinate system. We also assume that torsion (see the Appendix)
in the medium equals zero.

Let a point of the medium has coordinates $y^i$ in the ground state. After
deformation, this point has the coordinates
\begin{equation}                                        \label{eeldef}
  y^i\rightarrow x^i(y)=y^i+u^i(x)
\end{equation}
\begin{floatingfigure}{.45\textwidth}
\includegraphics[width=.4\textwidth]{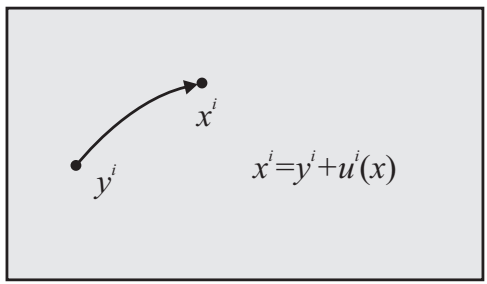}
 \caption{Elastic deformations}
 \label{feldef}
\end{floatingfigure}
in the initial coordinate system, see Fig.~\ref{feldef}. The inverse notation
is used in the elasticity theory. One usually writes
$x^i\rightarrow y^i=x^i+u^i(x)$.
These are equivalent because both coordinate systems $x^i$ and $y^i$
cover the whole $\MR^3$. However, in the theory of defects considered in the
next sections, the situation is different. Generally, the elastic medium
fills the whole Euclidean space only in the final state. Here and in what
follows, we assume that fields depend on coordinates $x$ that are coordinates
of points of the medium after the deformation and cover the whole Euclidean
space $\MR^3$. In the presence of dislocations, the coordinates
$y^i$ do not cover the whole $\MR^3$ in the general case because part of the
media may be removed or, conversely, added. Therefore, the system of coordinates
related to points of the medium after an elastic deformation and defect creation
is more preferable.

In the linear elasticity theory, relative deformations are assumed to be small
($\pl_j u^i\ll1$). The functions $u^i(x)=u^i(y(x))$ are then components of
a vector field that is called the displacement vector field and is the
basic variable in elasticity theory.

In the absence of defects, we assume that the displacement field is a smooth
vector field in the Euclidean space $\MR^3$. The presence of discontinuities
and singularities of the displacement field is interpreted as a presence of
defects in elastic media.

In what follows, we consider only static deformations with the displacement
field $u^i$ independent of time. Then the basic equations of equilibrium
for small deformations are (see, e.g., \cite{LanLif70})
\begin{align}                                        \label{estati}
  &\pl_j\s^{ji}+f^i=0,
\\                                                      \label{eHook}
  &\s^{ij}=\lm\dl^{ij}\e_k{}^k+2\m\e^{ij},
\end{align}
where $\s^{ij}$ is the stress tensor, which is assumed to be symmetric.
The tensor of small deformations $\e_{ij}$ is given by the symmetrized
partial derivative of the displacement vector:
\begin{equation}                                        \label{edefte}
  \e_{ij}=\frac12(\pl_i u_j+\pl_j u_i).
\end{equation}
Lowering and raising of the Latin indices is performed with the Euclidean
metric $\dl_{ij}$ and its inverse $\dl^{ij}$.
The letters $\lm$ and $\mu$ denote constants characterizing elastic properties
of media and are called Lam\'e coefficients. Functions $f^i(x)$ describe the
total density of nonelastic forces inside the medium. We assume in what
follows that such forces are absent: $f^j(x)=0$. Equation (\ref{estati}) is
Newton's law, and Eqn (\ref{eHook}) is Hook's law relating stresses
to deformations.

In a Cartesian coordinate system and for small deformations, the difference
between upper and lower indices disappears because raising and lowering of
indices is performed with the help of the Euclidean metric. One usually
forgets about this difference due to this reason, and this is fully justified.
But in the presence of defects, the notion of Cartesian coordinate system and
Euclidean metric is absent, and the indices are raised and lowered with the
help of Riemannian metric. Therefore, we distinguish the upper and lower indices
as is accepted in differential geometry, having the transition to elastic
media with defects in mind.

The main problem in the linear elasticity theory is the solution of the
second-order equations for the displacement vector that arise after
substitution of (\ref{eHook}) into (\ref{estati}) with some boundary
conditions. Many known solutions are in good agreement with experiment.
Therefore, one may say that equations (\ref{eHook}), (\ref{estati}) have
a solid experimental background.

We now look at the elastic deformations from the standpoint of differential
geometry. From the mathematical standpoint, map (\ref{eeldef}) is itself a
diffeomorphism of the Euclidean space $\MR^3$. The Euclidean metric
$\dl_{ij}$ is then induced by the map $y^i\rightarrow x^i$. It means that
in the deformed state, the metric in the linear approximation is given by
\begin{equation}                                        \label{emetri}
  g_{ij}(x)=\frac{\pl y^k}{\pl x^i}\frac{\pl y^l}{\pl x^j}
     \dl_{kl}\approx\dl_{ij}-\pl_iu_j-\pl_ju_i=\dl_{ij}-2\e_{ij},
\end{equation}
i.e., is defined by the tensor of small deformations (\ref{edefte}).
We note that in the linear approximation, $\e_{ij}(x)=\e_{ij}(y)$ and
$\pl u_j/\pl x^i=\pl u_j/\pl y^i$.

In Riemannian geometry, the metric uniquely defines the Levi--Civita connection
$\widetilde\G_{ij}{}^k(x)$ (Christoffel's symbols), Eqn (\ref{etigam}). We can
compute curvature tensor (\ref{ecurva}) for these symbols. This tensor
equals identically zero, $\widetilde R_{ijk}{}^l(x)=0$, because the curvature of
the Euclidean space is zero, and the map $y^i\rightarrow x^i$ is a
diffeomorphism. The torsion tensor is equal to zero for the same reason.
Thus, an elastic deformation of the medium corresponds to the trivial
Riemann--Cartan geometry, with zero curvature and torsion tensors.

The physical interpretation of metric (\ref{emetri}) is as follows. The
external observer fixes Cartesian coordinate system corresponding to the
ground undeformed state of the medium. The medium is then deformed, and
external observer discovers that the metric becomes nontrivial in this
coordinate system. If we assume that elastic perturbations in the medium
(phonons) propagate along extremals (lines of minimal length), then their
trajectories in the deformed medium are defined by Eqns (\ref{exteqr}).
Trajectories of phonons are now not straight lines because the
Christoffel's symbols are nontrivial ($\widetilde\G_{jk}{}^i\ne0$).
In this sense, metric (\ref{emetri}) is observable. Here, we see the
essential role of the Cartesian coordinate system $y^i$ defined by the
undeformed state, with which the measurement process is connected.

We assume that the metric $g_{ij}(x)$ given in the Cartesian coordinates
corresponds to some state of elastic media without defects. The displacement
vector is then defined by the system of equations (\ref{emetri}), and its
integrability conditions are the equality of the curvature tensor to zero,
in accordance with theorem \ref{tlocev} in the Appendix. In the linear
approximation, these conditions are known in elasticity theory as the
Saint--Venant integrability conditions.

We make a remark that is important for the following consideration.
For appropriate boundary conditions, the solution of the elasticity theory
equations (\ref{estati}), (\ref{eHook}) is unique. From the geometric
standpoint, this means that elasticity theory fixes diffeomorphisms. This
fact is used in the geometric theory of defects. Equations of
nonlinear elasticity theory written in terms of the metric or vielbein
are used for fixing the coordinate system.
\section{Dislocations                                  \label{sdislo}}
\begin{figure}[h,t]
\hfill\includegraphics[width=.35\textwidth]{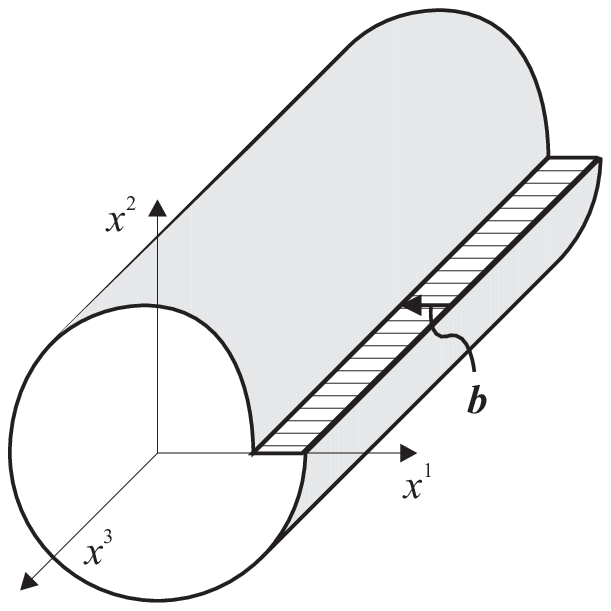}
\hfill \hspace*{.06\textwidth} \hfill
\includegraphics[width=.35\textwidth]{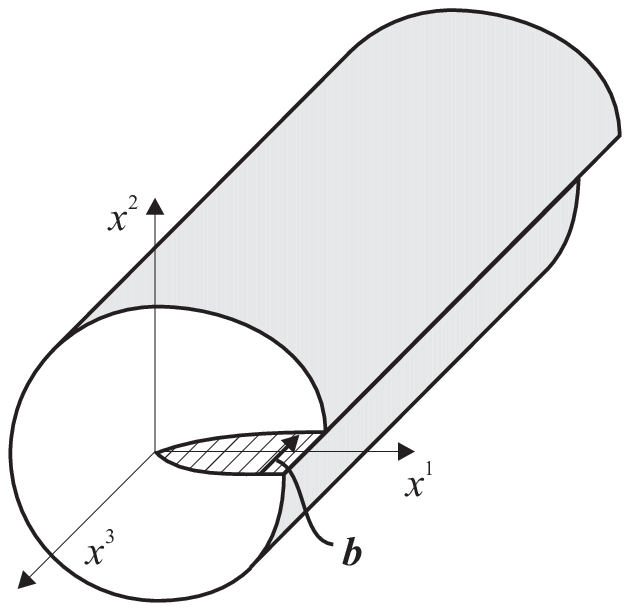}
\hfill {}
\\
\centering \caption{\label{fdislo} Straight linear dislocations.
            (\textit{a}) The edge
            dislocation. The Burgers vector $\Bb$ is perpendicular to the
            dislocation line. (\textit{b}) The screw dislocation. The Burgers
           vector $\Bb$ is parallel to the dislocation line.}
\end{figure}

We start with the description of linear dislocations in elastic media
(see, e.g., \cite{LanLif70,Kosevi81}). The simplest and most undespread
examples of linear dislocations are shown in Fig.~\ref{fdislo}.
We cut the medium along the half-plane $x^2=0$, $x^1>0$, move the upper part
of the medium located over the cut $x^2>0$, $x^1>0$ by the vector $\Bb$
towards the dislocation axis $x^3$, and glue the cutting surfaces.
The vector $\Bb$ is called the Burgers vector. In the general case, the
Burgers vector may not be constant on the cut. For the edge dislocation,
it varies from zero to some constant value $\Bb$ as it moves from the
dislocation axis. After the gluing, the media comes to the equilibrium state
called the edge dislocation, see Fig.~\ref{fdislo}\textit{a}.
If the Burgers vector is parallel to the dislocation line, it is called
the screw dislocation (Fig.~\ref{fdislo}\textit{b}).

The same dislocation can be made in different ways. For example,
if the Burgers vector is perpendicular to the cutting plane and directed
from it in the considered cases, then the produced cavity must be filled with
medium before gluing. It is easy to imagine that the edge dislocation
is also obtained as a result, but rotated by the angle $\pi/2$ around the
$x^3$ axis. This example shows that a dislocation is characterized not
by the cutting surface but by the dislocation line and the Burgers vector.

From the topological standpoint, the medium containing several dislocations
or even the infinite number of them is the Euclidean space $\MR^3$.
In contrast to the case of elastic deformations, the displacement vector
in the presence of dislocations is no longer a smooth function because
of the presence of cutting surfaces. At the same time, we assume that
partial derivatives of the displacement vector $\pl_j u^i$ (the distortion
tensor) are smooth functions on the cutting surface. This assumption
is justified physically because these derivatives define deformation
tensor (\ref{edefte}). In its turn, partial derivatives of the deformation
tensor must exist and be smooth functions in the equilibrium state everywhere
except, possibly, the dislocation axis, because otherwise equations
of equilibrium (\ref{estati}) have no meaning. We assume that the metric
and vielbein are smooth functions everywhere in $\MR^3$ except, may be,
dislocation axes, because the deformation tensor defines the induced metric
(\ref{emetri}).

The main idea of the geometric approach amounts to the following. To describe
single dislocations in the framework of elasticity theory we must solve
equations for the displacement vector with some boundary conditions on the
cuts. This is possible for small number of dislocations. But, with an
increasing number of dislocations, the boundary conditions become so
complicated that the solution of the problem becomes unrealistic. Besides,
one and the same dislocation can be created by different cuts which
leads to an ambiguity in the displacement vector field. Another shortcoming
of this approach is that it cannot be applied to the description of a
continuous distribution of dislocations because the displacement
vector field does not exist in this case at all because it must have
discontinuities at every point. In the geometric approach, the basic
variable is the vielbein which by assumption is a smooth function
everywhere except, possibly, dislocation axes. We postulate new equations
for the vielbein (see section \ref{sequil}). In the geometric approach, the
transition from a finite number of dislocations to their continuous
distribution is simple and natural. In that way, the smoothing of
singularities occurs on dislocation axes in analogy with smoothing of mass
distribution for point particles in passing to continuous media.

We now develop the formalism of the geometric approach. In a general
defect-present case, we do not have a preferred Cartesian
coordinate frame in the equilibrium because there is no symmetry.
Therefore, we consider arbitrary coordinates $x^\mu$, $\mu=1,2,3$, in
$\MR^3$. We use Greek letters for coordinates allowing
arbitrary coordinate changes. Then the Burgers vector can be expressed
as the integral of the displacement vector
\begin{equation}                                        \label{eBurge}
  \oint_Cdx^\mu\pl_\mu u^i(x)=-\oint_Cdx^\mu\pl_\mu y^i(x)=-b^i,
\end{equation}
where $C$ is a closed contour surrounding the dislocation axis, Fig.~\ref{fburco}.
\begin{floatingfigure}{.45\textwidth}
\includegraphics[width=.4\textwidth]{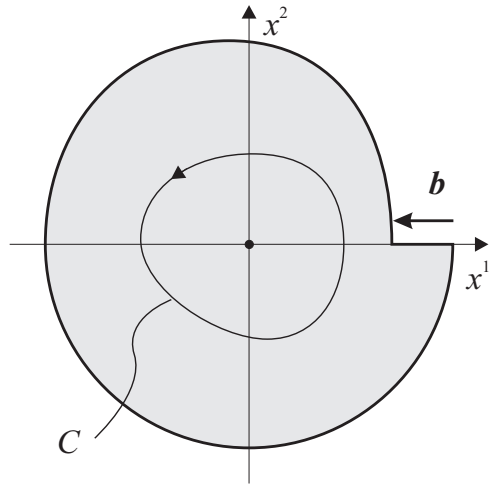}
 \caption{Section of the media with the edge dislocation. $C$ is the
 integration contour for the Burgers vector $\Bb$.}
 \label{fburco}
\end{floatingfigure}
This integral is invariant under arbitrary coordinate
transformations $x^\mu\rightarrow x^{\mu'}(x)$ and covariant under global
$\MS\MO(3)$-rotations of $y^i$. Here, components of the displacement vector
field $u^i(x)$ are considered with respect to the orthonormal basis in
the tangent space, $u=u^i e_i$. If components of the displacement vector
field are considered with respect to the coordinate basis $u=u^\mu\pl_\mu$,
the invariance of the integral (\ref{eBurge}) under general coordinate
changes is violated.

In the geometric approach, we introduce new independent variable -- the vielbein
-- instead of partial derivatives $\pl_\mu u^i$
\begin{equation}                                        \label{edevid}
  e_\mu{}^i(x)=\begin{cases} \pl_\mu y^i, &\text{outside the cut,}\\
               \lim\pl_\mu y^i, &\text{on the cut.}\end{cases}
\end{equation}
The vielbein is a smooth function on the cut by construction. We note that
if the vielbein was simply defined as the partial derivative $\pl_\mu y^i$,
then it would have the $\dl$-function singularity on the cut because functions
$y^i(x)$ have a jump. The Burgers vector can be expressed through the
integral over a surface $S$ having contour $C$ as the boundary
\begin{equation}                                        \label{eBurg2}
  \oint_Cdx^\mu e_\mu{}^i=\int\!\!\int_Sdx^\mu\wedge dx^\nu
  (\pl_\mu e_\nu{}^i-\pl_\nu e_\mu{}^i)=b^i,
\end{equation}
where $dx^\mu\wedge dx^\nu$ is the surface element. As a consequence of
the definition of the vielbein in (\ref{edevid}), the integrand is equal to
zero everywhere except at the dislocation axis. For the edge dislocation with
constant Burgers vector, the integrand has a $\dl$-function singularity at the
origin. The criterion for the presence of a dislocation is a violation of the
integrability conditions for the system of equations $\pl_\mu y^i=e_\mu{}^i$:
\begin{equation}                                        \label{eintco}
  \pl_\mu e_\nu{}^i-\pl_\nu e_\mu{}^i\ne0.
\end{equation}
If dislocations are absent, then the functions $y^i(x)$ exist and define
transformation to a Cartesian coordinates frame.

In the geometric theory of defects, the field $e_\mu{}^i$ is identified
with the vielbein. Next, we compare the integrand in (\ref{eBurg2}) with the
expression for the torsion in Cartan variables (\ref{ecurcv}). They differ
only by terms containing the $\MS\MO(3)$ connection. This is the ground for
the introduction of the following postulate. In the geometric theory of
defects, the Burgers vector corresponding to a surface $S$ is defined by
the integral of the torsion tensor:
\begin{equation*}
  b^i=\int\!\!\int_S dx^\mu\wedge dx^\nu T_{\mu\nu}{}^i.
\end{equation*}
This definition is invariant with respect to general coordinate transformations
of $x^\mu$ and covariant with respect to global rotations. Thus, the torsion
tensor has straightforward physical interpretation: it is equal to the surface
density of the Burgers vector.

The physical interpretation of the $\MS\MO(3)$ connection is given in section
\ref{sdiscl}, and now we show how this definition reduces to the expression
for the Burgers vector (\ref{eBurg2}) obtained within elasticity theory.
If the curvature tensor for the $\MS\MO(3)$ connection is zero, then,
according to theorem \ref{tlotrs}, the connection is locally trivial,
and there exists such $\MS\MO(3)$ rotation such that $\om_{\mu i}{}^j=0$.
In this case, we return to expression (\ref{eBurg2}).

If the $\MS\MO(3)$ connection is zero and vielbein is a smooth function,
then the Burgers vector corresponds uniquely to every contour. It can then
be expressed as a surface integral of the torsion tensor. The
surface integral depends only on the boundary contour but not on the
surface due to the Stokes theorem.
\begin{floatingfigure}{.45\textwidth}
\includegraphics[width=.4\textwidth]{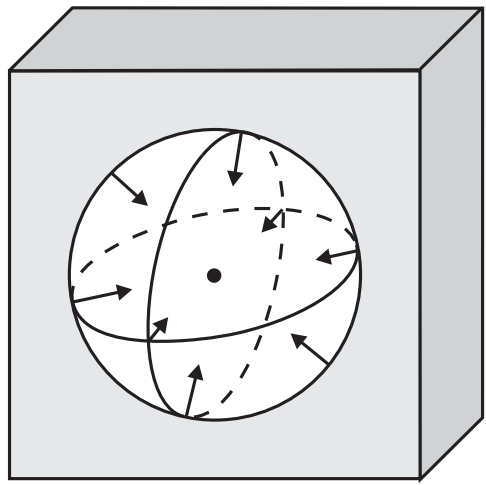}
 \caption{Point defect: a vacancy appears when a ball is cut out from the
 medium, and the boundary sphere is shrunk to a point.}
 \label{fpoide}
\end{floatingfigure}

We have shown that the presence of linear defects results in a nontrivial
torsion tensor. In the geometric theory of defects, the equality of the torsion
tensor to zero $T_{\mu\nu}{}^i=0$ is naturally considered the criterion
for the absence of dislocations. Then, under the name dislocation fall not
only linear dislocations but, in fact,
arbitrary defects in elastic media. For example, point defects; vacancies
and impurities, are also dislocations. In the first case we cut out a ball
from the Euclidean space $\MR^3$ and then shrink the boundary sphere to a
point (Fig.~\ref{fpoide}). In the case of impurity, a point of the Euclidean
space is blown up to a sphere and the produced cavity is filled with the medium.
Point defects are characterized by the mass of the removed or added media,
which is also defined by the vielbein \cite{KatVol92}
\begin{equation}                                        \label{emass}
  M=\rho_0\int\!\!\int\!\!\int_{\MR^3} d^3x\left(\det e_\mu{}^i
  -\det\overset\circ e_\mu{}^i\right),
  ~~~~\overset\circ e_\mu{}^i=\pl_\mu y^i,
\end{equation}
where $y^i(x)$ are the transition functions to a Cartesian coordinate frame
in $\MR^3$ and $\rho_0$ is the density of the medium which is supposed to
be constant. The mass is defined by the difference of two integrals, each of
them being separately divergent. The first integral equals to the volume
of the medium with defects and the second is equal to the volume of the
Euclidean space. The torsion tensor for a vacancy or impurity is zero
everywhere except at one point, where it has a $\dl$-function singularity.
For point defects, the notion of the Burgers vector is absent.

According to the given definition, the mass of an impurity is positive because
the matter is added to the media, and the mass of a vacancy is negative because
part of the medium is removed. The negative sign of the mass causes serious
problems for a physical interpretation of solutions of the equations of motion
or the Schr\"odinger equation. Hence, we make a remark.
Strictly speaking, the integral (\ref{emass})
should be called the ``bare'' mass because this expression does not account
for elastic stresses arising around a point dislocation. The effective
mass of such a defect must contain at least two contributions: the bare mass
and the free energy coming from elastic stresses. The question about the sign
of the effective mass is not solved yet and demands a separate analysis.

Surface defects may also exist in three-dimensional space, in addition to
point and line dislocations. In the geometric approach, all of them are called
dislocations because they correspond to a nontrivial torsion.
\section{Disclinations                                 \label{sdiscl}}
In the preceeding section, we related dislocations to a nontrivial
torsion tensor. For this, we introduced an $\MS\MO(3)$ connection.
Now we show that the curvature tensor for the $\MS\MO(3)$ connection
defines the surface density of the Frank vector characterizing other
well-known defects -- disclinations in the spin structure of media
\cite{LanLif70}.

\begin{figure}[h,t]
\hfill\includegraphics[width=.35\textwidth]{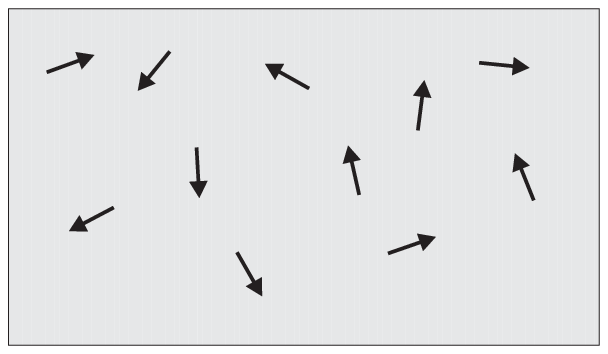}
\hfill  \hspace*{.06\textwidth} \hfill
\includegraphics[width=.35\textwidth]{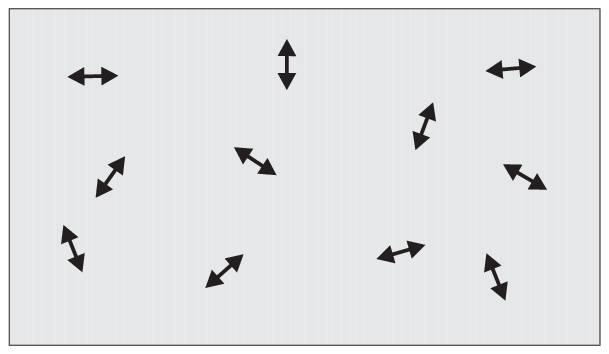}
\hfill {} \\
\centering \caption{\label{fspstr} Examples of media with the spin structure:
           (\textit{a}) ferromagnet, (\textit{b}) liquid crystal.}
\end{figure}

Let a unit vector field $n^i(x)$ $(n^in_i=1)$ be given at all points of
the medium. For example, $n^i$ has the meaning of the magnetic moment
located at each point of the medium for ferromagnets
(Fig.~\ref{fspstr}\textit{a}). For nematic liquid crystals, the unit vector
field $n^i$ with the equivalence relation $n^i\sim-n^i$ describes
the director field (Fig.~\ref{fspstr}\textit{b}).

We fix some direction in the medium $n_0^i$. Then the field $n^i(x)$
at a point $x$ can be uniquely defined by the field $\om^{ij}(x)=-\om^{ji}(x)$
taking values in the rotation algebra $\Gs\Go(3)$ (the rotation angle),
$$
  n^i=n_0^j S_j{}^i(\om),
$$
where $S_j{}^i\in\MS\MO(3)$ is the rotation matrix corresponding to the
algebra element $\om^{ij}$. Here, we use the following parameterization of
the rotation group $\MS\MO(3)$ by elements of its algebra (see, e.g.,
\cite{Katana04})
\begin{equation}                                        \label{elsogr}
  S_i{}^j=(e^{(\om\ve)})_i{}^j=\cos\om\,\dl_i^j
  +\frac{(\om\ve)_i{}^j}\om\sin\om
  +\frac{\om_i\om^j}{\om^2}(1-\cos\om)~          \in\MS\MO(3),
\end{equation}
where $(\om\ve)_i{}^j=\om^k\ve_{ki}{}^j$ and $\om=\sqrt{\om^i\om_i}$ is
the modulus of the vector $\om^i$. The pseudovector
$\om^k=\om_{ij}\ve^{ijk}/2$, where $\ve^{ijk}$ is the totally
antisymmetric third-rank tensor, $\ve^{123}=1$, is directed along the
rotation axis and its length equals the rotation angle. We
call the field $\om^{ij}$ spin structure of the media.

\begin{figure}[h,t]
\hfill\includegraphics[width=.35\textwidth]{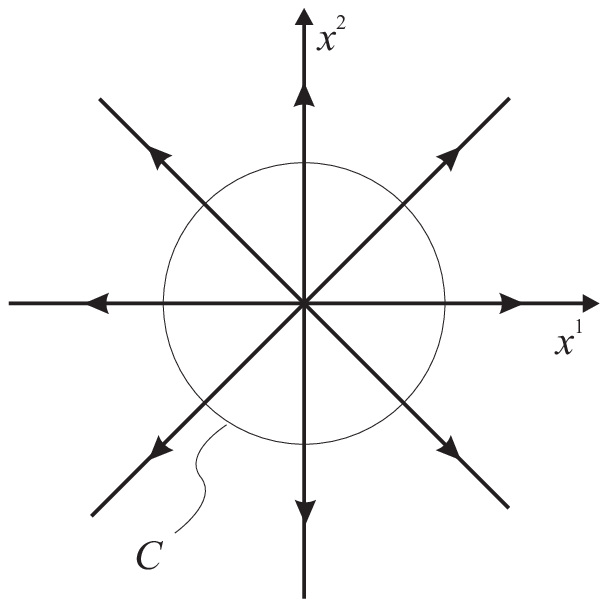}
\hfill  \hspace*{.06\textwidth} \hfill
\includegraphics[width=.35\textwidth]{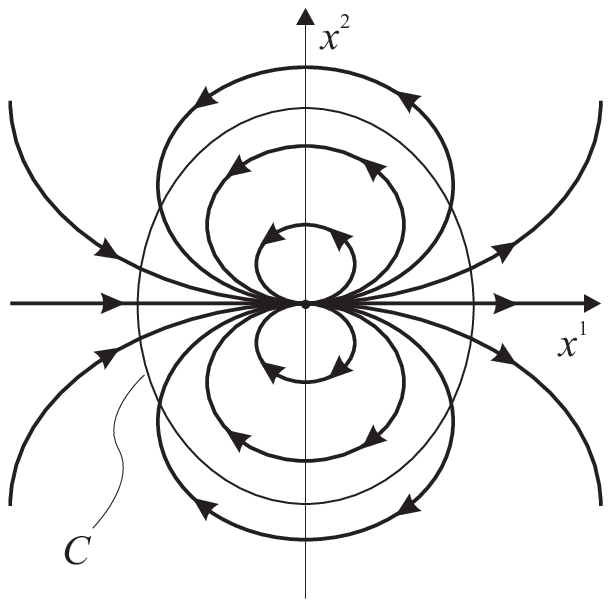}
\hfill {} \\
\centering \caption{\label{fdiscl} The vector field distributions on the
    plane $x^1,x^2$ for the linear disclinations parallel to the $x^3$ axis.
    (\textit{a}) $|\Theta|=2\pi$. (\textit{b}) $|\Theta|=4\pi$.}
\end{figure}

If a media has a spin structure, then it may have defects called
disclinations. For linear disclinations parallel to the $x^3$ axis,
the vector field $n$ lies in the perpendicular plain $x^1,x^2$.
The simplest examples of linear disclinations are shown in Fig.~\ref{fdiscl}.
Every linear disclination is characterized by the Frank vector
\begin{equation}                                        \label{etheta}
  \Th_i=\e_{ijk}\Om^{jk},
\end{equation}
where
\begin{equation}                                        \label{eomega}
  \Om^{ij}=\oint_Cdx^\mu\pl_\mu\om^{ij},
\end{equation}
and the integral is taken along closed contour $C$ surrounding the disclination
axis. The length of the Frank vector is equal to the total angle of
rotation of the field $n^i$ as it goes around the disclination.

The vector field $n^i$ defines a map of the Euclidean space to a sphere
$n:~\MR^3\rightarrow\MS^2$. For linear disclinations parallel to the
$x^3$ axis, this map is restricted to a map of the plane $\MR^2$ to
a circle $\MS^1$. In this case, the total rotation angle must obviously
be a multiple of $2\pi$.

\begin{figure}[h,t]
\hfill\includegraphics[width=.35\textwidth]{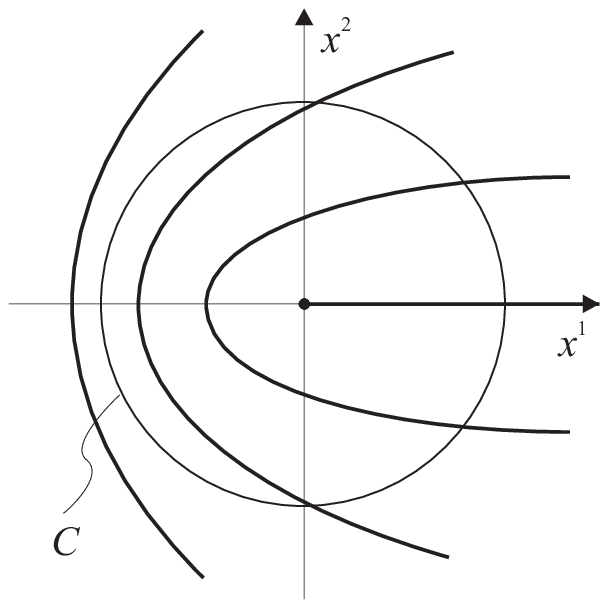}
\hfill  \hspace*{.06\textwidth} \hfill
\includegraphics[width=.35\textwidth]{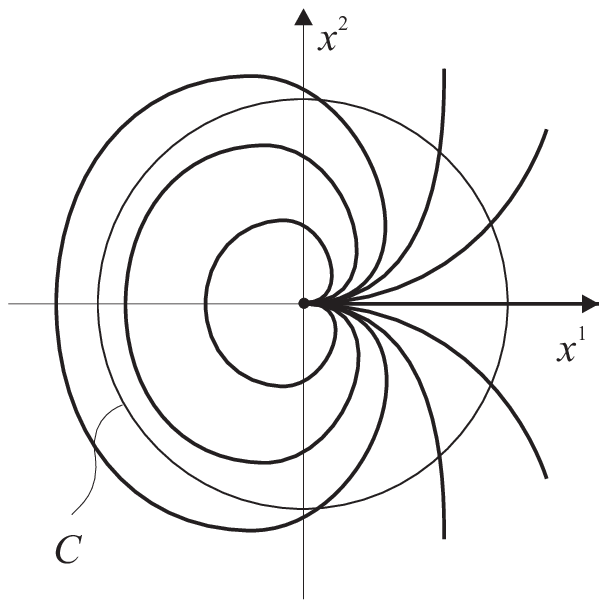}
\hfill {} \\
\centering \caption{\label{fdisc2} The director field distribution
    in the $x^1,x^2$ plane for the linear disclinations parallel to
    the $x^3$ axis. (\textit{a}) $|\Theta|=\pi$ and (\textit{b}) $|\Theta|=3\pi$.}
\end{figure}

For nematic liquid crystals, we have the equivalence relation $n^i\sim-n^i$.
Therefore, for linear disclinations parallel to the $x^3$ axis, the director
field defines a map of the plane into the projective line
$n:~\MR^2\rightarrow\MR\MP^1$. In this case, the length of the Frank vector
must be a multiple of $\pi$. The corresponding examples of disclinations
are shown in Fig.~\ref{fdisc2}.

As for the displacement field, the field $\om^{ij}(x)$, taking
values in the algebra $\Gs\Go(3)$, is not a smooth function in $\MR^3$ in
the presence of disclinations. We make a cut in $\MR^3$ bounded by
the disclination axis. Then the field $\om^{ij}(x)$ may be considered
smooth in the whole space except the cut. We assume that all partial
derivatives of $\om^{ij}(x)$ have the same limit as it approaches
the cut from both sides. Then we define the new field
\begin{equation}                                        \label{edesoc}
  \om_\mu{}^{ij}=\begin{cases}\pl_\mu\om^{ij}, &\text{outside the cut,}\\
                 \lim\pl_\mu\om^{ij}, &\text{on the cut.} \end{cases}
\end{equation}
The functions $\om_\mu{}^{ij}$ are smooth everywhere by construction except,
may be, on the disclination axis. Then the Frank vector may be
expressed as the surface integral
\begin{equation}                                        \label{eFrank}
  \Om^{ij}=\oint_Cdx^\mu\om_\mu{}^{ij}=
  \int\!\!\int_Sdx^\mu\wedge dx^\nu(\pl_\mu\om_\nu{}^{ij}-\pl_\nu\om_\mu{}^{ij}),
\end{equation}
where $S$ is an arbitrary surface having the contour $C$ as the boundary. If
the field $\om_\mu{}^{ij}$ is given, then the integrability conditions for
the system of equations $\pl_\mu\om^{ij}=\om_\mu{}^{ij}$ are
\begin{equation}                                        \label{eomcon}
    \pl_\mu\om_\nu{}^{ij}-\pl_\nu\om_\mu{}^{ij}=0.
\end{equation}
This noncovariant equality yields the criterion for the absence of
disclinations.

In the geometric theory of defects, we identify the field $\om_\mu{}^{ij}$
with the $\MS\MO(3)$ connection. In the expression for the curvature in
(\ref{ecucav}), the first two terms coincide with (\ref{eomcon}), and
we therefore postulate the covariant criterion of the absence of disclinations
as the equality of the curvature tensor for the $\MS\MO(3)$ connection to zero:
\begin{equation*}
    R_{\mu\nu}{}^{ij}=0.
\end{equation*}
Simultaneously, we give the physical interpretation of the curvature tensor
as the surface density of the Frank vector
\begin{equation}                                        \label{eOmega}
    \Om^{ij}=\int\!\!\int dx^\mu\wedge dx^\nu R_{\mu\nu}{}^{ij}.
\end{equation}
This definition reduces to the previous expression for the Frank vector
(\ref{eFrank}) in the case where rotation of the vector $n$ occures in
a fixed plane. In this case, rotations are restricted by the subgroup
$\MS\MO(2)\subset\MS\MO(3)$. The quadratic terms in the expression for
the curvature in (\ref{ecucav}) disappear because the rotation group $\MS\MO(2)$
is Abelian, and we obtain the previous expression for the Frank vector
(\ref{eFrank}).

Thus, we described the media with dislocations (defects of elastic media)
and disclinations (defects in the spin structure) in the framework of
Riemann--Cartan geometry, the torsion and curvature tensors being identified
with the surface density of dislocations and disclinations, respectively.
The relations between physical and geometrical notions
are summarized in the Table~\ref{tdegeo}.
\begin{table}[h]
\begin{center}
    \begin{tabular}{|l|r|r|}                                  \hline
    Existence of defects & $R_{\mu\nu}{}^{ij}$ &$T_{\mu\nu}{}^i$\\ \hline
    Elastic deformations &$0$&$0$\\ \hline
    Dislocations &$0$&$\ne0$\\ \hline
    Disclinations &$\ne0$&$0$\\ \hline
    Dislocations and disclinations&$\ne0$&$\ne0$\\
    \hline
    \end{tabular}
    \caption{\label{tdegeo} The relation between physical and geometrical
    notions in the geometric theory of defects.}
\end{center}
\end{table}

The same physical interpretation of torsion and curvature was considered in
\cite{Vercin90}. Several possible functionals for the free energy were also
considered. In the next section, we propose a new expression for
the free energy.
\section{Free energy                                   \label{sequil}}
Until now, we discussed only the relation between physical and geometrical
notions. To complete the construction of the geometric theory of defects,
we have to postulate equations of equilibrium describing static distribution
of defects in media. The vielbein $e_\mu{}^i$ and $\MS\MO(3)$ connection
$\om_\mu{}^{ij}$ are basic and independent variables in the geometric
approach. In contrast to previous geometric approaches, we completely abandon
the displacement field $u^i$ and spin structure $\om^{ij}$ as the fields
entering the system of equilibrium equations. In a general case of a
continuous distribution of defects, they simply do not exist. Nevertheless,
at some level and under definite circumstances they can be reconstructed,
and we discuss this in the following section.

The expression for the free energy was derived in \cite{KatVol92}.
We assume that equations of equilibrium must be covariant under
general coordinate transformations and local rotations, be at most of the
second order, and follow from a variational principle. The expression for
the free energy leading to the equilibrium equations must then be equal
to a volume integral of the scalar function (the Lagrangian) that is
quadratic in torsion and curvature tensors. There are three independent
invariants quadratic in the torsion tensor and three independent invariants
quadratic in the curvature tensor in three dimensions. It is possible to add
the scalar curvature and a ``cosmological'' constant $\Lm$.
We thus obtain a general eight-parameter Lagrangian
\begin{align}                                               \nonumber
  \frac1{e}L&=-\kappa R+\frac14T_{ijk}(\bt_1T^{ijk}+\bt_2T^{kij}
  +\bt_3T^j\dl^{ik})
\\                                                      \label{egenla}
  &+\frac14R_{ijkl}(\g_1R^{ijkl}+\g_2R^{klij}+\g_3R^{ik}\dl^{jl})
  -\Lm,~~~~e=\det e_\mu{}^i,
\end{align}
where $\kappa$, $\bt_{1,2,3}$ and $\g_{1,2,3}$ are some constants, and
we have introduced the trace of torsion tensor $T_j=T_{ij}{}^i$ and the Ricci
tensor $R_{ik}=R_{ijk}{}^j$. Here and in what follows, transformation of the
Greek indices into the Latin ones and vice versa is always performed using
the vielbein and its inverse. For example,
\begin{equation*}
  R_{ijkl}=R_{\mu\nu kl}e^\mu{}_i e^\nu{}_j,~~~~
  T_{ijk}=T_{\mu\nu k}e^\mu{}_i e^\nu{}_j.
\end{equation*}
The particular feature of three dimensions is that the full curvature
tensor is in a one-to-one correspondence with Ricci tensor
(\ref{ecurir}) and has three irreducible components. Therefore, the
Lagrangian contains only three independent invariants quadratic in
curvature tensor. We do not need to add the Hilbert--Einstein Lagrangian
$\widetilde R$, also yielding second-order equations, to the free
energy (\ref{egenla}) because of identity (\ref{eidcur}).

Thus, the most general Lagrangian depends on eight constants and leads to
very complicated equations of equilibrium. At present, we do not know
precisely what values of the constants describe this or that medium.
Therefore, we make physically reasonable assumptions to simplify matters.
Namely, we require that equations of equilibrium must admit the following
three types of solutions.
\begin{enumerate}
\item There are solutions describing the media with only dislocations,
\newline $R_{\mu\nu}{}^{ij}=0,~T_{\mu\nu}{}^i\ne0$.
\item There are solutions describing the media with only disclinations,
\newline $R_{\mu\nu}{}^{ij}\ne0,~T_{\mu\nu}{}^i=0$.
\item There are solutions describing the media without dislocations and
disclinations,
\newline $R_{\mu\nu}{}^{ij}=0,~T_{\mu\nu}{}^i=0$.
\end{enumerate}
It turns out that these simple assumptions reduce the number of independent
parameters from eight to two. We now prove this statement.
The Lagrangian (\ref{egenla}) yields the equilibrium equations
\begin{align}                                                \nonumber
  \frac1{e}\frac{\dl L}{\dl e_\mu{}^i}
  &=\kappa\left(Re^\mu{}_i-2R_i{}^\mu\right)+\bt_1\left(\nb_\nu T^{\nu\mu}{}_i
  -\frac14T_{jkl}T^{jkl}e^\mu{}_i+T^{\mu jk}T_{ijk}\right)
\\                                                           \nonumber
  &+\bt_2\left(-\frac12\nb_\nu(T_i{}^{\mu\nu}-T_i{}^{\nu\mu})
  -\frac14T_{jkl}T^{ljk}e^\mu{}_i-\frac12T^{j\mu k}T_{kij}
  +\frac12T^{jk\mu}T_{kij}\right)
\\                                                           \nonumber
  &+\bt_3\left(-\frac12\nb_\nu(T^\nu e^\mu{}_i-T^\mu e^\nu{}_i)
  -\frac14T_jT^je^\mu{}_i+\frac12T^\mu T_i+\frac12T^jT_{ij}{}^\mu\right)
\\                                                           \nonumber
  &+\g_1\left(-\frac14R_{jklm}R^{jklm}e^\mu{}_i+R^{\mu jkl}R_{ijkl}\right)
\\                                                           \nonumber
  &+\g_2\left(-\frac14R_{jklm}R^{lmjk}e^\mu{}_i+R^{kl\mu j}R_{ijkl}\right)
\\                                                      \label{etriad}
  &+\g_3\left(-\frac14R_{jk}R^{jk}e^\mu{}_i+\frac12R^{\mu j}R_{ij}
  +\frac12R^{jk}R_{jik}{}^\mu\right)+\Lm e^\mu{}_i=0,
\\                                                           \nonumber
  \frac1{e}\frac{\dl L}{\dl\om_\mu{}^{ij}}
  &=\kappa\left(\frac12T_{ij}{}^\mu+T_ie^\mu{}_j\right)
  +\bt_1\frac12T^\mu{}_{ji}
  +\bt_2\frac14\left(T_i{}^\mu{}_j-T_{ij}{}^\mu\right)
\\                                                           \nonumber
  &+\bt_3\frac14T_je^\mu{}_i
  +\g_1\frac12\nb_\nu R^{\nu\mu}{}_{ij}+\g_2\frac12\nb_\nu R_{ij}{}^{\nu\mu}
\\                                                      \label{econne}
  &+\g_3\frac14\nb_\nu\left(R^\nu{}_ie^\mu{}_j-R^\mu{}_ie^\nu{}_j\right)
  -(i\leftrightarrow j)=0,
\end{align}
where the covariant derivative acts with the $\MS\MO(3)$ connection on the
Latin indices and with the Christoffel symbols on the Greek ones. For example,
\begin{align*}
  \nb_\nu T^{\rho\mu}{}_i&=\pl_\nu T^{\rho\mu}{}_i
  +\widetilde\G_{\nu\s}{}^\rho T^{\s\mu}{}_i
  +\widetilde\G_{\nu\s}{}^\mu T^{\rho\s}{}_i-\om_{\nu i}{}^jT^{\rho\mu}{}_j,
\\
  \nb_\nu R^{\rho\mu}{}_{ij}&=\pl_\nu R^{\rho\mu}{}_{ij}
  +\widetilde\G_{\nu\s}{}^\rho R^{\s\mu}{}_{ij}
  +\widetilde\G_{\nu\s}{}^\mu R^{\rho\s}{}_{ij}-\om_{\nu i}{}^kR^{\rho\mu}{}_{kj}
  -\om_{\nu j}{}^kR^{\rho\mu}{}_{ik}.
\end{align*}

The first condition on the class of solutions of the equilibrium equations
is that they permit solutions describing the presence of only dislocations
in media. This means the existence of solutions with zero curvature tensor
corresponding to the absence of disclinations. Substitution of the condition
$R_{\mu\nu}{}^{ij}=0$ into Eqn (\ref{econne}) for the $SO(3)$ connection
yields
\begin{align}                                                \nonumber
  (12\kappa+2\bt_1-\bt_2-2\bt_3)T_i&=0,
\\                                                      \label{ecotor}
  (\kappa-\bt_1-\bt_2)T^*&=0,
\\                                                           \nonumber
  (4\kappa+2\bt_1-\bt_2)W_{ijk}&=0.
\end{align}
Here $T_i$, $T^*$, and $W_{ijk}$ are the irreducible components of the
torsion tensor,
\begin{equation*}
  T_{ijk}=W_{ijk}+T^*\e_{ijk}+\frac12(\dl_{ik}T_j-\dl_{jk}T_i),
\end{equation*}
where
\begin{align*}
  T^*&=\frac16T_{ijk}\e^{ijk}, &T_j&=T_{ij}{}^i,
\\
  W_{ijk}&=T_{ijk}-T^*\e_{ijk}-\frac12(\dl_{ik}T_j-\dl_{jk}T_i), &
  W_{ijk}\e^{ijk}&=W_{ij}{}^i=0.
\end{align*}

In a general case of dislocations all irreducible components of torsion
tensor differ from zero ($T_i$, $T^*$, $W_{ijk}\ne0$) and Eqns
(\ref{ecotor}) have a unique solution
\begin{equation}                                        \label{ebetak}
    \bt_1=-\kappa,~~~~\bt_2=2\kappa,~~~~\bt_3=4\kappa.
\end{equation}
For these coupling constants, the first four terms in Lagrangian
(\ref{egenla}) are equal to the Hilbert--Einstein Lagrangian
$\kappa\widetilde R(e)$ up to a total divergence due to identity
(\ref{eidcur}). Equation (\ref{etriad}) then reduces to the Einstein
equations with a cosmological constant
\begin{equation}                                        \label{eEinst}
  \widetilde R_{\mu\nu}-\frac12g_{\mu\nu}\widetilde R
  -\frac\Lm{2\kappa}g_{\mu\nu}=0.
\end{equation}
In this way, the first condition is satisfied.

According to the second condition, the equations of equilibrium must allow
solutions with zero torsion $T_{\mu\nu}{}^i=0$. In this case, the curvature
tensor has additional symmetry $R_{ijkl}=R_{klij}$, and Eqn (\ref{econne})
becomes
\begin{align}                                                \nonumber
  (\g_1+\g_2+\frac14\g_3)\nb_\nu\left(R^{S\nu}{}_ie^\mu{}_j
  -R^{S\mu}{}_ie^\nu{}_j-R^{S\nu}{}_je^\mu{}_i+R^{S\mu}{}_je^\nu{}_i\right)
\\                                                      \label{econre}
  +\frac16(\g_1+\g_2+4\g_3)\left(e^\nu{}_ie^\mu{}_j-e^\mu{}_ie^\nu{}_j\right)
  \nb_\nu R=0.
\end{align}
Here, we decompose the Ricci tensor onto irreducible components,
\begin{equation*}
  R_{ij}=R^S{}_{ij}+R^A{}_{ij}+\frac13R\dl_{ij},
\end{equation*}
where
\begin{equation*}
  R^S{}_{ij}=R^S{}_{ji},~~~~R^{Si}{}_i=0,~~~~R^A{}_{ij}=-R^A{}_{ji}.
\end{equation*}
Note that for zero torsion, the Ricci tensor is symmetrical: $R^A{}_{ij}=0$.
Contraction of Eqn (\ref{econre}) with $e_\mu{}^j$ leads to the equation
\begin{equation*}
  (\g_1+\g_2+\frac14\g_3)\nb_\nu R^{S\nu}{}_\mu
    +\frac13(\g_1+\g_2+4\g_3)\nb_\mu R=0.
\end{equation*}
In the general case of nonvanishing curvature, the covariant derivatives
$\nb_\nu R^{S\nu}{}_\mu$ and $\nb_\mu R$ differ from zero and are independent.
Therefore, we obtain two equations for the coupling constants,
\begin{equation*}
  \g_1+\g_2+\frac14\g_3=0,~~~~\g_1+\g_2+4\g_3=0.
\end{equation*}
which have a unique solution
\begin{equation}                                        \label{egamma}
  \g_1=-\g_2=\g,~~~~\g_3=0.
\end{equation}
In this case, Eqn (\ref{etriad}) for the vielbein corresponding to a
nonzero torsion also reduces to Einstein equations (\ref{eEinst}).

The last requirement for the existence of solutions with zero curvature and
torsion is satisfied only for the zero cosmological constant $\Lm=0$.

Therefore, the simple and physically reasonable requirements define the
two parameter Lagrangian \cite{KatVol92}
\begin{equation}                                        \label{eLagfi}
  \frac1eL=-\kappa\widetilde R+2\g R^A{}_{ij}R^{Aij},
\end{equation}
which is the sum of the Hilbert--Einstein Lagrangian for the vielbein and
the square of the antisymmetric part of the Ricci tensor. We note that
$\widetilde R(e)$ and $R^A{}_{ij}(e,\om)$ are constructed for different
metrical connections.

Other quadratic Lagrangians were considered, for example, in
\cite{KadEde83,Vercin90}. We note that they also contain the displacement
vector as an independent variable along with the vielbein.

Expression (\ref{eLagfi}) defines the free energy density in the geometric
theory of defects and leads to the equilibrium equations (the
Euler--Lagrange equations). In our geometric approach, the displacement
vector and spin structure do not enter expression (\ref{eLagfi}) for the
free energy.
\section{Gauge fixing                                 \label{sfixga}}
In the geometric approach, the vielbein $e_\mu{}^i$ and $\MS\MO(3)$
connection $\om_{\mu i}{}^j$ are the only variables. The displacement
field $u^i$ and the spin structure $\om_i{}^j$ can be introduced only
in those regions of media where defects are absent. As the consequence
of the absence of disclinations $R_{\mu\nu i}{}^j=0$, the $\MS\MO(3)$
connection is actually a pure gauge (\ref{eloctr}), i.e., the spin
structure $\om_i{}^j$ exists. If, in addition, dislocations are absent
($T_{\mu\nu}{}^i=0$) then there is the displacement field such that the
vielbein equals its partial derivatives (\ref{erepad}). In this and only
in this case can we introduce the displacement field and spin structure.
We show below that this can be done such that the equations of nonlinear
elasticity theory and the principal chiral $\MS\MO(3)$ field are fulfilled.

For free energy in (\ref{eLagfi}), the Euler--Lagrange equations are
covariant under general coordinate transformations in $\MR^3$ and local
$\MS\MO(3)$ rotations. This means that any solution of the equilibrium
equations is defined up to diffeomorphisms and local rotations. For the
geometric theory of defects to make predictions, we have to fix the coordinate
system (to fix the gauge in the language of gauge field theory). This allows
us to choose a unique representative from each class of equivalent solutions.
We say afterwards that this solution of the Euler--Lagrange equations
describes the distribution of defects in the laboratory coordinate system.

We start with gauge-fixing the diffeomorphisms. For this, we choose the
elastic gauge proposed in \cite{Katana03}. This question is of primary
importance, and we discuss it in detail.

Equations (\ref{estati}), (\ref{eHook}) of elasticity theory for $f^i=0$
yield the second-order equation for the displacement vector
\begin{equation}                                        \label{eqsepu}
  (1-2\s)\triangle u_i+\pl_i\pl_j u^j=0,
\end{equation}
where
\begin{equation*}
  \s=\frac\lm{2(\lm+\mu)}
\end{equation*}
is the Poisson ratio, ($-1\le\s\le1/2$), and $\triangle$ is the Laplace operator.
It can be rewritten in terms of the induced metric (\ref{emetri}), for which we
obtain a first-order equation. We choose precisely this equation as the
gauge condition fixing the diffeomorphisms. We note that the gauge condition is
not uniquely defined because the induced metric is nonlinear in the displacement
vector, and different equations for the metric may have the same linear
approximation. We give two possible choices,
\begin{align}                                           \label{eonead}
  g^{\mu\nu}\overset{\circ}{\nb}_\mu g_{\nu\rho}
  +\frac\s{1-2\s}g^{\mu\nu}\overset{\circ}{\nb}_\rho g_{\mu\nu}&=0,
\\                                                      \label{etwoad}
  \overset\circ g{}^{\mu\nu}\overset{\circ}{\nb}_\mu g_{\nu\rho}
  +\frac\s{1-2\s}\overset\circ\nb_\rho g^T&=0,
\end{align}
where we introduced the notation for the trace of metric
$g^T=\overset\circ g{}^{\mu\nu}g_{\mu\nu}$. Gauge conditions (\ref{eonead})
and (\ref{etwoad}) are understood in the following way. The metric
$\overset\circ g_{\mu\nu}$ is the Euclidean metric written in an arbitrary
coordinate system, for example, in the cylindrical or spherical coordinate
system. The covariant derivative $\overset\circ\nb_\mu$ is built from the
Christoffel symbols corresponding to the metric $\overset\circ g_{\mu\nu}$,
and $\overset\circ\nb_\mu\overset\circ g_{\nu\rho}=0$ as a consequence.
The metric $g_{\mu\nu}$ is the metric describing dislocations [an exact
solution of the equilibrium equations for the free energy (\ref{eLagfi})].
The gauge conditions differ because in the first and second cases the
contraction is performed with the metric of dislocation $g^{\mu\nu}$ and
the Euclidean metric $\overset\circ g{}^{\mu\nu}$, respectively, without
changing the linear approximation. Both gauge conditions yield Eqn
(\ref{eqsepu}) in the linear approximation in the displacement vector
(\ref{emetri}). This is most easily verified in Cartesian coordinates.

From the geometric standpoint, we have the following. The medium with
dislocations is diffeomorphic to the Euclidean space $\MR^3$ equipped
with two metrics $\overset\circ g_{\mu\nu}$ and $g_{\mu\nu}$. The metric
$\overset\circ g_{\mu\nu}$ is a flat Euclidean metric written in an
arbitrary coordinate system. The metric $g_{\mu\nu}$ is not flat and
describes the distribution of dislocations in the same coordinate system.
In fact, the metric $\overset\circ g_{\mu\nu}$ is used only to fix the
coordinate system in which the metric $g_{\mu\nu}$ is measured.

If the solution of the equilibrium equations satisfies one of the gauge
conditions (\ref{eonead}), (\ref{etwoad}), written, for example, in cylindrical
coordinate system then we say that the solution is found in the cylindrical
coordinates. We suppose here that the distribution of dislocations in
elastic media in the laboratory cylindrical coordinate system is described
by this particular solution. Analogously, we may seek solutions in a Cartesian,
spherical, or any other coordinate system.

Gauge conditions may also be written for the vielbein $e_\mu{}^i$, which
is defined by Eqn (\ref{etetra}). This involves additional arbitrariness
because the vielbein is defined up to local rotations. This invariance leads
to different linear approximations for the vielbein in terms of the
displacement vector. We consider two possibilities in Cartesian coordinates:
\begin{align}                                           \label{etrfip}
  e_{\mu i}&\approx\dl_{\mu i}-\pl_\m u_i,
\\                                                      \label{etrsep}
  e_{\mu i}&\approx\dl_{\mu i}-\frac12(\pl_\mu u_i+\pl_i u_\mu),
\end{align}
where the index is lowered with the help of the Kronecker symbol.
For these possibilities and gauge condition (\ref{etwoad}) we have two
gauge conditions for the vielbein
\begin{align}                                           \label{ethrad}
  \overset\circ g{}^{\mu\nu}\overset{\circ}{\nb}_\mu e_{\nu i}
  +\frac1{1-2\s}\overset\circ e{}^\mu{}_i\overset\circ\nb_\mu e^T&=0,
\\                                                      \label{efouad}
  \overset\circ g{}^{\mu\nu}\overset{\circ}{\nb}_\mu e_{\nu i}
  +\frac\s{1-2\s}\overset\circ e{}^\mu{}_i\overset\circ\nb_\mu e^T&=0,
\end{align}
where $e^T=\overset\circ e{}^\mu{}_ie_\mu{}^i$. These conditions differ
in the coefficient before the second term. We note that in a curvilinear
coordinate system, the covariant derivative $\overset\circ\nb_\mu$ must
also include the flat $\MS\MO(3)$ connection acting on indices $i,j$.
One can also write other possible gauge conditions having the same
linear approximation. The question of the correct choice is unanswered at
present and outside the scope of this review. At the moment, we want only
to demonstrate that the system of coordinates must be fixed, and that the
gauge condition depends on the Poisson ratio, which is the experimentally
observed quantity.

Gauge conditions (\ref{ethrad})--(\ref{efouad}) are first order
equations by themselves and have some arbitrariness. Therefore, to
fix a solution uniquely, we must impose additional boundary conditions
on the vielbein for any given problem.

If the defect-free case, $T_{\mu\nu}{}^i=0$, $R_{\mu\nu j}{}^i=0$, and the
equilibrium equations are satisfied because the Euler--Lagrange equations
for (\ref{eLagfi}) are satisfied. In this and only in this case,
we can introduce a displacement vector, and the elastic gauge reduces to
the equations of nonlinear elasticity theory. In the presence of defects,
a displacement field does not exist, and the elastic gauge simply defines
the vielbein.

In choosing the free energy functional, we required that the conditions
$R_{\mu\nu}{}^{ij}=0$ and $T_{\mu\nu}{}^i=0$ satisfy the Euler--Lagrange
equations. This is important because otherwise we would obtain an additional
condition on the displacement vector (the Euler--Lagrange equations) besides
the elasticity theory equations following from the elastic gauge.

We stress an important point once again. In the geometric theory of defects,
we assume that there is a preferred laboratory coordinate system in which
measurements are made. This coordinate system is related to the medium
without defects and elastic stresses and corresponds to the flat Euclidean
space $\MR^3$. Gauge conditions (\ref{eonead}), (\ref{etwoad}) and
(\ref{ethrad}), (\ref{efouad}) are written precisely in this Euclidean space
$\MR^3$ and contain a measurable quantity, the Poisson ratio $\s$.
This property essentially distinguishes the geometric theory of defects from
the models of gravity in which all coordinate systems are considered equivalent.

The elastic gauge is used to fix diffeomorphisms. The expression for the
free energy in (\ref{eLagfi}) is also invariant under local $\MS\MO(3)$
rotations, and they must also be fixed. For this, we recently proposed the
Lorentz gauge for the connection \cite{Katana04}
\begin{equation}                                        \label{elorga}
  \pl_\mu\om_{\mu j}{}^i=0.
\end{equation}
This gauge is written in the laboratory Cartesian coordinate system and has
deep physical meaning. That is, let disclinations be absent
($R_{\mu\nu j}{}^i=0$). Then the $\MS\MO(3)$ connection is a pure gauge:
\begin{equation*}
  \om_{\mu j}{}^i=\pl_\mu S^{-1}{}_j{}^k S_k{}^i,~~~~S_j{}^i\in\MS\MO(3).
\end{equation*}
In this case, the Lorentz gauge reduces to the principal chiral
$\MS\MO(3)$-field equations
\begin{equation*}
  \pl_\mu(\pl_\mu S^{-1}{}_j{}^k S_k{}^i)=0
\end{equation*}
for the spin structure $\om^{ij}(x)$. Principal chiral field models (see,
e.g., [64--68]) for different groups
\nocite{ZaMaNoPi80E,Rajara82,TakFad86E,Zakrze89,RybSan01E}
and in a different number of dimensions attract much interest in mathematical
physics because they admit solutions of topological soliton types and
find broad applications in physics.

Thus, the Lorentz gauge (\ref{elorga}) means the following. In the absence
of disclinations, equations of equilibrium are identically satisfied, and
there exists a field $\om^{ij}$ that satisfies equations for the principal
chiral field. By this we mean that the spin structure of the medium is described
by the model of the principal chiral field in the defect-free case.

The principal chiral field model is not the only one that can be used
for fixing local rotations. The Skyrme model \cite{Skyrme61} can also be used
for this purpose. The Euler--Lagrange equations for this model are not
difficult to rewrite in terms of the $\MS\MO(3)$ connection and use as
the gauge conditions.

There are other models for spin structures. For describing the distribution
of magnetic moments in ferromagnets or the director field in liquid crystals,
one uses the expression for the free energy depending on the vector $n$-field
itself \cite{LanLif82,LanLif70}. Lately, much attention is paid to
the Faddeev model of the $n$-field \cite{Faddee77}. The question whether there
are gauge conditions on the $\MS\MO(3)$ connection that yield these models
in the absence of disclinations is unanswered at present.

Thus, we pose the following problem in the geometric theory of defects:
to find the solution of the Euler--Lagrange equations for free energy
(\ref{eLagfi}) that satisfies the elastic gauge for the vielbein and
the Lorentz gauge for the $\MS\MO(3)$ connection. In sections \ref{swedie}
and \ref{swegeo}, we solve this problem for the wedge dislocation in the
framework of the classical elasticity theory and the geometric theory of
defects, respectively, and afterwards compare the obtained results.
\section{Asymmetric elasticity theory    }
In the preceeding section, we used the elasticity theory and the principal
chiral $\MS\MO(3)$-field model to fix the invariance of free energy
(\ref{eLagfi}) in the geometric theory of defects. This is not a unique
possibility, because other models may be used for gauge fixing. In the
present section, we show how another model -- asymmetric elasticity theory
--  can be used for fixing diffeomorphisms and local rotations.

At the beginning of the last century, the Cosserat brothers developed
the theory of elastic media, every point of which is characterized not
only be its position but also by its orientation in space \cite{CosCos09},
i.e., a vielbein is specified at every point (Fig.\ref{fcosme}).
\begin{floatingfigure}{.45\textwidth}
\includegraphics[width=.4\textwidth]{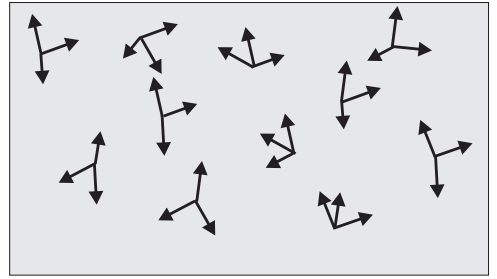}
 \caption{Every point of the Cosserat medium is characterized not only
 by its position but also by its orientation in space.}
 \label{fcosme}
\end{floatingfigure}
From physical standpoint, this means that every atom in the crystalline
structure is not a point but an extended object having orientation.
In this case, the stress tensor is no longer a symmetric tensor, and the
corresponding theory is called the asymmetric elasticity theory. Contemporary
exposition of this approach is given in \cite{Nowack85}. In the present
section, we show that the asymmetric theory of elasticity is naturally
incorporated into the geometric theory of defects.

The main variables in the asymmetric theory of elasticity are the displacement
vector $u^i(x)$ and the rotation angle $\om^i(x)$. The direction of the
pseudovector $\om^i$ coincides with the rotation axis of the medium element
and its length is equal to the angle of rotation. The angle of rotation
discussed in section \ref{sdiscl} is dual to the spin structure field
$\om^{ij}(x)$: $\om_{ij}=\ve_{ijk}\om^k$.

The Cosserat medium is characterized by the stress tensor $\s^{ij}(x)$
(the density of forces acting on the surface with normal $i$ in the
direction $j$) and the torque stress tensor  $\mu^{ij}(x)$ (the density
of torques acting on the surface with normal $i$ in the direction $j$).
The Cosserat medium is in equilibrium if forces and torques are balanced
at every point,
\begin{align}                                           \label{enewfo}
  &\pl_j\s^{ji}+f^i=0,
\\                                                      \label{emofor}
  &\ve^{ijk}\s_{jk}+\pl_j\mu^{ji}+m^i=0,
\end{align}
where $f^i(x)$ and $m^i(x)$ are the densities of nonelastic external forces
and torques. As a consequence of Eqn (\ref{emofor}), the stress tensor is
symmetric if and only if the condition $\pl_j\mu^{ji}+m^i=0$ is satisfied.

The displacement field and rotation angle uniquely define the
deformation tensor $\e_{ij}(x)$ and the twist tensor $\kappa_{ij}(x)$:
\begin{equation}                                        \label{edeizk}
\begin{split}
  \e_{ij}&=\pl_i u_j-\om_{ij},
\\
  \kappa_{ij}&=\pl_i\om_j.
\end{split}
\end{equation}
In general, the deformation and twist tensors have no symmetry
in their indices.

Hook's law in the Cosserat medium is changed to two linear relations
connecting the stress and torque tensors with the deformation and twist
tensors,
\begin{align}                                           \label{estrte}
  \s_{ij}&=2\mu\e_{\lbrace ij\rbrace}+2\al\e_{[ij]}+\lm\dl_{ij}\e_k{}^k,
\\                                                      \label{esizkr}
  \mu_{ij}&=2\g\kappa_{\lbrace ij\rbrace}+2\e\kappa_{[ij]}
  +\bt\dl_{ij}\kappa_k{}^k,
\end{align}
where $\mu,\lm$ are the Lam\'e coefficients, and $\al,\bt,\g,\e$ are four
new elastic constants characterizing the medium. Braces and square brackets
denote symmetrization and antisymmetrization of indices, respectively.

In the case where
\begin{equation}                                        \label{eresye}
  \om_{ij}=\frac12(\pl_i u_j-\pl_j u_i),
\end{equation}
the deformation tensor is symmetric and has the previous form (\ref{edefte}).
Equation (\ref{emofor}), together with (\ref{estrte}) and (\ref{esizkr}),
then reduces to the equation
\begin{equation*}
  (\g+\e)\ve^{ijk}\triangle\pl_j u_k+m^i=0.
\end{equation*}
The first term vanishes as a consequence of Eqn (\ref{eqsepu}). Thus,
for spin structure (\ref{eresye}) and $m^i=0$ we return to the symmetric
elasticity theory.

Equations (\ref{enewfo}), (\ref{emofor}), (\ref{estrte}), and (\ref{esizkr}),
together with the boundary conditions, define the equilibrium state of
Cosserat media. We now show how this model is included in the geometric
theory. First, we note that in the absence of defects ($T_{\mu\nu}{}^i=0$,
$R_{\mu\nu}{}^{ij}=0$), the fields $u^i$ and $\om^{ij}$ exist.
Then the vielbein and the $\MS\MO(3)$ connection are defined by the
deformation and twist tensors in the linear approximation:
\begin{align}                                           \label{edefta}
  &e_\mu{}^i=\pl_\mu y^j S_j{}^i(\om)
  \approx(\dl_\mu^j-\pl_\mu u^j)(\dl_j^i+\om_j{}^i)\approx\dl_\mu^i-\e_\mu{}^i,
\\                                                      \label{eizkrt}
  &\om_\mu{}^{ij}\approx\pl_\mu\om^{ij}=\ve^{ijk}\kappa_{\mu k}.
\end{align}

We note that relations (\ref{edeizk}) can be regarded as equations for the
displacement vector and rotation angle. The corresponding integrability
conditions were obtained in \cite{Sandru66}. These integrability conditions
are the linear approximations of equalities $T_{\mu\nu}{}^i=0$ and
$R_{\mu\nu}{}^{ij}=0$ defining the absence of defects.

If nonelastic forces and torques are absent ($f^i=0$, $m^i=0$),
then the asymmetric theory of elasticity reduces to second-order
equations for the displacement vector and rotation angle:
\begin{align}                                           \label{eqtrfi}
  &(\mu+\al)\triangle u^i+(\mu-\al+\lm)\pl^i\pl_j u^j-2\al\pl_j\om^{ji}=0,
\\                                                      \label{eqstco}
  &(\g+\e)\triangle\om^i+(\g-\e+\bt)\pl^i\pl_j\om^j
  +2\al\ve^{ijk}(\pl_j u_k-\om_{jk})=0.
\end{align}
We rewrite these equations for the vielbein and $\MS\MO(3)$ connection
\begin{align}                                           \label{etrgco}
  (\mu+\al)\overset\circ\nabla{}^\mu e_\mu{}^i+(\mu-\al+\lm)
  \overset\circ\nabla{}^i e^T-(\mu-\al)\om_\mu{}^{\mu i}&=0,
\\                                                      \label{esegco}
  \frac12(\g+\e)\ve^{ijk}\overset\circ\nabla{}^\mu\om_{\mu jk}
  +\frac12(\g-\e+\bt)\ve^{\mu jk}\overset\circ\nabla{}^i\om_{\mu jk}
  +2\al\ve^{i\mu j}e_{\mu j}&=0.
\end{align}
Of course, these are not the only equations that coincide with
Eqns (\ref{eqtrfi}) and (\ref{eqstco}) in the linear approximation.
At present, we do not have arguments for the unique choice. The derived
nonlinear equations of the asymmetric elasticity theory can be used as
gauge conditions in the geometric theory of defects. Here, we have six
equations for fixing diffeomorphisms (three parameters) and local
$\MS\MO(3)$ rotations (three parameters). Thus, the asymmetric theory of
elasticity is naturally embedded in the geometric theory of defects.

In section \ref{sfixga}, we considered the elastic gauge for the vielbein
and the Lorentz gauge for the $\MS\MO(3)$ connection. In this case,
the spin structure variables do not interact with elastic deformations
when defects are absent. In the asymmetric elasticity theory, the elastic
stresses directly influence the spin structure and vice versa.
\section{Wedge dislocation in the elasticity theory    \label{swedie}}
\begin{floatingfigure}{.45\textwidth}
\includegraphics[width=.4\textwidth]{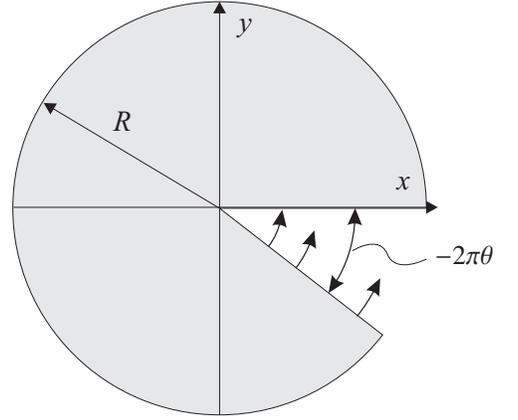}
 \caption{Wedge dislocation with the deficit angle $2\pi\theta$. For
 negative and positive $\theta$, the wedge is cut out or added, respectively.}
 \label{fwedis}
\end{floatingfigure}

By wedge dislocation, we understand an elastic medium that is topologically
the Euclidean space $\MR^3$ without the $z=x^3$ axis -- the core of
dislocation, obtained in the following way. We consider the infinite
elastic medium without defects and cut out the infinite wedge of angle
$-2\pi\theta$. For definiteness, we assume that the edge of the wedge
coincides with the $z$ axis (Fig.\ref{fwedis}).
The edges of the cut are then moved symmetrically one to the other and
glued together. After that, the medium moves to the equilibrium state under
the action of elastic forces. If the wedge is cut out from the medium, then
the angle is considered negative: $-1<\theta<0$. For positive $\theta$, the
wedge is added. Thus, the elastic medium initially occupies a domain
greater or lesser than the Euclidean space $\MR^3$, depending on the sign of
the deficit angle $\theta$; in the cylindrical coordinates $r,\vf,z$ this
domain is described by the inequalities
\begin{equation}                                        \label{ecylco}
  0\le r<\infty,~~0\le\vf\le 2\pi\al,~~-\infty<z<\infty,~~~~\al=1+\theta.
\end{equation}

We note that the wedge dislocation is often called disclination. In our
approach, this term seems unnatural because the wedge dislocation is related
to a nontrivial torsion. Moreover, the term disclination is used for
defects in the spin structure.

We now proceed with the mathematical formulation of the problem for the wedge
dislocation in the framework of elasticity theory. To avoid divergent
expressions arising for infinite media, we suppose that the wedge dislocation
is represented by the cylinder of a finite radius $R$. This problem has
translational symmetry along the $z$ axis and rotational symmetry in the
$x,y$ plane. Therefore, we use cylindrical coordinate system. Let
\begin{equation}                                        \label{ecovci}
  \hat u_i=(\hat u_r,\hat u_\vf,\hat u_z)
\end{equation}
be the components of the displacement covector with respect to orthonormal
basis in the cylindrical coordinate system. This covector satisfies the
equilibrium equation following the substitution of (\ref{eHook}) in
Eqn (\ref{estati}) in domain (\ref{ecylco}),
\begin{equation}                                        \label{qeeste}
  (1-2\s)\triangle\hat u_i+\overset\circ\nb_i\overset\circ\nb_j\hat u^j=0,
\end{equation}
where $\overset{\circ}{\nb}_i$ is the covariant derivative for the flat
Euclidean metric in the considered coordinate system.

For references, we write expressions for the divergence and Laplacian
for a covector field in the cylindrical coordinate system:
\begin{align*}                                               \nonumber
  \overset\circ\nb_i\hat u^i&=\frac1r\pl_r(r\hat u^r)
  +\frac1r\pl_\vf\hat u^\vf+\pl_z\hat u^z,
\\
  \triangle\hat u_r&=\frac1r\pl_r(r\pl_r\hat u_r)
  +\frac1{r^2}\pl^2_\vf\hat u_r+\pl^2_z\hat u_r-\frac1{r^2}\hat u_r
  -\frac2{r^2}\pl_\vf\hat u_\vf,
\\
  \triangle\hat u_\vf&=\frac1r\pl_r(r\pl_r\hat u_\vf)
  +\frac1{r^2}\pl^2_\vf\hat u_\vf+\pl^2_z\hat u_\vf
  -\frac1{r^2}\hat u_\vf+\frac2{r^2}\pl_\vf\hat u_r,
\\
  \triangle\hat u_z&=\frac1r\pl_r(r\pl_r\hat u_z)
  +\frac1{r^2}\pl^2_\vf\hat u_z+\pl^2_z\hat u_z.
\end{align*}
Taking the symmetry of the problem into account, we seek the solution
of Eqn (\ref{qeeste}) in the form
$$
  \hat u_r=u(r),~~~~\hat u_\vf=A(r)\vf,~~~~\hat u_z=0
$$
where $u(r)$ and $A(r)$ are two unknown functions depending only on the
radius. We impose the boundary conditions:
\begin{equation}                                        \label{ebocow}
  \left.\hat u_r\right|_{r=0}=0,~~~~
  \left.\hat u_\vf\right|_{r=0}=0,~~~~
  \left.\hat u_\vf\right|_{\vf=0}=0,~~~~
  \left.\hat u_\vf\right|_{\vf=2\pi\al}=-2\pi\theta r,~~~~
  \left.\pl_r\hat u_r\right|_{r=R}=0.
\end{equation}
The first four equations are geometrical and correspond to the process
of dislocation creation. The last condition has simple physical meaning:
the absence of external forces at the boundary of the medium. The unknown
function $A(r)$ is found from the second to the last boundary condition
(\ref{ebocow})
$$
  A(r)=-\frac{\theta}{1+\theta}r.
$$
Straightforward substitution then shows that $\vf$ and $z$
components of equation of equilibrium (\ref{qeeste}) are identically
satisfied, and the radial component reduces to the equation
$$
  \pl_r(r\pl_r u)-\frac ur=D,~~~~
  D=-\frac{1-2\s}{1-\s}\frac\theta{1+\theta}=\const.
$$
The general solution of this equation is
$$
  u=\frac D2 r\ln r+c_1 r+\frac{c_2}r,~~~~c_{1,2}=\const.
$$
The constant of integration $c_2=0$ due to the boundary condition at zero.
The constant $c_1$ is found from the last boundary condition in (\ref{ebocow}).
Finally, we obtain the known solution of the considered problem \cite{Kosevi81}
\begin{equation}                                        \label{ewedis}
\begin{split}
  \hat u_r&=\frac D2r\ln\frac r{\ex R},
\\
  \hat u_\vf&=-\frac\theta{1+\theta}r\vf.
\end{split}
\end{equation}
The letter $\ex$ in the expression for $\hat u_r$ denotes the base of the
natural logarithm. We note that the radial component of the displacement
vector diverges as $R\to\infty$. This means that the description of the
wedge dislocation requires considering a finite-radius cylinder.

The linear elasticity theory is applicable for small relative displacements,
which for a wedge dislocation are equal to
$$
  \frac{d\hat u_r}{dr}=-\frac\theta{1+\theta}\frac{1-2\s}{2(1-\s)}
  \ln\frac rR,~~~~
  \frac1r\frac{d\hat u_\vf}{d\vf}=-\frac\theta{1+\theta}.
$$
This means that we are able to expect correct results for the displacement
field for small deficit angles ($\theta\ll 1$) and near the boundary of the
cylinder ($r\sim R$).

We find the metric induced by the wedge dislocation in the linear
approximation in the deficit angle $\theta$. Calculations can be performed
using the general formulas (\ref{emetri}) or the known expression for
the variation of the form of the metric
\begin{equation}                                        \label{einmec}
  \dl g_{\mu\nu}=-\overset{\circ}{\nb}_\mu u_\nu-\overset{\circ}{\nb}_\nu u_\mu.
\end{equation}
After simple calculations, we obtain the expression for the
metric in the $x,y$ plane:
\begin{equation}                                        \label{emewed}
  ds^2=\left(1+\theta\frac{1-2\s}{1-\s}\ln\frac rR\right)dr^2
  +r^2\left(1+\theta\frac{1-2\s}{1-\s}\ln\frac rR
  +\theta\frac1{1-\s}\right)d\vf^2.
\end{equation}
This metric is compared with the metric obtained as the solution of
three-dimensional Einstein equations in section \ref{swegeo}.
\section{Edge dislocation in the elasticity theory     \label{sediel}}
\begin{figure}[h,b,t]
\hfill\includegraphics[width=.95\textwidth]{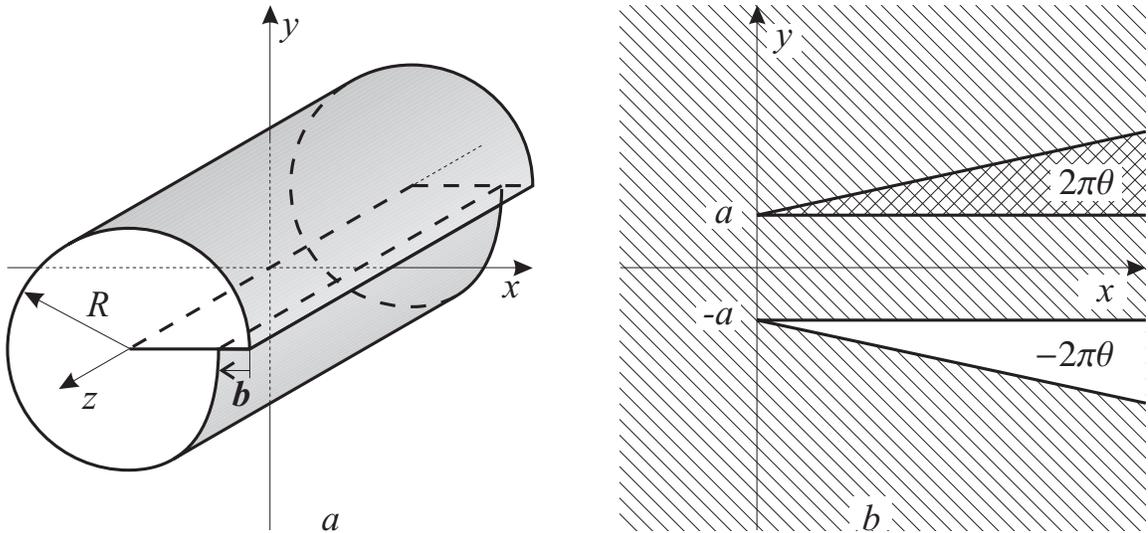}
\hfill {}
\\
\centering \caption{Edge dislocation with the Burgers vector $\Bb$
directed to the dislocation axis (\textit{a}). Edge dislocation as
the dipole of two wedge dislocations with positive and negative
deficit angles (\textit{b}).\label{feddis}}
\end{figure}

Wedge dislocations are relatively rarely met in nature because they
require a much amount of a medium to be added or removed, resulting in a
vast quantity of energy expenses. Nevertheless, their study is of great
importance because other linear dislocations can be expressed as a
superposition of wedge dislocations. In this respect, wedge dislocations
are elementary. We show this for an edge dislocation -- one of the most
widely spread dislocations, as an example. An edge dislocation with the
core coinciding with the $z$ axis is shown in Fig.\ref{feddis}{\it a}.
It appears as the result of cutting the medium over the half-plane $y=0$,
$x>0$, moving the lower edge of the cut towards the $z$ axis on a constant
(far from the core of the dislocation) Burgers vector $\Bb$, and subsequently
gluing the edges. To find the displacement vector field for the edge
dislocation, we may solve the corresponding boundary value problem for
the equilibrium equations (\ref{qeeste}) \cite{LanLif70}. However, we
follow another way, knowing the explicit form of the displacement vector
for a wedge dislocation. The edge dislocation is represented by the
dipole of two wedge dislocations with positive, $2\pi\theta$, and negative,
$-2\pi\theta$, deficit angles as shown in Fig.\ref{feddis}{\it b}. We
assume that the axes of the first and second wedge dislocations are
parallel to the $z$ axis and intersect the $x,y$ plane at points with
the respective coordinates $(0,a)$ and $(0,-a)$. The distance between
the wedge dislocation axes is equal to $2a$. It follows from the expression
for the displacement field (\ref{ewedis}) that far away from the origin
($r\gg a$), the displacement field for the wedge dislocations has the
following form in the first order in small $\theta$ and $a/r$:
\begin{align}                                           \label{efiwdi}
\begin{split}
  u_x^{(1)}&\approx-\theta\left[
  \frac{1-2\s}{2(1-\s)}x\ln\frac{r-a\sin\vf}{\ex R}
  -(y-a)\left(\vf-\frac{a\cos\vf}r\right)\right],
\\
  u_y^{(1)}&\approx-\theta\left[
  \frac{1-2\s}{2(1-\s)}(y-a)\ln\frac{r-a\sin\vf}{\ex R}
  +x\left(\vf-\frac{a\cos\vf}r\right)\right],
\end{split}
\\                                                      \label{esewdi}
\begin{split}
  u_x^{(2)}&\approx\phantom{-}\theta\left[
  \frac{1-2\s}{2(1-\s)}x\ln\frac{r+a\sin\vf}{\ex R}
  -(y+a)\left(\vf+\frac{a\cos\vf}r\right)\right],
\\
  u_y^{(2)}&\approx\phantom{-}\theta\left[
  \frac{1-2\s}{2(1-\s)}(y+a)\ln\frac{r+a\sin\vf}{\ex R}
  +x\left(\vf+\frac{a\cos\vf}r\right)\right].
\end{split}
\end{align}
It is sufficient to sum displacement fields (\ref{efiwdi}) and (\ref{esewdi})
to find the displacement field for the edge dislocation because
the elasticity theory equations are linear. After simple calculations,
up to the translation of the whole medium by a constant vector along
the $y$ axis, we obtain
\begin{equation}                                        \label{evedgd}
\begin{split}
  u_x&=\phantom{-}b\left[\arctg\frac yx
  +\frac1{2(1-\s)}\frac{xy}{x^2+y^2}\right],
\\
  u_y&=-b\left[\frac{1-2\s}{2(1-\s)}\ln\frac r{\ex R}
  +\frac1{2(1-\s)}\frac{x^2}{x^2+y^2}\right],
\end{split}
\end{equation}
where we have introduced the notation for the modulus of the Burgers vector
$$
  b=|\Bb|=-2a\theta.
$$
This result coincides with the expression for the displacement field
obtained by direct solution of the elasticity theory equations
\cite{LanLif70}. Thus, we have shown that an edge dislocation is the
dipole of two parallel wedge dislocations.

We next find the metric induced by the edge dislocation. Using
formulas (\ref{einmec}), we obtain the metric in the $x,y$ plane in the
linear approximation in $\theta$ and $a/r$:
\begin{equation}                                        \label{edgelt}
  dl^2=\left(1+\frac{1-2\s}{1-\s}\frac br\sin\vf\right)
  \left(dr^2+r^2 d\vf^2\right)-\frac{2b\cos\vf}{1-\s}drd\vf.
\end{equation}
We note that the induced metric for an edge dislocation does not depend
on $R$.
\section{Parallel wedge dislocations}
In the absence of disclinations ($R_{\mu\nu i}{}^j=0$), the
$\MS\MO(3)$ connection is a pure gauge, and equations of equilibrium
for the $\MS\MO(3)$ connection (\ref{econne}) are identically satisfied.
In this case, the explicit form of the $\MS\MO(3)$ connection is uniquely
determined by the spin structure $\om^{ij}$. The field $\om^{ij}$ satisfies
equations for the principal chiral field as a consequence of the Lorentz
gauge in (\ref{elorga}). Solution of this system of equations defines the
trivial $\MS\MO(3)$ connection. Thus in the absence of disclinations, the
problem is reduced to solution of the Einstein equations for the vielbein
in the elastic gauge and solution of the principal chiral field model
for the spin structure. After that, we can compute the torsion
tensor through the formulas (\ref{ecurcv}), which defines the surface
density of the Burgers vector.

Because disclinations are absent (the curvature tensor is equal to zero), we
have the space of absolute parallelism. The whole geometry is then defined
by the vielbein $e_\mu{}^i$, which uniquely defines the torsion tensor via
(\ref{ecurcv}) for a vanishing $\MS\MO(3)$ connection. Here, we assume that
a trivial $\MS\MO(3)$ connection is equal to zero. The vielbein $e_\mu{}^i$
satisfies the three-dimensional Einstein equations with a Euclidean signature
metric, which follow from the expression for the free energy (\ref{eLagfi})
for $R_{\mu\nu j}{}^i=0$,
\begin{equation}                                        \label{eindis}
  \widetilde R_{\mu\nu}-\frac12 g_{\mu\nu}\widetilde R=T_{\mu\nu}.
\end{equation}
Here, we have added the source of dislocations $T_{\mu\nu}$ to the right hand
side of Einstein equations (it is the energy-momentum tensor in gravity).

We note that without a source of dislocations, the model would be trivial.
Indeed, the scalar curvature and Ricci tensor are equal to zero:
$\widetilde R=0$, $\widetilde R_{\mu\nu}=0$ for $T_{\mu\nu}=0$ as a
consequence of Einstein equations (\ref{eindis}). Then, the full
curvature tensor without sources is also equal to zero because in
three-dimensional space, it is in a one-to-one correspondence with
Ricci tensor (\ref{ecurir}). The vanishing of the full curvature tensor means
the triviality of the model because defects are absent in this case. The
similar statement in three-dimensional gravity is well known. It is
usually formulated as: ``Three-dimensional gravity does not describe
dynamical, i.e., propagating degrees of freedom''.

For our purposes, we have to find the solution of Einstein equations
(\ref{eindis}) describing one wedge dislocation. The Einstein equations
are a system of nonlinear second-order partial differential equations.
Not too many exact solutions are known at present, even in a
three-dimensional space. The remarkable exact solution describing an
arbitrary static distribution of point particles is well known in
three-dimensional gravity for the Lorentz signature metric $(+--)$ [75--77].
\nocite{Starus63,Clemen76,DeJatH84} We find this solution
for the Euclidean signature metric and show that it describes an arbitrary
distribution of parallel wedge dislocations in the geometric theory of
defects. Hence, we first consider the more general case of an arbitrary number
of wedge dislocations and then analyze in detail one wedge dislocation
which is of interest to us. We do this deliberately because the solution in
a more general case does not involve essential complications. At the
same time, an arbitrary distribution of wedge dislocations is much more
interesting for applications. For example, the edge dislocation was shown
in the preceeding section to be represented by a dipole of two wedge
dislocations of different signs.

We consider elastic medium with arbitrary distributed but parallel
wedge dislocations. We choose the coordinate system such that
the $z=x^3$ axis is parallel to dislocations axes, and axes
$\lbrace x^\al\rbrace=\lbrace x,y\rbrace$, $\al=1,2$ are perpendicular
to the $z$ axis. Then the metric has the block-diagonal form
\begin{equation}                                        \label{eqmetr}
    ds^2=dl^2+N^2dz^2,
\end{equation}
where
\begin{equation*}
  dl^2=g_{\al\bt}dx^\al dx^\bt
\end{equation*}
is a two-dimensional metric on the $x,y$ plane. A two-dimensional metric
$g_{\al\bt}(x,y)$ and a function $N(x,y)$ are independent of $z$ due to
translational symmetry along the $z$ axis.

We can say this more simpler dropping the physical arguments. We consider the
block-diagonal metric of form (\ref{eqmetr}), which has translational
invariance along the $z$ axis. Afterwards, we show that the
corresponding solution of the Einstein equations indeed describes an
arbitrary distribution of parallel wedge dislocations.

The curvature tensor for metric (\ref{eqmetr}) has the components:
\begin{align*}
  \widetilde R_{\al\bt\g}{}^\dl&=R^{(2)}_{\al\bt\g}{}^\dl,~~~~~~~~~~
  \widetilde R_{\al z\g}{}^z=\frac1N\nb_\al\nb_\g N,
\\
  \widetilde R_{\al\bt\g}{}^z&=\widetilde R_{\al z\g}{}^\dl=0,
\end{align*}
where $R^{(2)}_{\al\bt\g}{}^\dl$ is the curvature tensor for the
two-dimensional metric $g_{\al\bt}$ and $\nb_\al$ is the two-dimensional
covariant derivative with the Christoffel symbols.

We choose the source of dislocations as
\begin{equation}                                        \label{edisso}
\begin{split}
  T_{zz}&=\frac{2\pi}{\sqrt{g^{(2)}}}\sum_{n=1}^M\theta_n\dl^{(2)}(\Br-\Br_n),
\\
  T_{\al\bt}&=T_{\al z}=T_{z\al}=0,
\end{split}
\end{equation}
where $\dl^{(2)}(\Br-\Br_n)=\dl(x-x_n)\dl(y-y_n)$ is the two-dimensional
$\dl$-function on the $x,y$ plane with the support at a point $\Br_n=(x_n,y_n)$.
The factor $g^{(2)}=\det g_{\al\bt}$ in front of the sum sign is due to the
property of the $\dl$-function, which is not a function but a tensor density
with respect to general coordinate transformations.
We show later that the solution of the Einstein equations with such a
source describes $M$ parallel wedge dislocations with deficit angles
$\theta_n$, which intersect the $x,y$ plane at the points $(x_n,y_n)$. In
three-dimensional gravity this source corresponds to particles of masses
$m_n=2\pi\theta_n$, being at rest at the points $\Br_n$.

Einstein equations (\ref{eindis}) then reduce to four equations,
\begin{align}                                        \label{eEinon}
  \nb_\al\nb_\bt N-g_{\al\bt}\nb^\g\nb_\g N&=0,
\\                                                      \label{eEintw}
    -\frac12N^3R^{(2)}&=\frac{2\pi}{\sqrt{g^{(2)}}}\sum_{n=1}^M
    \theta_n\dl^{(2)}(\Br-\Br_n),
\end{align}
where $R^{(2)}$ is the two-dimensional scalar curvature.

The metric of form (\ref{eqmetr}) is still invariant under coordinate
transformations in the $x,y$ plane. Using this residual symmetry,
we fix the conformal gauge on the plane (this is always possible locally)
$$
  g_{\al\bt}=e^{2\phi}\dl_{\al\bt},
$$
where $\phi(x,y)$ is some function.

In the conformal gauge, Eqn (\ref{eEinon}) becomes
\begin{equation*}
  \pl_\al\pl_\bt N=0.
\end{equation*}
For constant boundary conditions for $N$ on the boundary of the $x,y$ plane
this equation has the unique solution $N=\const$. Changing the scale of
$z$ coordinate, we can set $N=1$ without loss of generality. Then Eqn
(\ref{eEintw}) reduces to the Poisson equation
$$
  \triangle\phi=-2\pi\sum_n\theta_n\dl^{(2)}(\Br-\Br_n),
$$
which has the general solution
$$
  \phi=\sum_n\theta_n\ln|\Br-\Br_n|+\frac12\ln C,~~~~~~C=\const>0.
$$
Thus the metric in the $x,y$ plane is
\begin{equation}                                        \label{edisco}
  dl^2=C\prod_n|\Br-\Br_n|^{2\theta_n}(dr^2+r^2d\vf^2),~~~~
  0\le r<\infty,~~0\le\vf<2\pi,
\end{equation}
where the polar coordinates $r,\vf$ cover the whole plane $\MR^2$ and
not more (this is important !). Any solution of the Einstein equations
is defined up to choosing the coordinate system because the equations are
covariant. Using this, we set $C=1$, which is always possible by choosing
the scale of $r$.

This is indeed the exact solution of the Einstein equations describing
arbitrary distribution of parallel wedge dislocations. This statement
is made clear from the following consideration.

We note that transition to a continuous distribution of dislocations in
the geometric approach is simple. For this, we have to substitute a
continuous distribution of sources in the right hand side of the Einstein
equations instead of $\dl$-sources.

We consider one wedge dislocation with the source at the origin in more
detail to show that metric (\ref{edisco}) indeed describes an
arbitrary distribution of wedge dislocations
\begin{equation}                                        \label{esouwd}
  T_{zz}=\frac{2\pi}{\sqrt{g^{(2)}}}\theta\dl^{(2)}(x,y).
\end{equation}
The corresponding metric (\ref{edisco}) for $C=1$ is
\begin{equation}                                        \label{eonwed}
  dl^2=r^{2\theta}(dr^2+r^2d\vf^2).
\end{equation}
We pass to a new coordinate system
\begin{equation}                                        \label{ekoorv}
  r'=\frac1\al r^\al,~~~~~~\vf'=\al\vf,~~~~~~\al=1+\theta,
\end{equation}
in which the metric becomes Euclidean
\begin{equation}                                        \label{equlco}
  dl^2=dr'^2+r'^2d\vf'^2,
\end{equation}
but the range of the polar angle differs now from $2\pi$:
$0\le\vf'<2\pi\al$, and covers the $x,y$ plane with removed or added
angle $2\pi\theta$.

Because the metric coincides with the Euclidean one in the primed
coordinate system $r',\vf'$, we have the Euclidean plane with a removed or
added wedge because the angle $\vf'$ varies within the interval $(0,2\pi\al)$.
The transformation to coordinates $r,\vf$ in (\ref{ekoorv}) means the gluing of
the edges of the wedge that has appeared, which produces a cone. Therefore,
both metrics (\ref{eonwed}) and (\ref{equlco}) describe the same geometric
object -- conical singularity. The torsion and curvature tensors are obviously
zero everywhere except at the origin.

Creation of a conical singularity coincides exactly with creation of
the wedge dislocation in the geometric theory of defects. It is not
difficult to show that general solution (\ref{edisco}) describes
an arbitrary distribution of conical singularities with deficit angles
$\theta_n$ at points $\Br_n$. Thus, this solution describes
an arbitrary distribution of parallel wedge dislocations.

In the next section, we consider one wedge dislocation in detail.
For this, we perform one more coordinate transformation
\begin{equation}                                        \label{ekoofv}
  f=\al r',~~~~~~\vf=\frac1\al\vf'.
\end{equation}
Metric (\ref{equlco}) then becomes
\begin{equation}                                        \label{ecosth}
  dl^2=\frac1{\al^2}df^2+f^2d\vf^2,~~~~\al=1+\theta.
\end{equation}
This is one more frequently used form of the metric for a conical singularity.
\section{Wedge dislocation in the geometric approach  \label{swegeo}}
We now consider a wedge dislocation from the geometric standpoint.
From the qualitative standpoint, the creation of a wedge dislocation coincides
with the definition of conical singularity. However, there is a quantitative
difference because metric (\ref{ecosth}) depends only on the deficit
angle $\theta$ and cannot coincide with the induced metric (\ref{emewed})
obtained within the elasticity theory. The difference arises because
we require the displacement vector in the equilibrium to satisfy
equilibrium equations after removing the wedge and gluing its edges
(creating conical singularity) in the elasticity theory. At the same
time the $x,y$ plane for a conical singularity after the gluing can be
deformed in an arbitrary way. Formally, this manifests itself in that
metric (\ref{emewed}) obtained within the elasticity depends explicitly on
the Poisson ratio, which is absent in gravity theory.

We proposed the elastic gauge for solving this problem \cite{Katana03}.
We choose elastic gauge (\ref{efouad}) as the simplest one for a
wedge dislocation. This problem can be solved in two ways. First, the gauge
condition can be inserted in the Einstein equations directly. Second, we
can find the solution in any suitable coordinate system and then find
the coordinate transformation such that the gauge condition is satisfied.

It is easier to follow the second way because the solution for the metric
is known, Eqn (\ref{ecosth}). The vielbein can be associated with metric
(\ref{ecosth}) as
\begin{equation*}
  e_r{}^{\hat r}=\frac1\al,~~~~e_\vf{}^{\hat\vf}=f.
\end{equation*}
Here, a hat over an index means that it corresponds to the orthonormal
coordinate system, and an index without a hat is the coordinate one.
Components of this vielbein are the square roots of the corresponding metric
components and therefore have the symmetric linear approximation (\ref{etrsep}).
We transform the radial coordinate $f\rightarrow f(r)$ because a wedge
dislocation is symmetric under rotations in the $x,y$ plane.
After that transformation the vielbein components take the form
\begin{equation}                                        \label{erepfu}
  e_r{}^{\hat r}=\frac{f'}\al,~~~~e_\vf{}^{\hat\vf}=f,
\end{equation}
where the prime denotes differentiation with respect to $r$. We choose the
vielbein corresponding to the Euclidean metric as
\begin{equation}                                         \label{eflrep}
    \overset\circ e_r{}^{\hat r}=1,~~~~\overset\circ e_\vf{}^{\hat\vf}=r.
\end{equation}
It defines the Christoffel symbols $\overset\circ\G_{\mu\nu}{}^\rho$ and
$\MS\MO(3)$ connection $\overset\circ\om_{\mu i}{}^j$, which define the
covariant derivative. We write only nontrivial components
\begin{equation*}
\begin{split}
  \overset\circ\G_{r\vf}{}^\vf&=\overset\circ\G_{\vf r}{}^\vf=\frac1r,~~~~~~
  \overset\circ\G_{\vf\vf}{}^r=-r,
\\
  \overset\circ\om_{\vf\hat r}{}^{\hat\vf}&
  =-\overset\circ\om_{\vf\hat\vf}{}^{\hat r}=1.
\end{split}
\end{equation*}
Substitution of the vielbein into gauge condition (\ref{efouad})
yields the Euler differential equation for the transition function
\begin{equation*}
  \frac{f''}\al+\frac{f'}{\al r}-\frac f{r^2}
  +\frac\s{1-2\s}\left(\frac{f''}\al+\frac{f'}r-\frac f{r^2}\right)=0.
\end{equation*}
Its general solution depends on two arbitrary constants $C_{1,2}$,
\begin{equation}                                    \label{eftrra}
  f=C_1r^{\g_1}+C_2r^{\g_2},
\end{equation}
where the exponents $\g_{1,2}$ are defined by the quadratic equation
\begin{equation*}
  \g^2+2\g\theta b-\al=0,~~~~b=\frac\s{2(1-\s)},
\end{equation*}
which has real roots for $\theta>-1$ with different signs: positive
root $\g_1$ and negative $\g_2$. We recall that there are thermodynamical
constraints $-1\le\s\le1/2$ on the Poisson ratio \cite{LanLif70}.

To fix the constants, we impose boundary conditions on the vielbein:
\begin{equation}                                        \label{eborep}
  \left. e_r{}^{\hat r}\right|_{r=R}=1,~~~~
  \left. e_\vf{}^{\hat\vf}\right|_{r=0}=0.
\end{equation}
The first boundary condition corresponds to the last boundary condition
for displacement vector (\ref{ebocow}) (the absence of external forces
on the surface of the cylinder), and the second one corresponds to the
absence of the angular component of the deformation tensor at the core of
dislocation. Equations (\ref{eborep}) define the values of integration
constants
\begin{equation}                                        \label{ecoifr}
  C_1=\frac\al{\g_1 R^{\g_1-1}},~~~~C_2=0.
\end{equation}
The obtained vielbein defines the metric
\begin{equation}                                        \label{emeweg}
  dl^2=\left(\frac rR\right)^{2\g_1-2}
  \left(dr^2+\frac{\al^2r^2}{\g_1^2}d\vf^2\right),
\end{equation}
where
\begin{equation*}
  \g_1=-\theta b+\sqrt{\theta^2 b^2+1+\theta}.
\end{equation*}
This is the solution of the posed problem. The derived solution is valid for
all deficit angles $\theta$ and for all $0<r<R$. The obtained metric depends
on three constants: $\theta$, $\s$, and $R$. The dependence on the deficit
angle $\theta$ is due to its occurrence on the right hand side of Einstein
equations (\ref{eindis}). The dependence on the Poisson ratio comes from
elastic gauge (\ref{efouad}), and, finally, the dependence on the
cylinder radius comes from boundary condition (\ref{eborep}).

If a wedge dislocation is absent, then $\theta=0$, $\al=1$, $\g_1=1$, and
metric (\ref{emeweg}) goes to the Euclidean one $dl^2=dr^2+r^2d\vf^2$,
as expected.

We compare metric (\ref{emeweg}) obtained within the geometric approach
with the induced metric from elasticity theory in Eqn (\ref{emewed}). First,
it has a simpler form. Second, in the linear approximation in $\theta$
\begin{equation*}
  \g_1\approx1+\theta\frac{1-2\s}{2(1-\s)},
\end{equation*}
and metric (\ref{emeweg}) can be easily shown to coincide precisely
with metric (\ref{emewed}) obtained within the elasticity theory.
We see that induced metric (\ref{emewed}) provides only the linear
approximation for the metric obtained within the geometric theory of defects,
which, in addition, has a simpler form. Beyond the
perturbation theory, we see essential differences. In particular,
metric (\ref{emewed}) is singular at the origin, whereas metric
(\ref{emeweg}) obtained beyound the perturbation theory is regular.

The stress and deformation tensors are related by Hook's law (\ref{eHook}).
There is an experimental possibility to check formulas (\ref{emeweg})
because the deformation tensor is the linear approximation for the induced
metric. For this one has to measure the stress field for a single wedge
dislocation. In this way, the geometric theory of defects can be experimentally
confirmed or discarded.

The problem of reconstruction of the displacement field for a given
metric reduces to solution of differential equations (\ref{emetri}) with
metric (\ref{emeweg}) on the right hand side and boundary conditions
(\ref{ebocow}). We do not discuss this problem here. We note that in the
geometric theory of defects, a complicated stage of finding the displacement
vector where it exists is simply absent and is not necessary.

Two-dimensional metric (\ref{emeweg}) describes the conical singularity
in the elastic gauge. The relation between conical singularities and wedge
dislocations was established in [22,78--80].
\nocite{Holz88,Kohler95A,Kohler95B}
In these papers, the metric was used in the other gauges (coordinate systems).
\section{Elastic oscillations in media with dislocations    }
Elastic oscillations in elastic media without defects are described
by a time-dependent vector field $u^i(t,x)$ that satisfies the wave
equation (see, e.g., \cite{LanLif70})
\begin{equation}                                         \label{efleld}
  \rho_0\ddot u^i-\mu\triangle u^i-(\lm+\mu)\pl^i\pl_ju^j=0,
\end{equation}
where dots denote differentiation with respect to time and $\rho_0$ is
the mass density of medium, which is assumed to be constant. If the medium
contains defects, then the metric of the space becomes nontrivial,
$\dl_{ij}\rightarrow g_{\mu\nu}=e_\mu{}^ie_\nu{}^j\dl_{ij}$. We assume
that relative displacements for elastic oscillations are much smaller
than stresses induced by defects:
\begin{equation}                                         \label{edueco}
  \pl_\mu u^i\ll e_\mu{}^i.
\end{equation}
Then, in the first approximation, we assume that elastic oscillations
propagate in a Riemannian space with a nontrivial metric induced by
dislocations. Here, we discard changes in the metric due to elastic
oscillation themselves. Therefore, for elastic oscillations we postulate
the following equation, which is a covariant generalization of
(\ref{efleld}) for spatial variables:
\begin{equation}                                        \label{ephops}
  \rho_0\ddot u^i-\mu\tilde\triangle u^i-(\lm+\mu)
  \widetilde\nb^i\widetilde\nb_ju^j=0,
\end{equation}
where $u^i$ denote components of displacement vector field with respect
to an orthonormal space basis $e_i$ (see the Appendix),
$\tilde\triangle=\widetilde\nb^i\widetilde\nb_i$ is the covariant
Laplace--Beltrami operator built for the vielbein $e_\mu{}^i$, and
$\widetilde\nb_i$ is the covariant derivative. The
explicit form of the covariant derivative of the displacement field is
\begin{equation*}
  \widetilde\nb_iu^j=e^\mu{}_i\widetilde\nb_\mu u^j
  =e^\mu{}_i(\pl_\mu u^j+u^k\widetilde\om_{\mu k}{}^j),
\end{equation*}
where $\widetilde\om_{\mu k}{}^j$ is the $\MS\MO(3)$ connection
built for zero torsion.

We note that the displacement vector field describing elastic oscillations
is not the total displacement vector field of points of a medium with
dislocations. It was already said in section \ref{sdislo} that the
displacement field for dislocations can be introduced only in those
regions of media where defects are absent. If we denote it by $u^i_\Sd$,
then the total displacement field in these regions is defined by the sum
$u^i_\Sd+u^i$. There, the vielbein is defined only by the displacement field
for dislocations $e_\mu{}^i=\pl_\mu u^i_\Sd$. We note that the smallness of
relative deformations (\ref{edueco}) is also meaningful in those regions
of space where displacements $u^i_\Sd$ are not defined.

We now decompose the displacement field covariantly into transversal
$u^{\St i}$ and longitudinal parts $u^{\Sl i}$,
\begin{equation*}
  u^i=u^{\St i}+u^{\Sl i},
\end{equation*}
which are defined by the relations
\begin{align}                                           \label{edetrp}
  \widetilde\nb_iu^{\St i}&=0,
\\                                                      \label{edelop}
  \widetilde\nb_iu^\Sl_j-\widetilde\nb_ju^\Sl_i&=0.
\end{align}
The decomposition of a vector field into longitudinal and transversal parts
in three-dimensional space with an accuracy up a constant is unique. We recall
that Latin indices are lowered with the help of Kronecker symbols, $u^i=u_i$,
and this operation commutes with covariant differentiation.
Equation (\ref{edelop}) can be rewritten as
\begin{equation*}
  \widetilde\nb_iu^\Sl_j-\widetilde\nb_ju^\Sl_i
  =e^\mu{}_ie^\nu{}_j(\widetilde\nb_\mu u^\Sl_\nu-\widetilde\nb_\nu u^\Sl_\mu)
  =e^\mu{}_ie^\nu{}_j(\pl_\mu u^\Sl_\nu-\pl_\nu u^\Sl_\mu)=0,
\end{equation*}
because transformation of Latin indices into Greek ones commutes with
the covariant differentiation, and the Christoffel symbols are symmetrical
in the first two indices. The last equality means that the
1-form $dx^\mu u^\Sl_\mu$ is closed. It is easily confirmed that Eqn
(\ref{ephops}) for elastic oscillations is equivalent to two independent
equations for transverse and longitudinal oscillations,
\begin{equation}                                        \label{eelaos}
  \frac1{c^2_\St}\ddot u^{\St i}-\tilde\triangle u^{\St i}=0,~~~~
  \frac1{c^2_\Sl}\ddot u^{\Sl i}-\tilde\triangle u^{\St i}=0,
\end{equation}
where
\begin{equation*}
  c^2_\St=\frac\mu{\rho_0},~~~~~~c^2_\Sl=\frac{\lm+2\mu}{\rho_0}
\end{equation*}
are the squares of sound velocities for transverse and longitudinal
oscillations.

Particles arising after the secondary quantization of Eqns (\ref{eelaos})
are called phonons in solids. Therefore, strictly speaking, the problem of
scattering of phonons on dislocations is a quantum mechanical one.
In the present review, we consider only classical aspects of this problem.

Wave equations (\ref{eelaos}) contain second and first derivatives
of the displacement field. The latter are contained in the covariant
Laplace--Beltrami operator $\tilde\triangle$. The terms with second
derivatives can be written in the four-dimensional form
$g^{\al\bt}\pl_\al\pl_\bt$ where $g^{\al\bt}$ is the inverse metric to
\begin{equation}                                        \label{etigas}
  g_{\al\bt}=\begin{pmatrix}c^2 & 0 \\ 0 & -g_{\mu\nu}
             \end{pmatrix},
\end{equation}
where $c$ is either the transverse or longitudinal sound velocity.
Above, we used the following notations. Four-dimensional coordinates
are denoted by Greek letters from the beginning of the alphabet
$\lbrace x^\al\rbrace=\lbrace x^0=t,x^1,x^2,x^3\rbrace$, and letters
from the middle of Greek alphabet denote only space coordinates
$\lbrace x^\mu\rbrace=\lbrace x^1,x^2,x^3\rbrace$. This rule can be easily
remembered by the following inclusions
$\lbrace1,2,3\rbrace\subset\lbrace0,1,2,3\rbrace$ and
$\lbrace\mu,\nu,\dotsc\rbrace\subset\lbrace\al,\bt,\dotsc\rbrace$.
Christoffel symbols (\ref{etigam}) can be computed for
four-dimensional metric (\ref{etigas}), which defines a system of
ordinary nonlinear equations for extremals $x^\al(\tau)$ (lines of
extremal lengths that coincide with geodesics in Riemannian geometry),
where dots denote differentiation with respect to canonical parameter
$\tau$. For the block-diagonal metric in (\ref{etigas}), these equations
decompose:
\begin{align}                                           \label{efextr}
  \ddot x^0&=0,
\\                                                      \label{esextr}
  \ddot x^\mu&=-\widetilde\G_{\nu\rho}{}^\mu\dot x^\nu\dot x^\rho,
\end{align}
where $\widetilde\G_{\nu\rho}{}^\mu$ are the three-dimensional Christoffel
symbols constructed for the three-dimensional metric $g_{\mu\nu}$ which,
as we recall, depends only on spatial coordinates for a static distribution
of defects. Let $\lbrace x^\al(\tau)\rbrace$ be an arbitrary extremal
in four-dimensional space-time. For metric (\ref{etigas}), its
natural projection on space
$\lbrace x^\al(\tau)\rbrace\rightarrow\lbrace 0,x^\mu(\tau)\rbrace$ is
also an extremal but now for the spatial part of the metric $g_{\mu\nu}$.

Equations for extremals (\ref{efextr}), (\ref{esextr}) are invariant
under linear transformations of the canonical parameter $\tau$.
Therefore, the canonical parameter can be identified with time,
$\tau=t=x^0$, without loss of generality as a consequence of Eqn
(\ref{efextr}).

We assume that a particle moves in space along an extremal $x^\mu(t)$
with velocity $v$. This means that
\begin{equation*}
  g_{\mu\nu}\dot x^\mu\dot x^\nu=v^2.
\end{equation*}
The length of the tangent vector to the corresponding extremal
$\lbrace t,x^\mu(t)\rbrace$ in four-dimensional space-time is then equal to
\begin{equation*}
  g_{\al\bt}\dot x^\al\dot x^\bt=c^2-v^2.
\end{equation*}
Hence, if the particle moves in space along extremal with velocity
less than, equal to, or greater than the speed of sound, then its world
line in space-time is timelike, null, or spacelike, respectively.

We return now to propagation of phonons in media with defects.
As in geometric optics \cite{LanLif62}, there are useful notions of wave
fronts and rays in the analysis of the asymptotic form of solutions for
wave equations (\ref{eelaos}). We do not consider mathematical
aspects of this approach, which is nontrivial and complicated
\cite{GuiSte77}, and instead give only a physical description. In the eikonal
(high frequency) approximation, phonons propagate
along rays coinciding with null extremals for the four-dimensional
metric $g_{\al\bt}$. Forms of rays, which are identified with trajectories
of phonons, are defined by the three-dimensional metric $g_{\mu\nu}$.
This means that in the eikonal approximation, trajectories of transverse
and longitudinal phonons in a medium with defects are the same and are
defined by Eqn (\ref{esextr}). The difference reduces to
the velocities of propagation for transverse and longitudinal phonons
being different and equal to $c_\St$ and $c_\Sl$, respectively.
\section{Scattering of phonons on a wedge dislocation     }
Calculations in the present section coincide, in fact, with the analysis
performed in Section 3 of \cite{KatVol99}. The difference is
that in what follows, we use the metric written in the elastic gauge. This
is important because we assume that trajectories of phonons seen in an
experiment coincide with extremals for the metric precisely in this
gauge.

In the presence of one wedge dislocation, the space metric in the
cylindrical coordinates $r,\vf,z$ is
\begin{equation}                                        \label{ewemep}
  g_{\mu\nu}=
\begin{pmatrix}
  {\displaystyle\dfrac{\raise-.7ex\hbox{$r^{2\g-2}$}}{R^{2\g-2}}} & 0 & 0 \\[2mm]
  0 & {\displaystyle\dfrac{\raise-.7ex\hbox{$\al^2$}}{\g^2}
  \dfrac{\raise-.7ex\hbox{$r^{2\g}$}}{R^{2\g-2}}} & 0 \\[2mm]
  0 & 0 & 1
\end{pmatrix}
\end{equation}
where the nontrivial part of the metric in the $r,\vf$ plane was obtain earlier
[see Eqn (\ref{emeweg})]. Here, we change $\g_1$ to $\g$ for simplicity of
notation. The inverse metric is also diagonal,
\begin{equation*}
  g^{\mu\nu}=
\begin{pmatrix}
  {\displaystyle\dfrac{\raise-.7ex\hbox{$R^{2\g-2}$}}{r^{2\g-2}}} & 0 & 0 \\[2mm]
  0 & {\displaystyle\dfrac{\raise-.7ex\hbox{$\g^2$}}{\al^2}
  \dfrac{\raise-.7ex\hbox{$R^{2\g-2}$}}{r^{2\g}}} & 0 \\[2mm]
  0 & 0 & 1
\end{pmatrix}
\end{equation*}

Christoffel symbols for metric (\ref{ewemep}) are calculated according
to formulas (\ref{etigam}). As a result, only four Christoffel symbols
differ from zero:
\begin{align*}
  \widetilde\G_{rr}{}^r&=\frac{\g-1}r,
\\
  \widetilde\G_{\vf\vf}{}^r&=-\frac{\al^2 r}\g,
\\
  \widetilde\G_{r\vf}{}^\vf&=\widetilde\G_{\vf r}{}^\vf=\frac\g r.
\end{align*}

In the preceeding section, we showed that in the eikonal approximation,
phonons propagate along extremals $x^\mu(t)$ defined by Eqns
(\ref{esextr}). In the case considered here, these equations are
\begin{align}                                           \label{extrfw}
  \ddot r&=-\frac{\g-1}r\dot r^2+\frac{\al^2}\g r\dot\vf^2,
\\                                                      \label{extvfw}
  \ddot\vf&=-\frac{2\g}r\dot r\dot\vf,
\\                                                      \label{extzwe}
  \ddot z&=0,
\end{align}
where the dot denotes differentiation with respect to time $t$.
It follows from the last equation that phonons move along the $z$
axis with constant velocity, which corresponds to translational
invariance along $z$. This means that scattering on a wedge dislocation
is reduced to a two-dimensional problem in the $r,\vf$ plane, as could be
expected.

The system of equations for $r(t)$ and $\vf(t)$ in (\ref{extrfw}) and
(\ref{extvfw}) can be explicitly integrated. For this, we find two
first integrals. First, for any metric, there is the integral for
equations for extremals
\begin{equation*}
  g_{\mu\nu}\dot x^\mu\dot x^\nu=\const.
\end{equation*}
We then have the equality
\begin{equation}                                        \label{efinte}
  r^{2\g-2}\dot r^2+\frac{\al^2}{\g^2}r^{2\g}\dot\vf^2=C_0=\const>0.
\end{equation}

Second, the invariance of the metric under rotations about
the $z$ axis results in the existence of an additional integral. It is
constructed as follows. There is a Killing vector corresponding to the
invariance of the metric, which in cylindrical coordinates has the simple
form $k=\pl_\vf$. Straightforward verification proves that
\begin{equation*}
  g_{\mu\nu}k^\mu\dot x^\nu=\const.
\end{equation*}
In the considered case, this results in the identity
\begin{equation}                                        \label{einsew}
  r^{2\g}\dot\vf=C_1=\const.
\end{equation}

We analyze the form of an extremal $r=r(\vf)$. First derivatives
can be found from Eqns (\ref{efinte}) and (\ref{einsew})
\begin{align}                                           \label{erdotf}
  \dot r&=\pm r^{-2\g+1}\sqrt{C_0r^{2\g}-\frac{\al^2}{\g^2}C_1^2},
\\                                                      \label{evfdot}
  \dot\vf&=C_1 r^{-2\g}.
\end{align}
Admissible values of the radial coordinate $r$ for which the expression
under the square root is nonnegative are to be found later. From the above
equations, we obtain the equation defining the form of nonradial
($C_1\ne0$) extremals
\begin{equation}                                        \label{eqfexw}
  \frac{dr}{d\vf}=\frac{\dot r}{\dot\vf}
  =\pm r\sqrt{\frac{C_0}{C_1^2}r^{2\g}-\frac{\al^2}{\g^2}}.
\end{equation}
This equation can be easily integrated, and we finally obtain
explicit formulas defining the form of an extremal:
\begin{equation}                                        \label{efoexe}
  \left(\frac r{r_m}\right)^{2\g}\sin^2[\al(\vf+\vf_0)]=1,
\end{equation}
where
\begin{equation*}
  r_m=\left(\frac{C_1\al}{\sqrt{C_0} \g}\right)^{1/\g}=\const>0,~~~~\vf_0=\const.
\end{equation*}
The constant $r_m$ is positive and defines the minimal distance at which an
extremal approaches the core of the dislocation, i.e., $r\ge r_m$. Only for
these values of $r$ is the expression under the root in Eqn (\ref{eqfexw})
nonnegative. The integration constant $\vf_0$ is arbitrary and corresponds
to the invariance of the problem under rotations around the
core of dislocation.

Equations for extremals (\ref{extrfw}) and (\ref{extvfw}) also have
degenerate solutions:
\begin{equation}                                          \label{eradex}
  \frac1\g r^\g=\pm\sqrt{C_0}(t+t_0),~~~~\vf=\const,~~~~t_0=\const.
\end{equation}
These extremals correspond to the radial motion of phonons. Such trajectories
are unstable in the sense that there are no nonradial extremals near them.

We note that circular extremals $r=\const$ are absent as a consequence
of Eqn (\ref{extrfw}), although integrals of motion (\ref{efinte}) and
(\ref{einsew}) admit such a solution. This is because Eqn (\ref{extrfw})
was multiplied by $\dot r$ in deriving first integral (\ref{efinte}).

We now analyze the form of nonradial extremals (\ref{efoexe}). For any
extremal, the radius $r$ decreases first from infinity to the minimal
value $r_m$ and then increases from $r_m$ to infinity. We can assume here
without loss of generality that the argument of sine in (\ref{efoexe})
varies from 0 to $\pi$. Thus, we obtain the range of changes of the polar
angle
\begin{equation*}
  0\le\vf+\vf_0\le\frac\pi\al.
\end{equation*}
This means that the extremal comes from infinity at the angle $-\vf_0$
and goes to infinity at the angle $\pi/\al-\vf_0$. This corresponds
to the scattering angle
\begin{equation}                                        \label{escang}
  \chi=\pi-\frac\pi\al=\frac{\pi\theta}{1+\theta}.
\end{equation}
We note that the scattering angle depends only on the deficit angle $\theta$
and does not depend on the elastic properties of the medium. The scattering
angle has a simple physical interpretation. For positive $\theta$, the medium
is cut and moved apart. A wedge of the same medium without elastic stresses
is inserted in the cavity that has appeared. Afterwards, gluing is performed,
and the wedge is compressed. The compression coefficient for all circles
centered at the origin is equal to $1/(1+\theta)$ due to symmetry
considerations. Therefore, the scattering angle equals half the
deficit angle times the compression coefficient,
\begin{equation*}
  \chi=\frac{2\pi\theta}2\times\frac1{1+\theta}.
\end{equation*}
For $\theta=0$, the dislocation is absent, and the scattering angle is equal
to zero.

\begin{figure}[h,t]
\hfill\includegraphics[width=\textwidth]{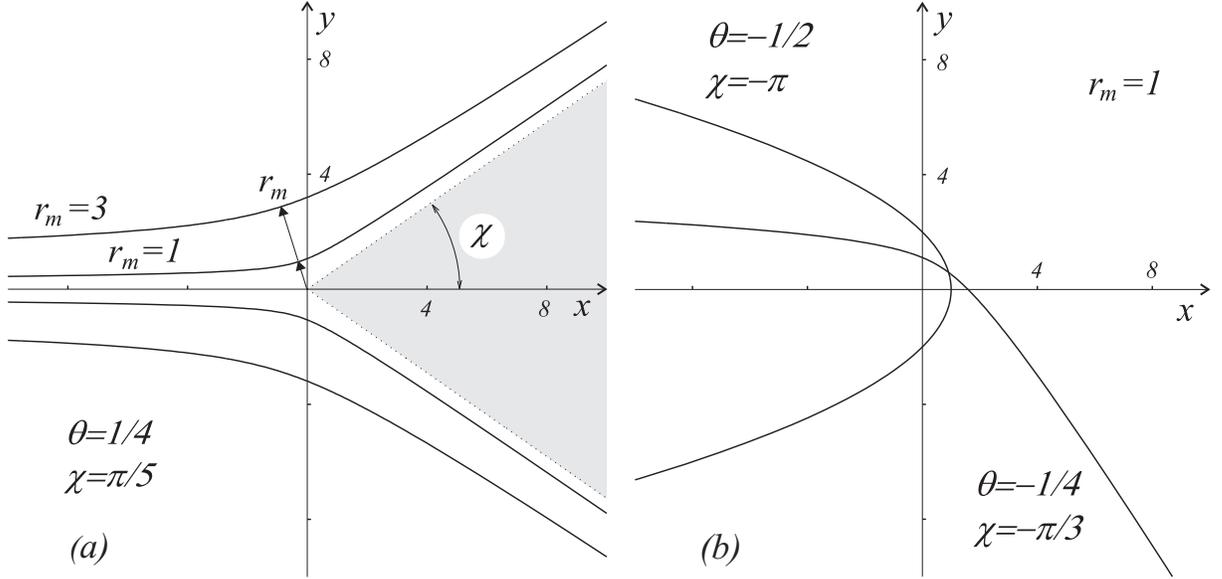}
\hfill {} \\
\centering \caption{\label{fexwpo} {\it(a)} Extremals for the wedge dislocation
with a positive ($\theta>0$) deficit angle. Two extremals and their reflections
with respect to the $x$ axis are shown for the same $\theta>0$ but different
$r_m$. {\it(b)} Extremals for the wedge dislocation with a negative
($-1/2\le\theta<0$) deficit angle. Two extremals are shown for the same
$r_m$ but different deficit angles $\theta$.}
\end{figure}

For positive deficit angles, the scattering angle is positive, which
corresponds to repulsion of phonons from the wedge dislocation.
Corresponding extremals are shown in Fig.\ref{fexwpo}{\it a}, and
they have asymptotes as $r\to\infty$.
We note that for positive deficit angles, no two points on the $r,\vf$
plane can be connected by an extremal, i.e., there is a domain to the
right of the wedge dislocation that cannot be reached at all
by a phonon falling from the left.

All extremals shown in the figures in the present section are calculated
numerically. The values of the deficit angle $\theta$, the scattering
angle $\chi$, and the minimal distance $r_m$ to the dislocation axis are
shown in the figures. For definiteness, we choose $\vf_0=\pi$ corresponding
to the fall of phonons from the left.

For negative deficit angles the scattering angle is negative and is
defined by the same formulas (\ref{escang}). This corresponds to the
attraction of phonons to the dislocation axis. In Fig.\ref{fexwpo}{\it b},
we show two extremals with the same parameter $r_m$ but
for two dislocations with different deficit angles. For $-1/2<\theta<0$, the
scattering angle varies from 0 to $2\pi$ (Fig.\ref{fexwpo}{\it b}).
For $\theta=-1/2$, the scattering angle is equal to $2\pi$. We note that
for negative deficit angles, phonons have no asymptotes as $r\to\infty$,
i.e., for $\theta=-1/2$, phonons fall from infinity
($x\to-\infty,y\to+\infty$) and return to the infinity
($x\to-\infty,y\to-\infty$).

\begin{figure}[h,t]
\hfill\includegraphics[width=\textwidth]{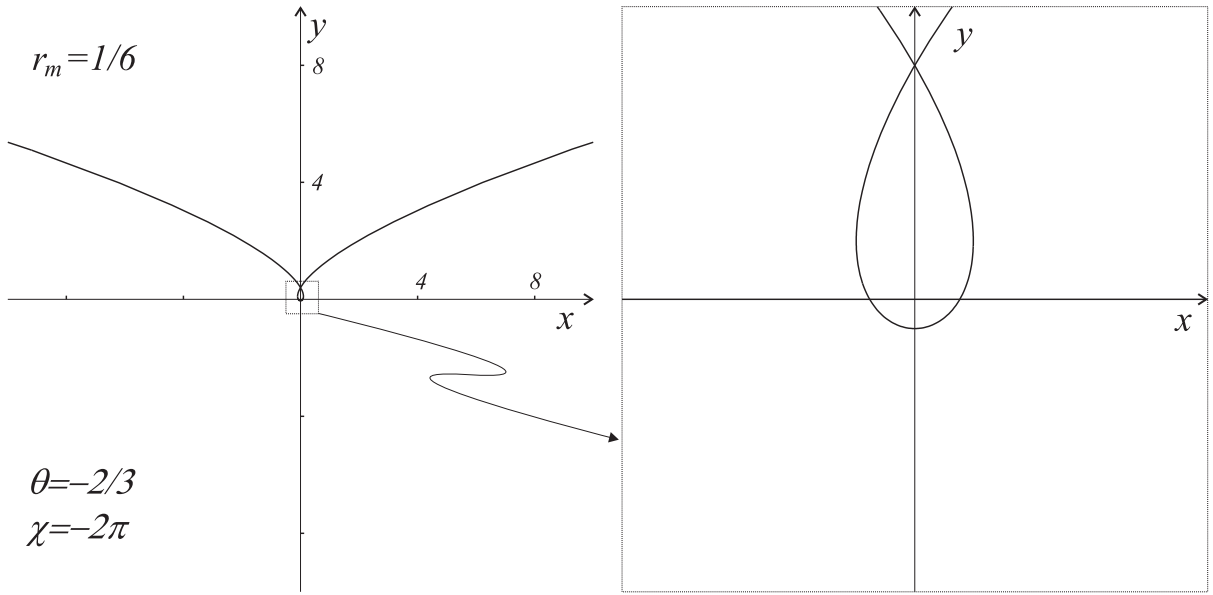}
\hfill {} \\
\centering \caption{\label{fwede2} For $\theta=-2/3$, an extremal makes
one rotation around the dislocation and then goes forward in the original
direction. The blown-up part of the trajectory is shown in the square
to the right.}
\end{figure}

\begin{figure}[h,t]
\hfill\includegraphics[width=\textwidth]{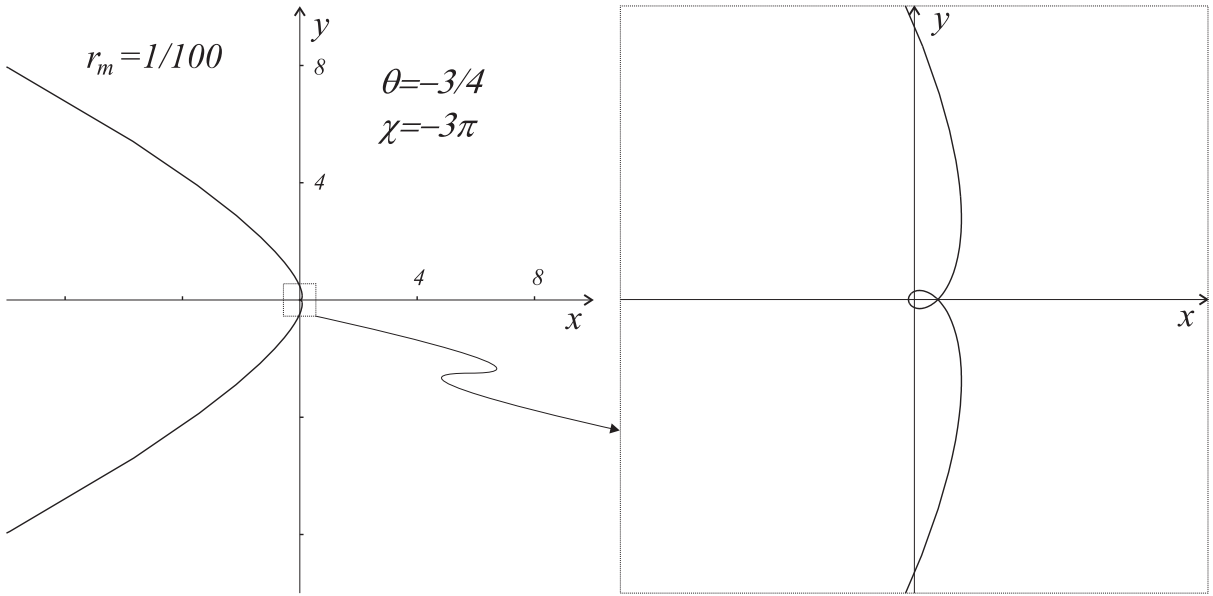}
\hfill {} \\
\centering \caption{\label{fwede3} For $\theta=-3/4$, an extremal makes
two turns around the dislocation and then goes back. The blown-up
part of the trajectory  is shown in the square to the right.}
\end{figure}

\begin{figure}[h,t]
\hfill\includegraphics[width=\textwidth]{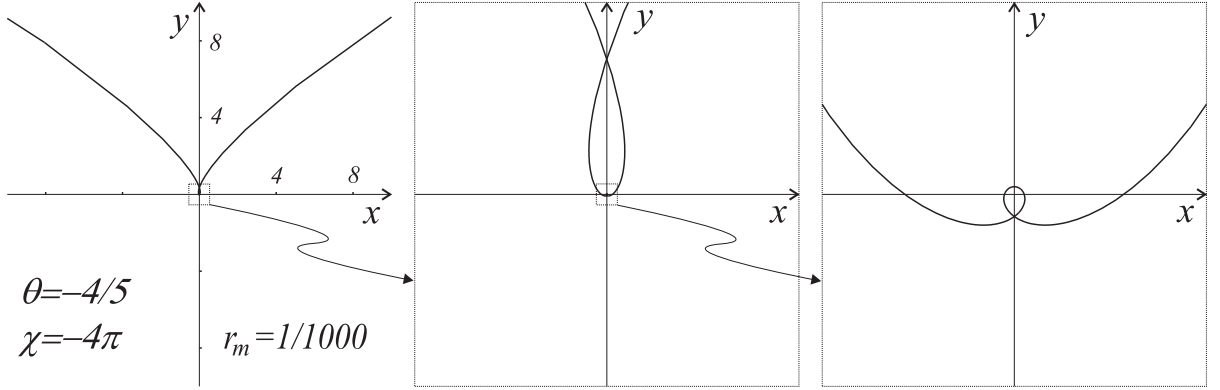}
\hfill {} \\
\centering \caption{\label{fwede4} For $\theta=-4/5$, an extremal
makes two and a half rotations around the dislocation and then goes
forward in the original direction. Subsequent blown-up parts
of the trajectory are shown in the two squares to the right.}
\end{figure}

When the deficit angle is sufficiently small ($-1<\theta<-1/2$), a phonon
makes one or several rotations around the dislocation and then goes to
infinity. Examples of such trajectories are shown in Figs.\
\ref{fwede2}--\ref{fwede4}.
\clearpage

We consider the asymptotic behavior of nonradial extremals as
$r\rightarrow\infty$. As a consequence of Eqn (\ref{erdotf}), far from
the core of dislocation, we obtain
\begin{equation*}
  \dot r\approx\pm\sqrt{C_0}r^{-\g+1}.
\end{equation*}
It follows from this equation that the dependence of the radius on time is
the same as for radial extremals (\ref{eradex}). Because $\g>0$,
an infinite value of $r$ corresponds to an infinite value of time $t$.
This means that the $r,\vf$ plane with the given metric is complete
at $r\to\infty$. The origin (the core of dislocation) is a singular point.
Only radial extremals fall into it at a finite moment of time.

Integrals of motion (\ref{efinte}) and (\ref{einsew}) have simple physical
meaning. Equations for extremals (\ref{esextr}) follow from the variational
principle for the Lagrangian
\begin{equation}                                        \label{elagex}
  L=\frac12g_{\mu\nu}\dot x^\mu\dot x^\nu,
\end{equation}
describing motion of a free massless point particle in a nontrivial metric
$g_{\mu\nu}(x)$. Here, the metric is considered as a given external field
and is not varied.

The energy corresponding to this integral is equal to
\begin{equation*}
  E=\frac12g_{\mu\nu}\dot x^\mu\dot x^\nu
  =\frac12\frac{r^{2\g-2}}{R^{2\g-2}}\dot r^2
  +\frac12\frac{\al^2}{\g^2}\frac{r^{2\g}}{R^{2\g-2}}\dot\vf^2+\frac12\dot z^2.
\end{equation*}
If the metric, as in our case, does not depend on time explicitly, then
the energy is conserved ($E=\const$) and its numerical value for the
motion in the $r,\vf$ plane is proportional to the integral of motion $C_0$.

For a wedge dislocation, the metric is independent on the polar angle
$\vf$, and the Lagrangian is invariant under rotations:
$\vf\rightarrow\vf+\const$. By the Noether theorem, the angular momentum
conservation law corresponds to this invariance,
\begin{equation*}
  J=-\frac{\al^2}{\g^2}\frac{r^{2\g}}{R^{2\g-2}}\dot\vf=\const.
\end{equation*}
As a consequence, the constant of integration $C_1$ is proportional to the
angular momentum.

We note that the behavior of extremals differs qualitatively from trajectories
of point particles moving in flat space with Euclidean metric
$\dl_{\mu\nu}$ in an external potential field $U(x)$. It can be easily
shown that trajectories of point particles described by the Lagrangian
\begin{equation*}
  L=\frac12\dl_{\mu\nu}\dot x^\mu\dot x^\nu-U
\end{equation*}
cannot coincide with extremals (\ref{esextr}) for any function $U(x)$.

In the present section, we have showed how the problem of scattering of phonons
on the simplest wedge dislocation is solved in the geometric approach.
The results of calculations can be checked experimentally. The more general
problem of scattering of phonons on an arbitrary distribution of wedge
dislocations, including the edge dislocation and continuous distribution
of dislocations, is solved in \cite{KatVol99}. Solution of this problem
in the geometric approach is simpler than in the framework of
classical elasticity theory, where we have to solve partial differential
equations with complicated boundary conditions. In \cite{KatVol99},
calculations were performed in the conformal gauge for the metric. For
comparison with experiment, the results must be recalculated in the
elastic gauge, which has physical meaning in the geometric approach.
\section{An impurity in the field of a wedge dislocation                 }
We consider elastic media with one wedge dislocation containing
one atom of impurity or vacancy. If we consider the influence of impurity
on the distribution of elastic stresses small compared with the elastic
stresses induced by the dislocation itself, then the motion of the impurity
can be considered as taking place in three-dimensional space with the
nontrivial metric (\ref{emeweg}). In the geometric approach, we assume
that the potential energy of the interaction equals zero, and all interactions
are due to changes in the kinetic energy, which depends explicitly
on the nontrivial metric.

We solve the corresponding quantum mechanical problem. We consider bounded
states of impurity moving inside a cylinder of radius $R$ in the presence
of a wedge dislocation. We assume that the cylinder axis coincides with the
core of dislocation. The stationary Schr\"odinger equation is
\begin{equation}                                        \label{eshrse}
  -\frac{\hbar^2}{2M}\tilde\triangle\Psi=E\Psi,
\end{equation}
where $\hbar$ is the Plank constant and $M$, $\Psi$, and $E$ are mass,
wave function, and energy of the impurity. The nontriviality of the interaction
with dislocation reduces to the nontrivial Laplace--Beltrami operator
\begin{equation*}
  \tilde\triangle\Psi=\frac1{\sqrt g}\pl_\mu(\sqrt g g^{\mu\nu}\pl_\nu\Psi),
\end{equation*}
where the metric was found earlier [see Eqn (\ref{emeweg})], and
$g=\det g_{\mu\nu}$.

Taking the symmetry of the problem into account, we solve the Schr\"odinger
equation (\ref{eshrse}) in cylindrical coordinates by the separation of
variables. Let
\begin{equation*}
  \Psi(r,\vf,z)=Z(z)\sum_{m=-\infty}^\infty\psi_m(r)\ex^{im\vf},
\end{equation*}
where we have the two following possibilities for the normalization
function $Z(z)$. If the impurity moves freely along the $z$ axis with
momentum $\hbar k$, then
\begin{equation*}
  Z(z)=\frac1{\sqrt{2\pi}}\ex^{ikz}.
\end{equation*}
If its motion is restricted by the planes $z=0$ and $z=z_0$, then
\begin{equation*}
  Z(z)=\sqrt{\frac2{z_0}}\sin(k_lz),~~~~~~k_l=\frac{\pi l}{z_0}.
\end{equation*}
In what follows, we drop the integer-valued subscript $l$ indicating
the restricted motion, having both possibilities in mind.

The condition for the constant $m$ (the eigenvalue of the projection of the
momentum on the $z$ axis) to be an integer appears due to the periodicity
condition
\begin{equation*}
  \Psi(r,\vf,z)=\Psi(r,\vf+2\pi,z).
\end{equation*}

Then for the radial wave function, $\psi_m(r)$ we then obtain the equation
\begin{equation}                                        \label{eradeq}
  \frac{R^{2\g-2}}{r^{2\g-1}}\pl_r(r\pl_r\psi_m)
  +\left(\frac{2ME}{\hbar^2}-\frac{\g^2}{\al^2}\frac{R^{2\g-2}}{r^{2\g}}m^2
  -k^2\right)\psi_m=0.
\end{equation}
We introduce the new radial coordinate
\begin{equation*}
  \rho=\frac{r^\g}{\g R^{\g-1}}.
\end{equation*}
This is the transformation in (\ref{eftrra}) and (\ref{ecoifr}) up to a constant.
The radial equation is then
\begin{equation}                                        \label{erashe}
  \frac1\rho\pl_\rho(\rho\pl_\rho\psi_m)
  +\left(\bt^2-\frac{\nu^2}{\rho^2}\right)\psi_m=0,
\end{equation}
where
\begin{equation*}
  \bt^2=\frac{2ME}{\hbar^2}-k^2,~~~~~~\nu=\frac{|m|}\al>0.
\end{equation*}
This is the Bessel equation. We solve it with the boundary condition
\begin{equation}                                        \label{eboube}
  \psi_m|_{\rho=R/\g}=0,
\end{equation}
which corresponds to the motion of an impurity inside the cylinder with
an impenetrable  boundary. The general solution of the Bessel equation
(\ref{erashe}) is
\begin{equation*}
  \psi_m=c_m J_\nu(\bt\rho)+d_m N_\nu(\bt\rho),~~~~~~c_m,d_m=\const,
\end{equation*}
where $J_\nu$ and $N_\nu$ are Bessel and Neumann functions of order $\nu$
\cite{JaEmLo60}. The boundedness of the wave function on the axis
of the cylinder requires $d_m=0$. The constants of integration $c_m$ are
found from the normalization condition
\begin{equation*}
  \int_0^R dr\, r|\psi_m|^2=1.
\end{equation*}
Boundary condition (\ref{eboube}) yields the equation for $\bt$
\begin{equation}                                        \label{eqenlv}
  J_\nu(\bt R/\g)=0,
\end{equation}
which defines the energy levels of bounded states. It is well known
that for real $\nu>-1$ and $R/\g$, this equation has only real roots.
Positive roots form an infinite countable set and all of them are simple
\cite{JaEmLo60}. This provides the inequality
\begin{equation*}
  \bt^2=\frac{2ME}{\hbar^2}-k^2\ge0.
\end{equation*}
We label the positive roots of Eqn (\ref{eqenlv}) by the index
$n=1,2,\dotsc$ (principle quantum number): $\bt\rightarrow\bt_n(m,\al,\g,R)$.
Then the spectrum of bounded states is
\begin{equation}                                        \label{espekr}
  E_n=\frac{\hbar^2}{2M}(k^2+\bt_n^2).
\end{equation}

For large radii $\bt\rho\gg1$ and $\bt\rho\gg\nu$ we have the asymptotic form:
\begin{equation*}
  J_\nu(\bt\rho)\approx\sqrt{\frac2{\pi\bt\rho}}
  \cos\left(\bt\rho-\frac{\nu\pi}2-\frac\pi4\right).
\end{equation*}
As a result, we obtain explicit expression for the spectrum
\begin{equation}                                        \label{espebt}
  \bt_n=\frac{\g\pi}R\left(n+\frac{|m|}{2\al}-\frac14\right).
\end{equation}

In the absence of defect, $\al=1$, $\g=1$, $\rho=r$, and the radial
functions $\psi_m$ are expressed through the Bessel functions of integer
order $\nu=|m|$. In this case, the spectrum of bounded states depends
only on the sizes of the cylinder. In the presence of the wedge
dislocation, Bessel functions have a noninteger order in general.
In this case, the spectrum of energy levels acquires dependence on
the deficit angle $\theta$ and the Poisson ratio $\sigma$ characterizing
elastic properties of the medium.

If the mass of impurity and vacancy is defined by integral
(\ref{emass}), then $M>0$ for impurity and the energy eigenvalues are
positive. For vacancy, $M<0$ and the energy eigenvalues are negative.
In this case, the energy spectrum is not bounded from below, which causes
serious problems for physical interpretation. It seems that one has
to insert in the Schr\"odinger equation not the bare mass (\ref{emass})
but the effective mass, with the contribution of elastic stresses arising
around a vacancy taken into account. This question presently remains
unanswered.

The presentation in the present section is close to that in \cite{AzeMor98}.
In contrast to that paper, we use the elastic gauge for the metric.
Therefore, our results depend not only on the deficit angle of
a wedge dislocation but also on the elastic properties of the medium.

Calculations of the energy levels of an impurity in the field of a wedge
dislocation are actually equivalent to the calculations of bounded-state
energies in the Aharonov--Bohm effect \cite{AhaBoh59} (see
reviews \cite{Feinbe62,Skarzh81}). The difference reduces only to
changing the order of the Bessel functions,
\begin{equation*}
  \nu=\left|m-\frac\Phi{\Phi_0}\right|,
\end{equation*}
where $\Phi_0=2\pi\hbar c/e$ is the magnetic flux quantum and
$e$ is the electron charge.

The considered example shows how the influence of the presence of
defects is taken into account in the geometric approach in the first
approximation. If calculations in some problem were performed in
elastic media without defects, then to take the influence of defects
into account we have to replace the flat Euclidean metric with
a nontrivial metric describing the given distribution of defects.
This problem may appear complicated mathematically because we have
to solve the three-dimensional Einstein equations to find the metric.
However, there are no principal difficulties: the effect of dislocations
reduces to a change in the metric.
\section{Conclusion}
The geometric theory of defects describes defects in elastic media
(dislocations) and defects in the spin structure (disclinations) from
a single point of view. This model can be used for describing
single defects as well as their continuous distribution. The geometric
theory of defects is based on the Riemann-Cartan geometry. By definition,
torsion and curvature tensors are equal to surface densities of Burgers
and Frank vectors, respectively.

Equations defining the static distribution of defects are covariant and
have the same form as equations of gravity models with dynamical torsion.
To choose a solution uniquely, one must fix the coordinate system.
For this, the elastic gauge for the vielbein and the Lorentz gauge for
the $\MS\MO(3)$ connection are proposed. In the defect-free case the
displacement vector field and the field of the spin structure can be
introduced. Equations of equilibrium are then identically satisfied, and
the gauge conditions reduce to the equations of elasticity theory and
the principal chiral $\MS\MO(3)$-field. In this way, the geometric theory
of defects incorporates elasticity theory and the model of principal
chiral field.

In a certain sense, the elastic gauge represents the equations of the
nonlinear elasticity theory. Nonlinearity is introduced in elasticity theory
in two ways. First, the deformation tensor is defined through the induced
metric
\begin{equation*}
  \e_{ij}=\frac12(\dl_{ij}-g_{ij})
\end{equation*}
instead of being defined by linear relation (\ref{edefte}). Then the stress
tensor is given by an infinite series in the displacement vector. Second,
Hook's law can be modified assuming nonlinear dependence of the stress
tensor on the deformation tensor. Hence, the elastic gauge condition is the
equations of the nonlinear elasticity theory where the deformation tensor is
assumed to be defined through the induced metric and Hook's law is kept linear.
A generalization to the nonlinear dependence of the deformation tensor on the
stress tensor is obvious.

As an example, we considered the wedge dislocation from the standpoint of the
elasticity theory and the geometric theory of defects. We showed that the
elasticity theory reproduces only the linear approximation of the geometric
approach. In contrast to the induced metric obtained within the exact solution
of the linear elasticity theory, the expression for the metric obtained as
the exact solution of the Einstein equations in the elastic gauge is simpler
and is defined on the whole space and for all deficit angles. The obtained
expression for the metric can be checked experimentally.

Two problems are considered as applications of the geometric theory of
defects. The first is the scattering of phonons on a wedge dislocation.
In the eikonal approximation, the problem is reduced to the analysis of
extremals for the metric describing a given dislocation. Equations for
extremals are integrated explicitly, and the scattering angle is found.
The second of the considered problems is the construction of the wave functions
and energy spectrum of impurity in the presence of a wedge dislocation.
This requires solving the Schr\"odinger equation. This problem is
mathematically equivalent to solving the Schr\"odinger equation for
bound states in the Aharonov--Bohm effect \cite{Skarzh81}. The
explicit dependence of the spectrum on the deficit angle and elastic
properties of the medium is found in the presence of a wedge dislocation.

The geometric theory of static distribution of defects can also be
constructed for membranes, i.e., on the plane $\MR^2$. For this,
one has to consider the Euclidean version \cite{Katana97} of two-dimensional
gravity with torsion [88--90].
\nocite{VolKat86,KatVol86,KatVol90}
This model is favored by its integrability [91--94].
\nocite{Katana89B,Katana90,Katana91,Katana93A}

The developed geometric construction in the theory of defects can be inverted,
and we can consider the gravitational interaction of masses in the universe as
the interaction of defects in elastic ether. Point masses and cosmic strings
\cite{VilShe94,HinKib95} then correspond to point defects (vacancies and
impurities) and wedge dislocations. In this interpretation of gravity, we
have a question about the elastic gauge, which has direct physical meaning
in the geometric theory of defects. If we take the standpoint of the theory
of defects, then the elastic properties of ether correspond to some value
of the Poisson ratio, which can be measured experimentally.

It seems interesting and important for applications to include time in the
considered static approach for describing motion of defects in the medium.
Such a model is lacking at present. From the geometric standpoint this
generalization can be easily performed, at least in principle. It is
sufficient to change the Euclidean space $\MR^3$ to the Minkowski space
$\MR^{1,3}$ and to write a suitable Lagrangian quadratic in curvature and
torsion, which corresponds to the true gravity model with torsion. One of the
arising problems is the physical interpretation of the additional components
of the vielbein and Lorentz connection that contain the time index. The
physical meaning of the time component of the vielbein
$e_0{}^i\rightarrow\pl_0u^i=v^i$ is simple -- this is the velocity of a point
of the medium. This interpretation is natural from the physical standpoint
because the motion of continuously distributed dislocations means a flowing
of the medium. In fact, the liquid can be imagined as the elastic media with
a continuous distribution of moving dislocations. This means that the
dynamical theory of defects based on the Riemann--Cartan geometry must
include hydrodynamics. It is not clear at present how this can happen.
Physical interpretation of the other components of the vielbein and the
Lorentz connection with the time index also remains unclear.

The author is sincerely grateful to I V Volovich for numerous discussions
on the problems considered in this paper.
This work is supported by the RFBR (grant 05-01-00884) and the program of
support for leading academic schools (grant AS-1542.2003.1).
\section*{Appendix. Some differential geometry         \label{scodte}}
In the appendix, we briefly present the main facts from differential geometry
and introduce the notation used in this review. The description
is given in local coordinates, which is sufficient for our purposes.
We recommend \cite{DuNoFo98E} as a textbook on differential geometry.

If the metric $g$ and the affine connection $\G$ are given, then we say that
the geometry is defined on a differentiable manifold $\MM$, $\dim\MM=m$.
We assume that all fields on the manifold are given by smooth $\CC^\infty(\MM)$
functions except, possibly, some singular points, and do not stipulate that
in what follows. We also assume that the manifold $\MM$ is topologically
trivial, i.e., diffeomorphic to the Euclidean space $\MR^m$.

In a local coordinate system $x^\mu$, $\mu=1,\dotsc,m$, the metric is given
by a nondegenerate symmetric covariant second rank tensor $g_{\mu\nu}(x)$,
which defines the scalar product of vector fields $X=X^\mu\pl_\mu$,
$Y=Y^\mu\pl_\mu$
\begin{equation}                                        \label{escprv}
  (X,Y)=X^\mu Y^\nu g_{\mu\nu},~~~~g_{\mu\nu}=g_{\nu\mu},
  ~~~~\det g_{\mu\nu}\ne0.
\end{equation}
In general, the scalar product may be not positive definite. If the
scalar product is positive definite, we say that a Riemannian metric is
given on a manifold. By definition, the metric is a covariant tensor
field, i.e., it transforms under coordinate changes
$x^\mu\rightarrow x^{\mu'}(x)$ by the tensor law
\begin{equation*}
  g_{\mu'\nu'}=\frac{\pl x^\mu}{\pl x^{\mu'}}
  \frac{\pl x^\nu}{\pl x^{\nu'}}g_{\mu\nu}.
\end{equation*}
This means that the scalar product of vector fields $(X,Y)$ is a scalar
field. In a similar way, contractions with the metric $g_{\mu\nu}$ and its
inverse $g^{\mu\nu}$, $g^{\mu\nu}g_{\nu\rho}=\dl^\mu_\rho$, allows one
to build scalar fields from higher-rank tensors or to lower their rank.
The metric is also used for lowering and raising of tensor indices.

An affine connection on a manifold in a local coordinate system is given by
the set of coefficients $\G_{\mu\nu}{}^\rho(x)$ that transform under the
diffeomorphisms as
\begin{equation}                                        \label{econts}
  \G_{\mu'\nu'}{}^{\rho'}=\frac{\pl x^\mu}{\pl x^{\mu'}}
  \frac{\pl x^\nu}{\pl x^{\nu'}}
  \G_{\mu\nu}{}^\rho\frac{\pl x^{\rho'}}{\pl x^\rho}
  +\frac{\pl^2x^\rho}{\pl x^{\mu'}\pl x^{\nu'}}\frac{\pl x^{\rho'}}{\pl x^\rho}.
\end{equation}
These coefficients do not constitute a tensor field because of the presence
of an inhomogeneous term in (\ref{econts}). An affine connection on a manifold
defines covariant derivatives of tensor fields. In particular, covariant
derivatives of a vector field and 1-form $A=dx^\mu A_\mu$ have the form
\begin{align}                                              \label{ecodev}
  \nabla_\mu X^\nu&=\pl_\mu X^\nu+X^\rho\G_{\mu\rho}{}^\nu,
\\                                                     \label{ecodec}
  \nabla_\mu A_\nu&=\pl_\mu A_\nu-\G_{\mu\nu}{}^\rho A_\rho.
\end{align}
The covariant derivative of a scalar field coincides with the partial derivative
$\nb_\mu\vf=\pl_\mu\vf$. The covariant derivative of higher-rank tensors is
built in a similar way and contains one term with the plus and minus sign
for each contravariant and covariant index, respectively. One can easily
check that the covariant derivative of a tensor of an arbitrary type
$(r,s)$ is a tensor field of type $(r,s+1)$, i.e., it has one additional
covariant index. We note that the covariant derivative of a product of
tensor fields may contain arbitrary contractions of indices. For example,
$$
  \pl_\mu(X^\nu A_\nu)=\nb_\mu(X^\nu A_\nu)
  =(\nb_\mu X^\nu)A_\nu+X^\nu(\nb_\mu A_\nu).
$$

Because the inhomogeneous term in (\ref{econts}) is symmetric in indices $\mu'$
and $\nu'$ the antisymmetric part of the affine connection
$2\G_{[\mu\nu]}{}^\rho$ forms
a tensor field of type $(1,2)$, which is called the torsion tensor
\begin{equation}                                        \label{etorsi}
    T_{\mu\nu}{}^\rho=\G_{\mu\nu}{}^\rho-\G_{\nu\mu}{}^\rho.
\end{equation}

In general, the connection $\G_{\mu\nu}{}^\rho$ has no symmetry in its
indices and is not related to the metric $g_{\mu\nu}$ in any way because
these notions define different geometric operations on a manifold $\MM$.
We then say that the affine geometry is given on $\MM$. We emphasize
that the metric and the affine connection are defined arbitrarily and are
completely independent geometric notions. Therefore, in the construction
of physical models, they can be considered independent fields having
different physical interpretations.

If the affine geometry is given on a manifold, then we can construct the
nonmetricity tensor $Q_{\mu\nu\rho}$ that is by definition equal to
the covariant derivative of the metric:
\begin{equation}                                       \label{enonme}
  -Q_{\mu\nu\rho}=\nb_\mu g_{\nu\rho}=\pl_\mu g_{\nu\rho}
  -\G_{\mu\nu}{}^\s g_{\s\rho}-\G_{\mu\rho}{}^\s g_{\nu\rho}.
\end{equation}
By construction, the nonmetricity tensor is symmetric with respect to
the permutation of the last two indices: $Q_{\mu\nu\rho}=Q_{\mu\rho\nu}$.
We note that we need both objects, the metric and the connection to define
the nonmetricity.

Thus, for a given metric and connection we constructed two tensor fields:
the torsion and the nonmetricity tensor. We prove that for a given metric,
torsion, and nonmetricity tensor we can uniquely reconstruct the
corresponding affine connection. Equation (\ref{enonme}) can always be
solved for the connection $\G$. Indeed, the linear combination
$$
  \nb_\mu g_{\nu\rho}+\nb_\nu g_{\rho\mu}-\nb_\rho g_{\mu\nu}
$$
yields the expression for the affine connection with all lowered indices:
\begin{align}                                           \nonumber
  \G_{\mu\nu\rho}=\G_{\mu\nu}{}^\s g_{\s\rho}&=
  \frac12(\pl_\mu g_{\nu\rho}+\pl_\nu g_{\rho\mu}-\pl_\rho g_{\mu\nu})
  +\frac12(T_{\mu\nu\rho} - T_{\nu\rho\mu} + T_{\rho\mu\nu} )
\\                                                      \label{elicon}
  &+\frac12(Q_{\mu\nu\rho} + Q_{\nu\rho\mu} - Q_{\rho\mu\nu} ).
\end{align}
The right hand side of this equality is symmetric in the indices $\mu$ and
$\nu$ except one term $T_{\mu\nu\rho}/2$, and this is in accord with the
definition of the torsion tensor (\ref{etorsi}). Thus, to define the affine
geometry on a manifold $\MM$, it is necessary and sufficient to define
three tensor fields: metric, torsion, and nonmetricity. We stress once again
that all three object can be specified in a completely arbitrary way, and they
can be considered different dynamical variables in models of mathematical
physics.

It is easy to compute the number of independent components of connection,
torsion, and nonmetricity tensors:
$$
  [\G_{\mu\nu}{}^\g]=m^3,~~~~~~
  [T_{\mu\nu}{}^\rho]=\frac{m^2(m-1)}2,~~~~~~
  [Q_{\mu\nu\rho}]=\frac{m^2(m+1)}2.
$$
This shows that the total number of independent components of the torsion
and nonmetricity tensors equals the number of components of the affine
connection.

We now consider particular cases of the affine geometry.

In the attempt to unite gravity and electromagnetism, H.~Weyl considered
the nonmetricity tensor of a special type \cite{Weyl18A}
\begin{equation}                                        \label{ewevec}
  Q_{\mu\nu\rho}=W_\mu g_{\nu\rho},
\end{equation}
where $W_\mu$ is the Weyl form identified with the electromagnetic potential
(here, torsion was assumed to be identically equal to zero). We say that the
{\em Riemann--Cartan--Weyl geometry} is defined on a manifold if there are
given a metric, torsion, and nonmetricity of special type (\ref{ewevec}).

If the nonmetricity tensor is identically zero, $Q_{\mu\nu\rho}=0$, but the
metric and torsion are nontrivial, then we say that the {\em Riemann--Cartan
geometry} is given on a manifold. As a consequence of (\ref{elicon}), the
affine connection is defined uniquely by the metric and torsion in this case:
\begin{equation}                                        \label{emecon}
  \G_{\mu\nu\rho}=
  \frac12(\pl_\mu g_{\nu\rho}+\pl_\nu g_{\mu\rho}-\pl_\rho g_{\mu\nu})
  +\frac12(T_{\mu\nu\rho} - T_{\nu\rho\mu} + T_{\rho\mu\nu} ).
\end{equation}
This connection is called metrical because the covariant derivative
of the metric is identically equal to zero:
\begin{equation}                                        \label{emetco}
  \nb_\mu g_{\nu\rho}=\pl_\mu g_{\nu\rho}-\G_{\mu\nu}{}^\s g_{\s\rho}-
                       \G_{\mu\rho}{}^\s g_{\nu\s} = 0.
\end{equation}
The metricity condition provides commutativity of covariant differentiation
and the raising and lowering of indices.

If the torsion tensor identically vanishes, $T_{\mu\nu}{}^\rho=0$, and
nonmetricity have special form (\ref{ewevec}), then we say that the
{\em Riemann--Weyl geometry} is given.

If both nonmetricity and torsion tensors are identically equal to zero,
$Q_{\mu\nu\rho}=0$, $T_{\mu\nu\rho}=0$, and the metric is nontrivial, then
we say that {\em Riemannian geometry} is given on a manifold. In this
case, the metrical connection is symmetric with respect to the first
two indices and uniquely defined by the metric
\begin{equation}                                        \label{etigam}
  \widetilde\G_{\mu\nu\rho}=\frac12(\pl_\mu g_{\nu\rho}
               +\pl_\nu g_{\mu\rho}-\pl_\rho g_{\mu\nu}).
\end{equation}
This connection is called the {\em Levy-Civita connection} or
{\em Christoffel symbols}.

We use the tilde to denote the geometrical objects constructed only for the
metric and zero torsion and nonmetricity, i.e., in a (pseudo-)Riemannian
geometry. The prefix pseudo- is used when the metric is not positive definite.

From the expression for Christoffel symbols (\ref{etigam}), we see that
\begin{equation}                                        \label{ederga}
  \pl_\mu g_{\nu\rho}
  =\widetilde\G_{\mu\nu\rho}+\widetilde\G_{\mu\rho\nu}.
\end{equation}
This means that for the Christoffel symbols to be zero in some coordinate
system it is necessary and sufficient that the metric components be
constant in these coordinates. In another coordinate system, they
may be nontrivial because Christoffel symbols are not
components of a tensor. For example, the Christoffel symbols for
Euclidean space are zero in the Cartesian coordinates but differ from
zero, e.g., in spherical or cylindrical coordinate systems.

In the case where the nonmetricity tensor and torsion are identically zero,
and there is a coordinate system in which the metric is equal to the
diagonal unit matrix in the neighborhood of every point, and, consequently,
the Christoffel symbols are zero, the geometry is called locally Euclidean.
The corresponding coordinate system is called {\em Cartesian}.

We say that a vector field $X=X^\mu\pl_\mu$ is transported parallel
along a curve $x^\mu(t)$, $t\in\MR$, if
\begin{equation*}
  \dot x^\nu\nb_\nu X^\mu=0,
\end{equation*}
where $\dot x^\mu=dx^\mu/dt$ is the tangent vector to the curve. Multiplying
this equation by $dt$ we obtain
\begin{equation*}
  \dl X^\mu=\dl x^\nu\pl_\nu X^\mu=-\dl x^\nu\G_{\nu\rho}{}^\mu X^\rho,
\end{equation*}
where $\dl x^\nu=dt\dot x^\nu$. We say that under the parallel transport
from a point $x^\nu$ to a neighboring point $x^\nu+\dl x^\nu$, the vector
$X^\mu$ acquires a differential $\dl X^\mu$, which is linear in $\dl x^\nu$
and the components of the vector field. In a similar way, we define
parallel transport of arbitrary tensor fields along a curve $x^\mu(t)$.
The result of parallel transport between two points depends in general
on a curve connecting these points. We note that parallel transport along a
curve is defined only by an affine connection and has no relation to a metric.

There are two types of distinguished curves $x^\mu(t)$ in an affine geometry:
geodesics and extremals. A geodesic line is a curve such that a vector
tangent to it remains tangent under parallel transport along the curve.
With the parameter $t$ along the curve chosen canonically, a geodesic
line is defined by the system of ordinary nonlinear differential equations
\begin{equation}                                        \label{egeoeq}
  \ddot x^\mu=-\G_{\nu\rho}{}^\mu\dot x^\nu\dot x^\rho.
\end{equation}
Although a geodesic line is defined only by the symmetric part of the affine
connection $\G_{\lbrace\mu\nu\rbrace}{}^\rho$, the latter depends nontrivially
on torsion and nonmetricity. Explicit expression (\ref{elicon}) yields
\begin{equation*}
  \G_{\lbrace\mu\nu\rbrace}{}^\rho=\widetilde\G_{\mu\nu}{}^\rho
  +\frac12(T^\rho{}_{\mu\nu}+T^\rho{}_{\nu\mu})
  +\frac12(Q_{\mu\nu}{}^\rho+Q_{\nu\mu}{}^\rho-Q^\rho{}_{\mu\nu}).
\end{equation*}

The second type of distinguished curves in an affine geometry are extremals
or lines of extremal length connecting two points. These lines are
exclusively defined by the metric. With the parameter along the curve chosen
canonically, extremals are defined by an equation similar to (\ref{egeoeq}),
\begin{equation}                                     \label{exteqr}
  \ddot x^\mu=-\widetilde\G_{\nu\rho}{}^\mu\dot x^\nu\dot x^\rho.
\end{equation}
However, we now have not a general affine connection on the right-hand side
but Christoffel symbols constructed only from the metric. We see that
geodesics and extremals are in general different curves on a manifold.
In a Riemann--Cartan geometry, geodesics and extremals coincide if and
only if the torsion tensor is antisymmetric in all three indices.
In Riemannian geometry, geodesics and extremals always coincide.

The primary role in differential geometry is played by the curvature tensor
of the affine connection, which arises in different contexts. In local
coordinates, it is defined by the expression
\begin{equation}                                        \label{ecurva}
  R_{\mu\nu\rho}{}^\s=\pl_\mu\G_{\nu\rho}{}^\s-\G_{\mu\rho}{}^\lm
                     \G_{\nu\lm}{}^\s-(\mu\leftrightarrow\nu),
\end{equation}
where the parenthesis $(\mu\leftrightarrow\nu)$ denote the previous terms with
the permuted indices $\mu$ and $\nu$. It can be easily verified that curvature
(\ref{ecurva}) is indeed a tensor field. We note that the curvature tensor
has no relation to a metric and is defined entirely by the connection.
In an affine geometry, the curvature tensor with all of its indices lowered
has no symmetry under permutations of indices except antisymmetry in the
first two indices.

Contraction of the curvature tensor in the two indices yields the Ricci
tensor $R_{\mu\nu}=R_{\mu\rho\nu}{}^\rho$, which is also defined entirely
by the connection. In Riemannian geometry, the Ricci tensor for the Levi-Civita
connection is symmetric with respect to the permutation of indices:
$\widetilde R_{\mu\nu}=\widetilde R_{\nu\mu}$. In a Riemann--Cartan geometry,
this symmetry is generally absent for a nonzero torsion tensor.

If a metric is also given on a manifold, then we can construct the scalar
curvature $R=R_{\mu\nu}g^{\mu\nu}$.

The affine connection is called locally trivial if in the neighborhood of
every point one can choose the coordinate system in which all components
of the connection are zero. Now we formulate two important theorems.
\begin{theorem}
For the local triviality of the affine connection it is necessary and
sufficient that its torsion and curvature tensors are equal to zero on $\MM$.
\end{theorem}
The proof of this theorem is reduced to an analysis of the transformation
rule for the affine connection components in (\ref{econts}). If in the new
coordinate system the symmetric part of the connection components is equal to
zero, then the system of differential equations for the transition functions
appears. The local integrability condition for this system of equations
is provided by the equality of the curvature tensor to zero.

A more thorough statement is proven in \cite{Wolf72}.
\begin{theorem}  \label{tlocev}
If for a given affine geometry the torsion, nonmetricity, and curvature tensors
are equal to zero on the whole manifold, then this manifold is isometric
either to the whole (pseudo-)Euclidean space $\MR^m$ or to the quotient space
$\MR^m/\MG$, where $\MG$ is a discreet transformation group acting freely.
\end{theorem}
In the last theorem, we suppose that both the metric and affine connection
are given on a manifold.

The Riemann--Cartan geometry defined by the metric and torsion provides the
basis for the geometric theory of defects. In the analysis of such models,
it is more convenient to introduce the Cartan variables: the vielbein and
$\MS\MO(m)$ connection instead of the metric and torsion. We assume here
that a metric on a manifold is positive definite. If a metric were not
positively definite then the $\MS\MO(p,q)$ connection would appear,
where $p+q=m$.

For a Riemannian metric, the orthonormal vielbein  $e_\mu{}^i(x)$,
$i=1,\dotsc,m$, is defined by the system of quadratic equations
\begin{equation}                                        \label{etetra}
  g_{\mu\nu}=e_\mu{}^ie_\nu{}^j\dl_{ij},
\end{equation}
where $\dl_{ij}$ is the Kronecker symbol. This relation uniquely defines
metric for a given vielbein. Conversely, system of equations (\ref{etetra})
defines the vielbein for a given metric up to local $\MS\MO(m)$ rotations.
We distinguish Greek and Latin indices because different transformation groups
act on them. From definition (\ref{etetra}), we see
that $\det e_\mu{}^i\ne0$. Components of the inverse vielbein $e^\mu{}_i$,
$e^\mu{}_i e_\nu{}^i=\dl^\mu_\nu$, can be considered components of
$m$ orthonormal vector fields $e_i=e^\mu{}_i\pl_\mu$ on a manifold $\MM$
with respect to the metric $g$
\begin{equation*}
  (e_i,e_j)=e^\mu{}_i e^\nu{}_j g_{\mu\nu}=\dl_{ij}.
\end{equation*}
Components of the vielbein $e_\mu{}^i$ define $m$ orthonormal 1-forms
$e^i=dx^\mu e_\mu{}^i$ on a manifold.

It is known that any manifold $\MM$ can be equipped with a Riemannian
metric (see, for e.g., \cite{ChChLa00}). At the same time, the global
existence of a vielbein provides, in particular, an orientation of the
manifold $\MM$. Thus, a vielbein may exist globally only on orientable
manifolds. There are also other topological restrictions for the global
existence of a vielbein that we do not discuss here.

The coordinate basis $\pl_\mu$ of the tangent space in every point of $\MM$
is called holonomic. Components of tensors of an arbitrary type may also be
considered with respect to the unholonomic basis $e_i$ of tangent spaces
defined by the vielbein. For example, a vector field has the components
\begin{equation*}
  X=X^\mu\pl_\mu=X^i e_i,~~~~X^i=X^\mu e_\mu{}^i.
\end{equation*}
We always assume that the transformation of Greek indices into Latin
ones and vice versa is performed with the help of the vielbein.

We now define the $\MS\MO(m)$ connection $\om_{\mu i}{}^j(x)$. This is
the name of the connection of the principal fiber bundle with the
structure Lie group $\MS\MO(m)$ and the base $\MM$. If the Riemann-Cartan
geometry is given on $\MM$, and the vielbein is defined, then we can
define the $\MS\MO(m)$ connection by the relation
\begin{equation}                                        \label{etecov}
  \nb_\mu e_\nu{}^i=\pl_\mu e_\nu{}^i-\G_{\mu\nu}{}^\rho e_\rho{}^i
                    +e_\nu{}^j\om_{\mu j}{}^i=0.
\end{equation}
We see that under coordinate changes, the components of $\om_{\mu j}{}^i$
transform as a covector field in the Greek index. Thus, they define a
1-form on $\MM$. For a given vielbein, this relation provides a
one-to-one correspondence between components of the connections
$\G_{\mu\nu}{}^\rho$ and $\om_{\mu j}{}^i$. The $\MS\MO(m)$ connection
defines covariant derivatives for components of tensor fields relative
to an unholonomic basis. For example,
\begin{equation}                                           \label{ecodve}
  \nb_\mu X^i=\pl_\mu X^i+X^j\om_{\mu j}{}^i,~~~~
  \nb_\mu X_i=\pl_\mu X_i-\om_{\mu i}{}^j X_j.
\end{equation}
The covariant derivative is naturally defined for a tensor field having
both Greek and Latin indices. Taking the covariant derivative of Eqn
(\ref{etetra}) leads to the antisymmetry of the components
$\om_\mu{}^{ij}=-\om_\mu{}^{ji}$. This means that the
1-form $dx^\mu\om_{\mu j}{}^i$ takes values in the Lie algebra $\Gs\Go(m)$,
and this indeed corresponds to the $\MS\MO(m)$ connection.

In the general case of an affine geometry, the Cartan variables can also
be defined by relations (\ref{etetra}) and (\ref{etecov}). For this,
we have to replace the Kronecker symbol on the right-hand side of
(\ref{etetra}) with an arbitrary nondegenerate symmetric matrix $\eta_{ij}$.
In this case, relation (\ref{etecov}) defines a linear $\MG\ML(m,\MR)$
connection. In the Riemann--Cartan geometry with the Lorentzian-signature
metric, we would have the Lorentz $\MS\MO(1,m-1)$ connection.

We consider local rotations with a matrix $S_j{}^i(x)\in\MS\MO(m)$ to
show that the components $\om_{\mu j}{}^i$ do define an
$\MS\MO(m)$ connection in the Riemann--Cartan geometry. By definition,
the components of vector fields and 1-forms transform under local
rotations according to the rule
\begin{equation}                                        \label{elotrv}
  X^{\prime i}=X^j S_j{}^i,~~~~X^{\prime}_i=S^{-1}{}_i{}^j X_j.
\end{equation}
In order that covariant derivatives (\ref{ecodve}) have the tensor
transformation rule under local rotations, it is necessary and sufficient
that the components of the $\MS\MO(m)$ connection transform according to
the rule
\begin{equation}                                        \label{elocta}
  \om^\prime_{\mu i}{}^j=S^{-1}{}_i{}^k\om_{\mu k}{}^lS_l{}^j
  +\pl_\mu S^{-1}{}_i{}^kS_k{}^j.
\end{equation}
This is the transformation law for the $\MS\MO(m)$ connection, indeed.
The same transformation law follows from definition (\ref{etecov}) if the
vielbein $e_\mu{}^i$ transforms as a vector with respect to the index $i$,
and the Christoffel symbols remain unchanged.

Of course, we can introduce the metrical and $\MS\MO(m)$ connections on
$\MM$ in an independent way. If we require afterwards that the
$\MS\MO(m)$ connection acts on tensor field components relative to
an unholonomic basis defined by the vielbein, we obtain a one-to-one
correspondence between the components of the connections (\ref{etecov}).

We now express the components of the metric connection
$\G_{\mu\nu}{}^\rho$ through the vielbein $e_\mu{}^i$ and
$\MS\MO(m)$ connection $\om_{\mu j}{}^i$ with the help of relation
(\ref{etecov}) and substitute them into the definitions of torsion
(\ref{etorsi}) and curvature (\ref{ecurva}). As a result, we obtain
expressions for torsion and curvature in terms of the Cartan variables
\begin{align}                                           \label{ecurcv}
  T_{\mu\nu}{}^i&=\pl_\mu e_\nu{}^i-e_\mu{}^j\om_{\nu j}{}^i
                    -(\mu\leftrightarrow\nu),
\\                                                      \label{ecucav}
  R_{\mu\nu j}{}^i&=\pl_\mu \om_{\nu j}{}^i-\om_{\mu j}{}^k
                       \om_{\nu k}{}^i-(\mu\leftrightarrow\nu),
\end{align}
where
$$
    T_{\mu\nu}{}^i=T_{\mu\nu}{}^\rho e_\rho{}^i,~~~~
    R_{\mu\nu j}{}^i=R_{\mu\nu\rho}{}^\s e^\rho{}_j e_\s{}^i.
$$
Torsion (\ref{ecurcv}) and curvature (\ref{ecucav}) are 2-forms on the
manifold $\MM$ with the values in the vector space and Lie algebra
$\Gs\Go(m)$, respectively.

An $\MS\MO(m)$ connection $\om_{\mu j}{}^i$ is called locally trivial if
every point has a neighborhood containing this point such
that it has the form
\begin{equation}                                        \label{eloctr}
  \om_{\mu j}{}^i=\pl_\mu S^{-1}{}_j{}^kS_k{}^i.
\end{equation}
Obviously, after local rotation with the matrix $S^{-1}{}_j{}^i$,
all components of the connection become zero. This connection is also called
a pure gauge. One can easily verify that the curvature of a locally trivial
connection identically vanishes. The inverse statement is also valid.
\begin{theorem}                                    \label{tlotrs}
An $\MS\MO(m)$ connection is locally trivial if and only if its curvature
form is identically zero on $\MM$.
\end{theorem}
This theorem is valid for any structure Lie group. The proof is reduced to an
analysis of transformation rule (\ref{elocta}). If the left-hand side of
this equation is zero, then we have a system of equations for the field
$S_j{}^i(x)$ for which vanishing of curvature tensor is the local
integrability condition.

The space with zero curvature tensor $R_{\mu\nu j}{}^i=0$ is called the space
of absolute parallelism because the parallel displacement of a vector does
not depend on the path connecting two fixed points of the manifold.

We can perform a local $\MS\MO(m)$ rotation for the locally trivial
$\MS\MO(m)$ connection such that it becomes zero: $\om_{\mu j}{}^i=0$. Then,
the zero-torsion equality becomes
\begin{equation*}
  \pl_\mu e_\nu{}^i-\pl_\nu e_\mu{}^i=0.
\end{equation*}
This equation is the local integrability condition for the system of equations
\begin{equation}                                           \label{erepad}
  \pl_\mu y^i =e_\mu{}^i
\end{equation}
for $m$ functions $y^i(x)$. A solution of this system of equations yields the
transition functions to Cartesian coordinates. Thus, the equalities of
curvature and torsion tensors to zero are the necessary and sufficient
conditions for the existence of the fields $S_j{}^i(x)$ and $y^i(x)$, i.e.,
the existence of such a local rotation and coordinate system where
the connection vanishes and the metric becomes Euclidean. We note that
equality of torsion tensor to zero alone is not enough for the existence
of a Cartesian coordinate system.

Three-dimensional space is considered in the geometric theory of defects.
We make two remarks concerning this. In lower dimensions, the algebraic
structure of the curvature tensor with all lowered indexes becomes much simpler.
In two dimensions the full curvature tensor in the Riemann--Cartan geometry
is in the one-to-one correspondence with its scalar curvature:
\begin{equation*}
  R_{ijkl}=\frac12(\dl_{ik}\dl_{jl}-\dl_{il}\dl_{jk})R.
\end{equation*}
In three dimensional space, the full curvature tensor is in the one-to-one
correspondence with its Ricci tensor
\begin{equation}                                     \label{ecurir}
  R_{ijkl}=\dl_{ik}R_{jl}-\dl_{il}R_{jk}-\dl_{jk}R_{il}+\dl_{jl}R_{ik}
  -\frac12(\dl_{ik}\dl_{jl}-\dl_{il}\dl_{jk})R.
\end{equation}
These formulas are also correct for nonzero torsion.

Concluding this appendix, we write the identity that is valid in the
Riemann--Cartan geometry in an arbitrary number of dimensions:
\begin{equation}                                        \label{eidcur}
  R(e,\om)+\frac14T_{ijk}T^{ijk}-\frac12T_{ijk}T^{kij}
  -T_iT^i-\frac2{e}\pl_\mu(eT^\mu)=\widetilde R(e),~~~~e=\det e_\mu{}^i.
\end{equation}
The Riemannian scalar curvature on the right-hand side of this equality
is constructed only from the vielbein for zero torsion.


\begin{thebibliography}{10}

\bibitem{Kondo52}
K.~Kondo.
\newblock On the geometrical and physical foundations of the theory of
  yielding.
\newblock In {\em Proc. 2nd Japan Nat. Congr. Applied Mechanics}, pages~41--47,
  Tokyo, 1952.

\bibitem{Nye53}
J.~F. Nye.
\newblock Some geometrical relations in dislocated media.
\newblock {\em Acta Metallurgica}, 1:153, 1953.

\bibitem{BiBuSm55}
B.~A. Bilby, R.~Bullough, and E.~Smith.
\newblock Continuous distributions of dislocations: a new application of the
  methods of non-{R}iemannian geometry.
\newblock {\em Proc. Roy. Soc. London}, A231:263--273, 1955.

\bibitem{Kroner58}
E.~Kr\"oner.
\newblock {\em Kontinums Theories der Versetzungen und Eigenspanungen}.
\newblock Spriger--Verlag, Berlin -- Heidelberg, 1958.

\bibitem{SedBer67}
L.~I. Sedov and V.~L. Berditchevski.
\newblock A dynamical theory of dislocations.
\newblock In E.~Kr\"oner, editor, {\em Mechanics of Generalized Continua,
  {UITAM} symposium}, pages 214--238, 1967.

\bibitem{Kleman80A}
M.~Kl\'eman.
\newblock The general theory of dislocations.
\newblock In Nabarro F.~R. N., editor, {\em Dislocations In Solids, Vol.~5},
  pages 243--297, Amsterdam, 1980. North-Holland Publishing Company.

\bibitem{Kroner81}
E.~Kr\"oner.
\newblock Continuum theory of defects.
\newblock In R.~Balian et~al., editor, {\em Less Houches, Session XXXV, 1980 --
  Physics of Defects}, pages 282--315. North-Holland Publishing Company, 1981.

\bibitem{DzyVol88}
I.~E. Dzyaloshinskii and G.~E. Volovik.
\newblock {\em Ann.\ Phys.}, 125:67, 1988.

\bibitem{KadEde83}
A.~Kadi\'c and D.~G.~B. Edelen.
\newblock {\em A gauge theory of dislocations and disclinations}.
\newblock Springer--Verlag, Berlin -- Heidelberg, 1983.

\bibitem{KunKun86}
I.~A. Kunin and B.~I. Kunin.
\newblock Gauge theories in mechanics.
\newblock In {\em Trends in Application of Pure Mathematics to Mechanics.
  Lecture Notes in Physics, V.249.}, pages 246--249, Berlin -- Heidelberg,
  1986. Springer--Verlag.

\bibitem{Kleine89}
H.~Kleinert.
\newblock {\em Gauge fields in condenced matter}, volume~2.
\newblock World Scientific, Singapore, 1990.

\bibitem{HeMcMiNe95}
F.~W. Hehl, J.~D. McCrea, E.~W. Mielke, and Y.~Ne'eman.
\newblock Metric-affine gauge theory of gravity: Field equations, {N}oether
  identities, world spinors, and breaking of dilaton invariance.
\newblock {\em Phys.\ Rep.}, 258(1\&2):1--171, 1995.

\bibitem{Cartan22}
E.~Cartan.
\newblock Sur une generalisation de la notion de courburu de {R}iemann et les
  aspases a torsion.
\newblock {\em Compt.\ Rend.\ Acad.\ Sci.\ (Paris)}, 174:593--595, 1922.

\bibitem{Malysh00}
C.~Malyshev.
\newblock The {$T(3)$}-gauge model, the {E}instein-like gauge equation, and
  {V}olterra dislocations with modified asymptotics.
\newblock {\em Ann.\ Phys.}, 286:249--277, 2000.

\bibitem{Lazar00}
M.~Lazar.
\newblock Dislocation theory as a 3-dimensional translation gauge theory.
\newblock {\em Ann.\ Phys.\ (Leipzig)}, 9:461--473, 2000.

\bibitem{Lazar02}
M.~Lazar.
\newblock An elastoplastic theory of dislocations as a physical field theory
  with torsion.
\newblock {\em J.\ Phys.\ A}, 35:1983--2004, 2002.

\bibitem{Lazar03}
M.~Lazar.
\newblock A nonsingular solution of the edge dislocation in the gauge theory of
  dislocations.
\newblock {\em J.\ Phys.\ A}, 35:1983--2004, 2002.

\bibitem{Frank58}
F.~C. Frank.
\newblock On the theory of liquid crystals.
\newblock {\em Discussions Farad.\ Soc.}, 25:19--28, 1958.

\bibitem{DzyVol78}
I.~E. Dzyaloshinskii and G.~E. Volovik.
\newblock On the concept of local invariance in the theory of spin glasses.
\newblock {\em J. Physique}, 39(6):693--700, 1978.

\bibitem{Hertz78}
J.~A. Hertz.
\newblock Gauge model for spin-glasses.
\newblock {\em Phys.\ Rev.}, B18(9):4875--4885, 1978.

\bibitem{RivDuf82}
N.~Rivier and D.~M. Duffy.
\newblock Line defects and tunneling modes in glasses.
\newblock {\em J.\ Physique}, 43(2):293--306, 1982.

\bibitem{KatVol92}
M.~O. Katanaev and I.~V. Volovich.
\newblock Theory of defects in solids and three-dimensional gravity.
\newblock {\em Ann.\ Phys.}, 216(1):1--28, 1992.

\bibitem{Katana03}
M.~O. Katanaev.
\newblock Wedge dislocation in the geometric theory of defects.
\newblock {\em Theor.\ Math.\ Phys.}, 135(2):733--744, 2003.

\bibitem{Katana04}
M.~O. Katanaev.
\newblock One-dimensional topologically nontrivial solutions in the {S}kyrme
  model.
\newblock {\em Theor.\ Math.\ Phys.}, 138(2):163--176, 2004.

\bibitem{FurMor94}
C.~Furtado and F.~Moraes.
\newblock On the binding of electrons and holes to disclinations.
\newblock {\em Phys.\ Lett.}, A188:394--396, 1994.

\bibitem{FudaMoBeBe94}
C.~Furtado, B.~G.~C. da~Cunha, F.~Moraes, E.~R. Bezerra~de Mello, and V.~B.
  Bezzerra.
\newblock Landau levels in the presence of disclinations.
\newblock {\em Phys.\ Lett.}, A195:90--94, 1994.

\bibitem{Moraes95B}
F.~Moraes.
\newblock Enhancement of the magnetic moment of the electron due to a
  topological defect.
\newblock {\em Mod.\ Phys.\ Lett.}, A10(31):2335--2338, 1995.

\bibitem{AzeMor98}
S.~Azevedo and F.~Moraes.
\newblock Topological {A}haronov--{B}ohm effect around a disclination.
\newblock {\em Phys.\ Lett.\ A}, 246:374--376, 1998.

\bibitem{BaJoMoMo98}
A.~P. Balachandran, V.~John, A.~Momen, and F.~Moraes.
\newblock Anomalous defects and their quantized transverse conductivities.
\newblock {\em Int.\ J.\ Mod.\ Phys.\ A}, 13(5):841--861, 1998.

\bibitem{FurMor99}
C.~Furtado and F.~Moraes.
\newblock Landau levels in the presence of a screw dislocation.
\newblock {\em Europhys.\ Lett.}, 45(3):279--282, 1999.

\bibitem{BaScTu98}
R.~Bausch, R.~Schmitz, and L.~A. Turski.
\newblock Single-particle quantum states in a crystal with topological defects.
\newblock {\em Phys.\ Rev.\ Lett.}, 80(11):2257--2260, 1998.

\bibitem{BaScTu99}
R.~Bausch, R.~Schmitz, and L.~A. Turski.
\newblock Quantum motion of electrons in topologically distorted crystal.
\newblock {\em Ann.\ Phys.\ (Leipzig)}, 8(3):181--189, 1999.

\bibitem{FurMor00}
C.~Furtado and F.~Moraes.
\newblock Harmonic oscillator interacting with conical singularities.
\newblock {\em J.\ Phys.\ A}, 33:5513--5519, 2000.

\bibitem{AzeMor00A}
S.~Azevedo and F.~Moraes.
\newblock Two-dimensional scattering by disclinations in monolayer graphite.
\newblock {\em J.\ Phys.\ C}, 12:7421--7424, 2000.

\bibitem{AzePer00}
S.~Azevedo and J.~Pereira.
\newblock Double {A}haronov--{B}ohm effect in a medium with a disclination.
\newblock {\em Phys.\ Lett.\ A}, 275:463--466, 2000.

\bibitem{FuBeMo00B}
C.~Furtado, V.~B. Bezerra, and F.~Moraes.
\newblock Berry's quantum phase in media with dislocations.
\newblock {\em Europhys.\ Lett.}, 52(1):1--7, 2000.

\bibitem{deFuMo01}
C.~A. de~Lima~Ribeiro, C.~Furtado, and F.~Moraes.
\newblock On the localization of electrons and holes by a disclination core.
\newblock {\em Phys.\ Lett.\ A}, 288:329--334, 2001.

\bibitem{VieAze01}
S.~R. Vieira and S.~Azevedo.
\newblock Double {A}haronov--{B}ohm effect in a medium with a linear
  topological defect.
\newblock {\em Phys.\ Lett.}, 288:29--32, 2001.

\bibitem{Azeved01B}
S.~Azevedo.
\newblock A charged particle with magnetic moment in a medium with a
  disclination.
\newblock {\em J.\ Phys.\ A}, 34:6081--6085, 2001.

\bibitem{FuBeMo01}
C.~Furtado, V.~B. Bezerra, and F.~Moraes.
\newblock Quantum scattering by a magnetic flux screw dislocation.
\newblock {\em Phys.\ Lett.}, A289:160--166, 2001.

\bibitem{deFuBeMo01}
G.~de~A~Marques, C.~Furtado, V.~B. Bezerra, and F.~Moraes.
\newblock Landau levels in the presence of topological defects.
\newblock {\em J.\ Phys.\ A}, 34:5945--5954, 2001.

\bibitem{FudeAz02}
C.~Furtado, C.~A. de~Lima~Ribeiro, and S.~Azevedo.
\newblock Aharonov--{B}ohm effect in the presence of a density of defects.
\newblock {\em Phys.\ Lett.}, A296:171--175, 2002.

\bibitem{Azeved02A}
S.~Azevedo.
\newblock Topological {A}haronov--{B}ohm effect in a two-dimensional harmonic
  oscillator.
\newblock {\em Phys.\ Lett.}, A293:283--286, 2002.

\bibitem{Azeved02B}
S.~Azevedo.
\newblock Bound particle with magnetic moment in a space with topological
  defect.
\newblock {\em Int.\ J.\ Quantum Chem.}, 90:1596--1599, 2002.

\bibitem{Azeved02C}
S.~Azevedo.
\newblock Bound charge moving in a magnetic field in a space with a topological
  defect.
\newblock {\em Mod.\ Phys.\ Lett.}, A17:1263--1268, 2002.

\bibitem{Azeved03}
S.~Azevedo.
\newblock Charged particle with magnetic moment in the background of line
  topological defect.
\newblock {\em Phys.\ Lett.}, A307:65--68, 2003.

\bibitem{DeBeBe99}
A.~De~S.~Barbosa, E.~R. Bezerra De~Mello, and V.~B. Bezerra.
\newblock Coulomb interaction between two charged particles on a cone.
\newblock {\em Int.\ J.\ Mod.\ Phys.\ A}, 14(22):3565--3580, 1999.

\bibitem{AzFuMo98}
S.~Azevedo, C.~Furtado, and F.~Moraes.
\newblock Charge localization around disclinations in monolayer graphite.
\newblock {\em phys.\ stat.\ sol.\ (b)}, 207:387--392, 1998.

\bibitem{Moraes95A}
F.~Moraes.
\newblock Casimir effect around disclination.
\newblock {\em Phys.\ Lett.}, A204:399--404, 1995.

\bibitem{BeBeGr98}
E.~R. Bezerra~de Mello, V.~B. Bezerra, and Yu.~V. Grats.
\newblock Self-forces in the spacetime of multiple cosmic strings.
\newblock {\em Class.\ Quantum Grav.}, 15:1915--1925, 1998.

\bibitem{Azeved01A}
S.~Azevedo.
\newblock Harmonic oscillator in a space with a linear topological defect.
\newblock {\em Phys.\ Lett.}, A288:33--36, 2001.

\bibitem{FuBeMo00}
C.~Furtado, V.~B. Bezerra, and F.~Moraes.
\newblock Aharonov--{B}ohm effect for bound states in {K}aluza--{K}lein theory.
\newblock {\em Mod.\ Phys.\ Lett.}, A15(4):253--258, 2000.

\bibitem{Tanaka00}
I.~Tanaka.
\newblock Geometrical aspect of pinning in superconducting material.
\newblock {\em Phys.\ Lett.\ A}, 277:262--266, 2000.

\bibitem{DeLMor02}
V.~A. De~Lorenci and E.~S.~Jr. Moreira.
\newblock Classical self-forces in a space with a topological defect.
\newblock {\em Phys.\ Rev.}, D65:085013, 2002.

\bibitem{AzeMor00B}
S.~Azevedo and F.~Moraes.
\newblock Self-force on a point charge and linear source in the space of a
  screw dislocation.
\newblock {\em Phys.\ Lett.}, 267:208--211, 2000.

\bibitem{MirMor03}
J.~A. Miranda and F.~Moraes.
\newblock Geometric approach to viscous fingering on a cone.
\newblock {\em J.\ Phys.\ A}, 36:863--874, 2003.

\bibitem{Moraes96}
F.~Moraes.
\newblock Geodesics around a dislocation.
\newblock {\em Phys.\ Lett.}, A214(3,4):189--192, 1996.

\bibitem{dePaMo98}
A.~de~Padua, F.~Parisio-Filho, and F.~Moraes.
\newblock Geodesics around line defects in elastic solids.
\newblock {\em Phys.\ Lett.\ A}, 238:153--158, 1998.

\bibitem{KatVol99}
M.~O. Katanaev and I.~V. Volovich.
\newblock Scattering on dislocations and cosmic strings in the geometric theory
  of defects.
\newblock {\em Ann.\ Phys.}, 271:203--232, 1999.

\bibitem{MirRiv02}
M.~F. Miri and N.~Rivier.
\newblock Continuum elasticity with topological defects, including dislocations
  and extra-matter.
\newblock {\em J.\ Phys.\ A}, 35:1727--1739, 2002.

\bibitem{LanLif70}
L.~D. Landau and E.~M. Lifshits.
\newblock {\em Theory of Elasticity}.
\newblock Pergamon, Oxford, 1970.

\bibitem{Kosevi81}
A.~M. Kosevich.
\newblock {\em Physical mechanics of real crystals}.
\newblock Naukova dumka, Kiev, 1981.
\newblock [in Russian].

\bibitem{Vercin90}
A.~Ver\c{c}in.
\newblock Metric--torsion gauge theory of continuum line defects.
\newblock {\em Int.\ J.\ Theor.\ Phys.}, 29(1):7--21, 1990.

\bibitem{ZaMaNoPi80E}
V.~E. Zakharov, S.~V. Manakov, S.~P. Novikov, and L.~P. Pitaevskii.
\newblock {\em The Inverse Scattering Method}.
\newblock Nauka, Moscow, 1980.
\newblock [in Russian]; English transl.: S.~P.~Novikov, S.~V.~Manakov,
  L.~P.~Pitaevskii, and V.~E.~Zakharov {\it The Inverse Scattering Method}
  Plenum, New York (1984).

\bibitem{Rajara82}
R.~Rajaraman.
\newblock {\em Solitons and Instantons in Quantum Field Theory}.
\newblock North--Holland, Amsterdam, 1982.

\bibitem{TakFad86E}
L.~A. Takhtadzhyan and L.~D. Faddeev.
\newblock {\em The Hamiltonian Methods in the Theory of Solitons}.
\newblock Nauka, Moscow, 1986.
\newblock [in Russian]; English transl.: L.~D.~Faddeev and L.~A.~Takhtajan {\it
  The Hamiltonian Methods in the Theory of solitons,} Berlin, Springer (1987).

\bibitem{Zakrze89}
W.~J. Zakrzewski.
\newblock {\em Low Dimensional Sigma Models}.
\newblock Adam Hilger, Bristol -- Philadelphia, 1989.

\bibitem{RybSan01E}
Yu.~P. Rybakov and V.~I. Sanyuk.
\newblock {\em Multidimensional Solitons}.
\newblock Russian Univ. of People's Friendship Publ., Moscow, 2001.
\newblock [in Russian].

\bibitem{Skyrme61}
T.~H.~R. Skyrme.
\newblock Nonlinear field theory.
\newblock {\em Proc.\ Roy.\ Soc. London}, A260:127--138, 1961.

\bibitem{LanLif82}
L.~D. Landau and E.~M. Lifshits.
\newblock {\em Electrodynamics of continuous media}.
\newblock Nauka, Moscow, second edition, 1982.
\newblock [in Russian].

\bibitem{Faddee77}
L.~D. Faddeev.
\newblock In search of multidimensional solitons.
\newblock In {\em Nonlocal. Nonlinear, and Nonrenormalizable Field Theories},
  pages 207--223, Dubna, 1977. Joint Inst.\ Nucl.\ Res.
\newblock [in Russian].

\bibitem{CosCos09}
E.~Cosserat and F.~Cosserat.
\newblock {\em Th\'eorie des corps d\'eformables}.
\newblock Hermann, Paris, 1909.

\bibitem{Nowack85}
W.~Nowacki.
\newblock {\em Theory of asymmetric elasticity}.
\newblock Pergamon Press, Oxford, 1985.

\bibitem{Sandru66}
N.~Sandru.
\newblock On some problems of the linear theory of the asymmetric elasticity.
\newblock {\em Int.\ J.\ Eng.\ Sci.}, 4(1), 1966.

\bibitem{Starus63}
A.~Staruszkiewicz.
\newblock Gravitational theory in three-dimensional space.
\newblock {\em Acta Phys.\ Polon.}, 24(6(12)):735--740, 1963.

\bibitem{Clemen76}
G.~Clement.
\newblock Field--theoretic particles in two space dimensions.
\newblock {\em Nucl.\ Phys.\ B}, 114:437--448, 1976.

\bibitem{DeJatH84}
S.~Deser, R.~Jackiw, and G.~'t~Hooft.
\newblock Three-dimensional {E}instein gravity: Dynamics of flat space.
\newblock {\em Ann.\ Phys.}, 152(1):220--235, 1984.

\bibitem{Holz88}
A.~Holz.
\newblock Geometry and action of arrays of disclinations in crystals and
  relation to (2+1)-dimensional gravitation.
\newblock {\em Class.\ Quantum Grav.}, 5:1259--1282, 1988.

\bibitem{Kohler95A}
C.~Kohler.
\newblock Point particles in 2+1-dimensional gravity as defects in solid
  continua.
\newblock {\em Class.\ Quantum Grav.}, 12:L11--L15, 1995.

\bibitem{Kohler95B}
C.~Kohler.
\newblock Line defects in solid continua and point particles in
  $(2+1)$-dimensional gravity.
\newblock {\em Class.\ Quantum Grav.}, 12:2977--2993, 1995.

\bibitem{LanLif62}
L.~D.~Landau and E.~M.~Lifshitz.
\newblock {\em The Classical Theory of Fields}.
\newblock Pergamon, New York, 1962.

\bibitem{GuiSte77}
V.~Guillemin and S.~Sternberg.
\newblock {\em Geometric Asymptotics}.
\newblock American Mathematical Society, Providence, Phode Island, 1977.

\bibitem{JaEmLo60}
E.~Janke, F.~Emde, and F.~L\"osch.
\newblock {\em Tafeln H\"oherer Funktionen}.
\newblock B. G. Teubner Verlagsgesellschaft, Stuttgart, sechste edition, 1960.

\bibitem{AhaBoh59}
Y.~Aharonov and D.~Bohm.
\newblock Significance of electromagnetic potentials in the quantum theory.
\newblock {\em Phys.\ Rev.}, 115(3):485--491, 1959.

\bibitem{Feinbe62}
E.~L.~Feinberg.
\newblock On a ``special role'' of electromagnetic potentials in quantum
          mechanics.
\newblock {\em UFN}, 78(No.1):53--64, 1962.
\newblock [in Russian.]

\bibitem{Skarzh81}
V.~D.~Skarzhinskii.
\newblock The Aharonov--Bohm effect: theoretical calculations and interpolation.
\newblock {\em FIAN Proc.}, 167:139--161, 1986.
\newblock [in Russian.]

\bibitem{Katana97}
M.~O. Katanaev.
\newblock Euclidean two-dimensional gravity with torsion.
\newblock {\em J.\ Math.\ Phys.}, 38(3):946--980, 1997.

\bibitem{VolKat86}
I.~V. Volovich and M.~O. Katanaev.
\newblock Quantum strings with a dynamical geometry.
\newblock {\em JETP Lett.}, 43(5):267--269, 1986.

\bibitem{KatVol86}
M.~O. Katanaev and I.~V. Volovich.
\newblock String model with dynamical geometry and torsion.
\newblock {\em Phys.\ Lett.}, 175B(4):413--416, 1986.

\bibitem{KatVol90}
M.~O. Katanaev and I.~V. Volovich.
\newblock Two-dimensional gravity with dynamical torsion and strings.
\newblock {\em Ann.\ Phys.}, 197(1):1--32, 1990.

\bibitem{Katana89B}
M.~O. Katanaev.
\newblock New integrable model -- two-dimensional gravity with dynamical
  torsion.
\newblock {\em Sov.\ Phys.\ Dokl.}, 34(3):982--984, 1989.

\bibitem{Katana90}
M.~O. Katanaev.
\newblock Complete integrability of two-dimensional gravity with dynamical
  torsion.
\newblock {\em J.\ Math.\ Phys.}, 31(4):882--891, 1990.

\bibitem{Katana91}
M.~O. Katanaev.
\newblock Conformal invariance, extremals, and geodesics in two-dimensional
  gravity with torsion.
\newblock {\em J.\ Math.\ Phys.}, 32(9):2483--2496, 1991.

\bibitem{Katana93A}
M.~O. Katanaev.
\newblock All universal coverings of two-dimensional gravity with torsion.
\newblock {\em J.\ Math.\ Phys.}, 34(2):700--736, 1993.

\bibitem{VilShe94}
A.~Vilenkin and E.~Shellard.
\newblock {\em Cosmic Strings and Other Topological Defects}.
\newblock Cambridge University Press, Cambridge, 1994.

\bibitem{HinKib95}
M.~B. Hindmarsh and T.~W.~B. Kibble.
\newblock Cosmic strings.
\newblock {\em Rep.\ Prog.\ Phys.}, 58:477, 1995.

\bibitem{DuNoFo98E}
B.~A.~Dubrovin, S.~P.~Novikov, A.~T.~Fomenko.
\newblock {\em Modern geometry: Methods and Applications}.
\newblock Nauka, Moscow, 1998, fourth edition. [In Russian];
English transl.\ prev.\ ed.: B.~A.~Dubrovin,
              A.~T.~Fomenko, and S.~P.~Novikov {\it Modern Geometry: Methods
              and Applications,} Part 1, {\it The Geometry of Surfaces,
              Transformation Groups, and Fields,} Springer, New York (1992).

\bibitem{Weyl18A}
H.~Weyl.
\newblock Gravitation und {E}lektrizit\"at.
\newblock {\em Sitz.\ Preuss.\ Akad.\ Wiss.}, page 465, 1918.

\bibitem{Wolf72}
J.~A. Wolf.
\newblock {\em Spaces of constant curvature}.
\newblock University of California, Berkley, California, 1972.

\bibitem{ChChLa00}
S.~S. Chern, W.~H. Chen, and K.~S. Lam.
\newblock {\em Lectures on Differential Geometry}.
\newblock World Scientific, Singapore, 2000.

\end{thebibliography}
\end{document}